\documentclass[11pt, a4paper]{article}
\pdfoutput=1
\usepackage{jcappub}

\usepackage{booktabs}
\usepackage{multirow}
\usepackage{amssymb}
\usepackage[export]{adjustbox}
\usepackage{floatrow}
\newfloatcommand{capbtabbox}{table}[][3.5in]
\newfloatcommand{capbfigbox}{figure}[][2.3in]
\usepackage{blindtext}
\usepackage{amsmath}
\usepackage{booktabs}
\usepackage{aas_macros}

\def\lsim{\mathrel{\raise.3ex\hbox{$<$\kern-.75em\lower1ex\hbox{$\sim$}}}}
\def\gsim{\mathrel{\raise.3ex\hbox{$>$\kern-.75em\lower1ex\hbox{$\sim$}}}}

\begin{document}

\hspace*{110mm}{\large \tt FERMILAB-PUB-18-627-A}
\vskip 0.2in

\title{Constraints on Decaying Dark Matter from the Isotropic Gamma-Ray Background}

\author{Carlos Blanco$^{a,b}$}\note{ORCID: http://orcid.org/0000-0001-8971-834X}
\emailAdd{carlosblanco2718@uchicago.edu}
\author{and Dan Hooper$^{b,c,d}$}\note{ORCID: http://orcid.org/0000-0001-8837-4127}
\emailAdd{dhooper@fnal.gov}

\affiliation[a]{University of Chicago, Department of Physics, Chicago, IL 60637}
\affiliation[b]{University of Chicago, Kavli Institute for Cosmological Physics, Chicago, IL 60637}
\affiliation[c]{Fermi National Accelerator Laboratory, Center for Particle Astrophysics, Batavia, IL 60510}
\affiliation[d]{University of Chicago, Department of Astronomy and Astrophysics, Chicago, IL 60637}

\abstract{If the dark matter is unstable, the decay of these particles throughout the universe and in the halo of the Milky Way could contribute significantly to the isotropic gamma-ray background (IGRB) as measured by Fermi. In this article, we calculate the high-latitude gamma-ray flux resulting from dark matter decay for a wide range of channels and masses, including all contributions from inverse Compton scattering and accounting for the production and full evolution of cosmological electromagnetic cascades. We also make use of recent multi-wavelength analyses that constrain the astrophysical contributions to the IGRB, enabling us to more strongly restrict the presence any component arising from decaying dark matter. Over a wide range of decay channels and masses (from GeV to EeV and above), we derive stringent lower limits on the dark matter's lifetime, generally in the range of $\tau \sim (1-5)\times 10^{28}$ s.}

\maketitle

\section{Introduction}

Although dark matter particles must be very long-lived, it is possible that they decay on timescales much longer than the age of the universe. In particular, observations of the cosmic microwave background (CMB) and large scale structure indicate that the dark matter's lifetime is no less than $\tau_{X} \sim 2 \times 10^{19}$ s, even if its decays products are invisible~\cite{Poulin:2016nat}. If the dark matter decays produce potentially detectable particles, much stronger constraints can be obtained (for a review, see Ref.~\cite{Ibarra:2013cra}). Searches for dark matter decay products in the form of gamma-rays~\cite{Cohen:2016uyg,Ando:2015qda,Hutsi:2010ai,Murase:2015gea,Murase:2012xs,Ackermann:2015lka,Ackermann:2012rg,Kalashev:2016cre,Cirelli:2012ut,Esmaili:2015xpa,Liu:2016ngs}, X-rays~\cite{Boyarsky:2007ge,Yuksel:2007xh,Perez:2016tcq}, neutrinos~\cite{Murase:2012xs,PalomaresRuiz:2007ry} and cosmic rays~\cite{Ibarra:2013zia} have each been carried out. In this study, we revisit these constraints, focusing on searches for dark matter decay products in the spectrum of the isotropic gamma-ray background (IGRB) as measured by the Fermi Collaboration~\cite{Ackermann:2014usa}.

If the dark matter is unstable, its decays will produce a number of different contributions to the IGRB. In addition to gamma rays that are directly produced through the decay of such particles, the electron and position decay products will generate an additional population of photons through inverse Compton scattering. Furthermore, over cosmological distances, a significant fraction of high-energy gamma rays scatter with the cosmic infrared, optical and microwave backgrounds, producing $e^+ e^-$ pairs which then go on to generate additional gamma rays as part of an electromagnetic cascade. In this study, we calculate each of these components and derive constraints on their presence among the photons that make up the IGRB.

Our analysis differs from previous work in a number of significant ways. In particular, building on previous studies~\cite{Cholis:2013ena,Ando:2015qda}, the analysis presented here makes use of the results of recent multi-wavelength investigations that have identified the origin of the majority of the isotropic gamma-ray background~\cite{hooper2016radio,Linden:2016fdd}. This information enables us to more strongly restrict any contribution from decaying dark matter and to place more stringent constraints on the dark matter's lifetime. Additionally, whereas other groups have sometimes adopted a region-of-interest designed to emphasize dark matter decays taking place in the halo of the Milky Way (see, for example, Ref.~\cite{Cohen:2016uyg}), we take as our dataset the entire high-latitude ($|b| > 20^{\circ}$) gamma-ray sky. This choice is motivated in part by a desire to make our analysis as robust as possible to uncertainties related to diffuse Galactic backgrounds, the Milky Way's dark matter halo profile, and features of Galactic cosmic-ray transport. We present robust limits on the dark matter's lifetime for decays to a wide range of final states ($u\bar{u}$, $d\bar{d}$, $s\bar{s}$, $c\bar{c}$, $b\bar{b}$, $t\bar{t}$, $e^+ e^-$, $\mu^+ \mu^-$, $\tau^+ \tau^-$, $W^+ W^-$, $ZZ$, $hh$, $hZ$, $\gamma \gamma$, $\gamma Z$ and $\gamma h$) and masses (10 GeV to 1 EeV). Our results are complementary and, in many cases, more stringent than those presented in other recent studies~\cite{Cohen:2016uyg,Ando:2015qda,Liu:2016ngs} and to those derived from other data sets~\cite{Ibarra:2013zia,Ackermann:2015lka,Acciari:2018sjn,Liu:2018uzy,Blanco:2017sbc,Abeysekara:2017jxs,Kalashev:2016cre,Rott:2014kfa,Dugger:2010ys,Huang:2011xr}.

The remainder of this article is structured as follows. In Sec.~\ref{igrbsec}, we describe our treatment of the IGRB, including a summary of the multi-wavelength procedure that has been used to constrain its origin. In Sec.~\ref{decay} we describe our calculation of the gamma-ray spectrum from decaying dark matter, including the contributions from inverse Compton scattering and cosmological electromagnetic cascades. In Sec.~\ref{results} we present our main results, placing constraints on the dark matter's lifetime for a wide range of masses and final states. In Sec.~\ref{comparison} we compare our results to those presented in other studies and based on a variety of different observations. Finally, we summarize our results and conclusions in Sec.~\ref{conclusion}.

\section{The Isotropic Gamma-Ray Background}
\label{igrbsec}

The Fermi Collaboration has reported their measurement of an isotropic gamma-ray background (IGRB) at energies ranging from 100 MeV to 820 GeV~\cite{Ackermann:2014usa,Abdo:2010nz}.  Although previously detected by SAS-2~\cite{1978ApJ...222..833F} and EGRET~\cite{Sreekumar:1997un}, Fermi's measurement of the IGRB has provided a more detailed description of its characteristics and led to a more complete understanding of its origin.

It has long been speculated that the majority of the IGRB is produced by a large number of unresolved sources, such as active galactic nuclei (AGN)~\cite{Stecker:1993ni,1993MNRAS.260L..21P, Salamon:1994ku,Stecker:1996ma,Mukherjee:1999it,Narumoto:2006qg,Giommi:2005bp,Dermer:2006pd,Pavlidou:2007dv,Inoue:2008pk} and star-forming galaxies~\cite{Pavlidou:2002va,Thompson:2006qd,Fields:2010bw,Makiya:2010zt}. In addition to such astrophysical objects, it was appreciated that the annihilations or decays of dark matter particles could also contribute to this observed emission~\cite{Stecker:1978du,Gunn:1978gr,Gao:1991rz,Ullio:2002pj}. Through an extensive campaign of multi-wavelength observations, the origin of most of the IGRB has been identified. In particular, Fermi's detection of gamma-ray emission from both non-blazar AGN~\cite{Inoue:2011bm} and star-forming galaxies~\cite{Ackermann:2012vca}, combined with the observed correlations of the emission at gamma-ray and radio/infrared wavelengths, has revealed that these source classes each contribute significantly to the IGRB. More recent studies have shown that the combination of these source classes dominates the observed IGRB~\cite{hooper2016radio,Linden:2016fdd}. In contrast, a smaller fraction of this emission originates from blazars~\cite{Cuoco:2012yf,Harding:2012gk,Collaboration:2010gqa,Ajello:2011zi,SiegalGaskins:2010nh},\footnote{Although blazars collectively produce a greater flux of gamma rays than non-blazar AGN or star-forming galaxies, most of the emission from blazars has been resolved into individual sources, and thus (by definition) does not contribute to the IGRB.} along with contributions from other source classes, such as merging galaxy clusters~\cite{Keshet:2002sw,Gabici:2002fg,Gabici:2003kr}. Truly diffuse processes, such as cascades generated in the propagation of ultra-high energy cosmic rays~\cite{Ahlers:2011sd,Gelmini:2011kg} or cosmic-ray interactions with circum-galactic gas~\cite{Feldmann:2012rx} are also thought to contribute at a modest level.

\begin{figure}[h]
\includegraphics[scale=0.47]{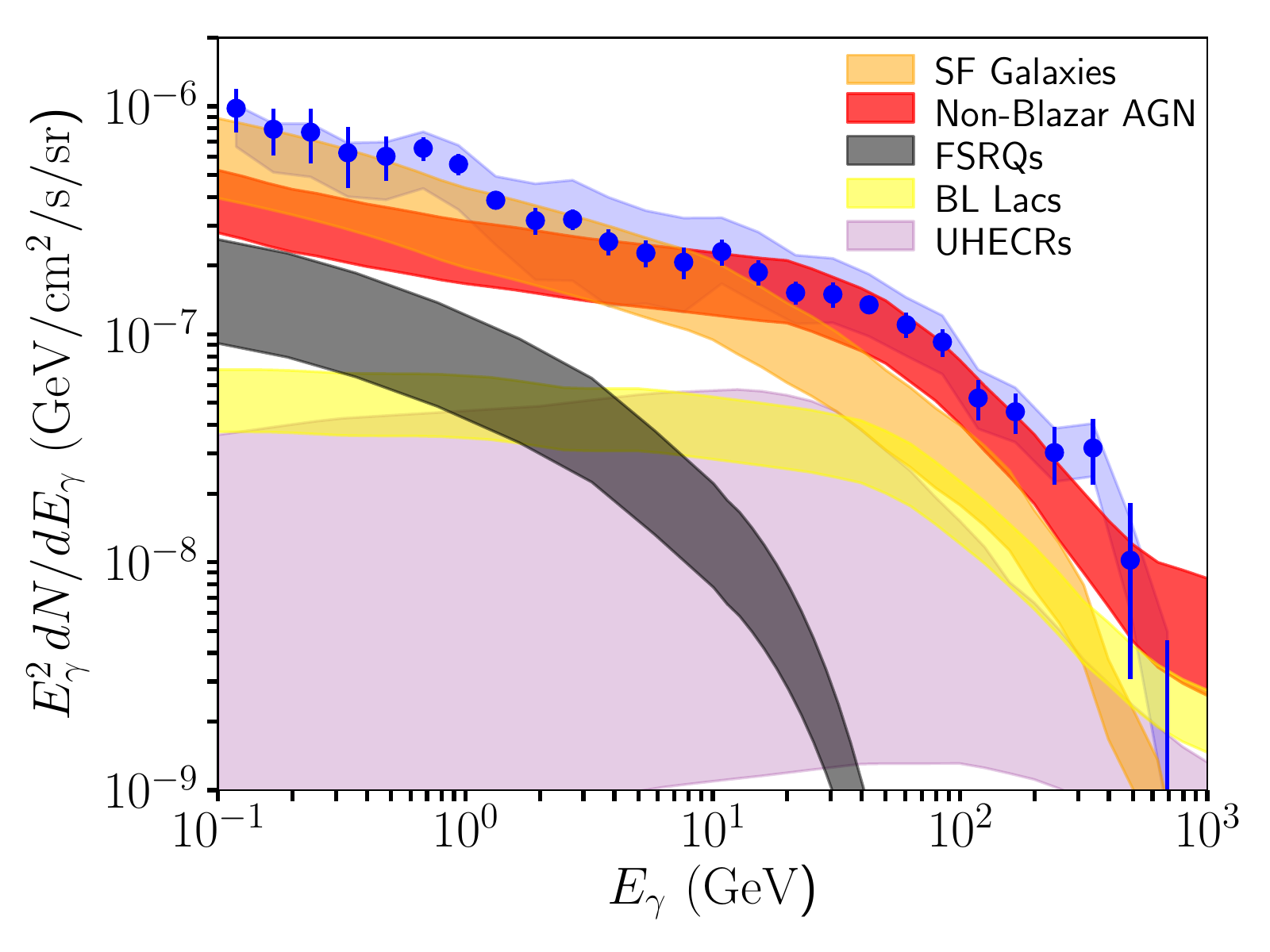} 
\includegraphics[scale=0.47]{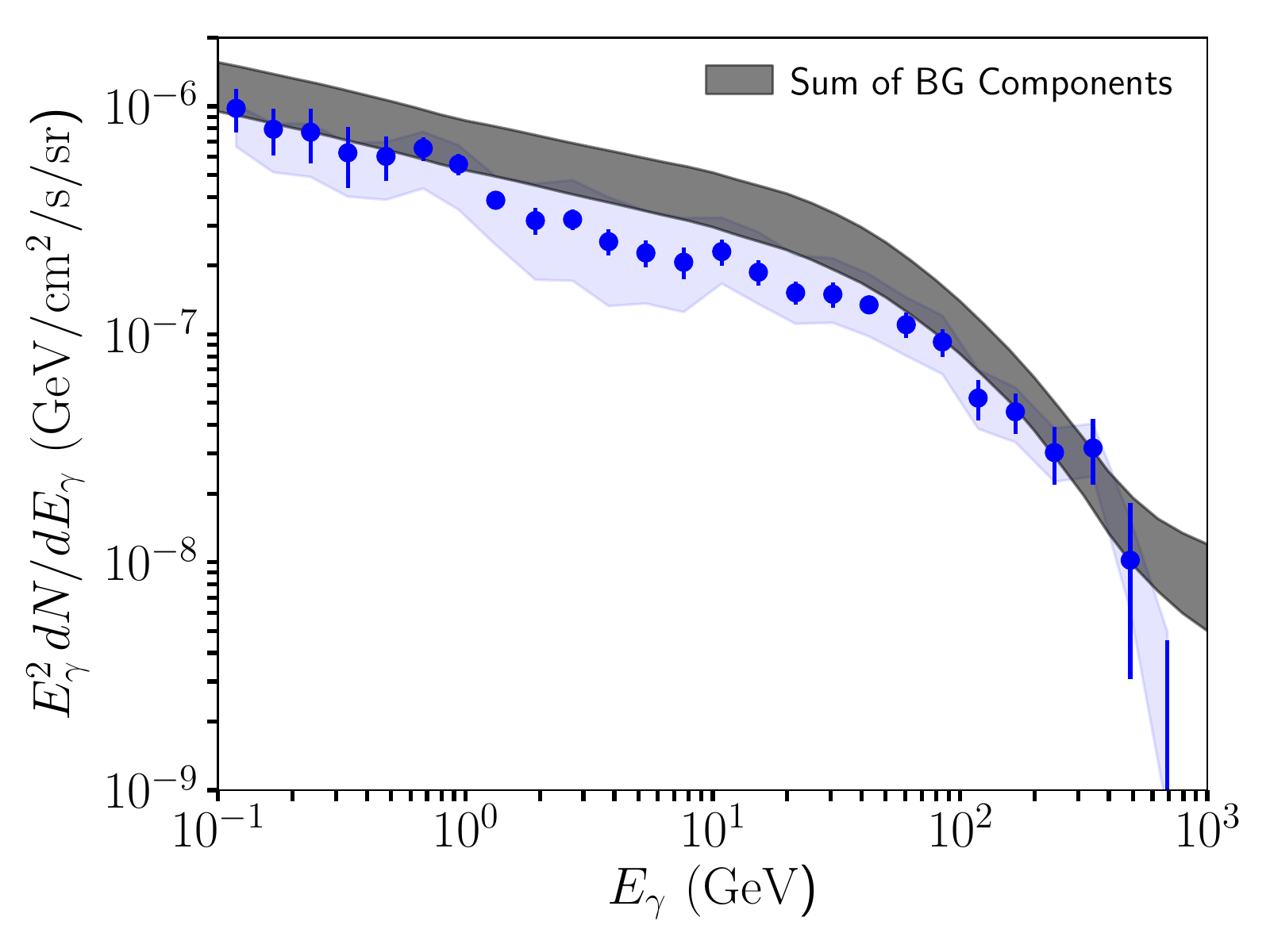} 
\caption{Left Frame: The isotropic gamma-ray background (IGRB) as measured by Fermi~\cite{Ackermann:2014usa} (error bars and surrounding blue band), compared to the contributions from star-forming galaxies, non-blazar active galactic nuclei, flat-spectrum radio quasars, BL Lacs, and ultra-high energy cosmic-ray propagation. Right Frame: When these contributions are combined, we find that they collectively predict a spectral shape and overall normalization that is approximately consistent with that of the observed IGRB, leaving relatively little room for any exotic component, such as that from decaying dark matter. See text for details.}
\label{igrb}
\end{figure}

In the left frame of Fig.~\ref{igrb}, we plot the contributions to the IGRB from several classes of astrophysics sources. At the lowest energies, this spectrum is dominated by emission from unresolved star-forming galaxies~\cite{Linden:2016fdd}, whereas non-blazar AGN produce the bulk of this emission above $\sim$10 GeV~\cite{hooper2016radio}.\footnote{The contribution to the IGRB from non-blazar AGN shown here is slightly different from that presented in Ref.~\cite{hooper2016radio}, in particular at the highest energies. This is because the results shown in Ref.~\cite{hooper2016radio} assume an average luminosity density for this source class in the local universe, whereas our Fig.~\ref{igrb} takes into account the stochastic uncertainties associated with this quantity.} We also show the contributions from BL Lac and flat-spectrum radio quasar (FSRQ) blazars, as calculated in Ref.~\cite{Cholis:2013ena} (see also Refs.~\cite{Ajello:2013lka,2012ApJ...751..108A,2010ApJ...720..435A}), as well as the contribution from electromagnetic cascades generated through ultra-high energy cosmic-ray propagation~\cite{Ahlers:2011sd}. When these contributions and their uncertainties are combined, we arrive at the total flux shown in the right frame of this figure. This total predicted flux is in good overall agreement with the IGRB as measured by the Fermi Collaboration~\cite{Ackermann:2014usa}. The error bars shown denote the statistical uncertainties associated with this measurement, while the blue band represents the systematic errors associated with the modeling of the Galactic diffuse emission, the cosmic-ray background subtraction, and Fermi's effective area~\cite{2012ApJS..203....4A}. Although these systematic errors are expected to be correlated from bin-to-bin, the Fermi Collaboration has not quantified this correlation. With this ambiguity in mind, we have calculated the $\chi^2$ of the data compared to the prediction shown in the right frame of Fig.~\ref{igrb} treating the systematic errors in two different ways. First, treating the systematics as entirely uncorrelated and combining the statistical and systematic errors in quadrature, we arrive at $\chi^2=24.5$ (over 26 degrees-of-freedom). If we instead treat the systematic error band as fully correlated, moving the measured spectrum collectively upward or downward as a single degree-of-freedom, we find an overall fit of $\chi^2=9.4$. As this later value is much smaller than the number of degrees-of-freedom in the fit, we conclude that these errors are almost certainly not correlated in this simplistic way. Throughout this study, we will present our results as derived under each of these sets of assumptions (uncorrelated, or fully correlated). That being said, we encourage the reader to treat the constraints derived with uncorrelated systematics as our main results.

\section{Gamma Rays From Dark Matter Decay}
\label{decay}

\subsection{The Galactic Contribution}

The high-latitude gamma-ray sky receives contributions from several different processes associated with the decay of dark matter particles. First, dark matter particles decaying in the halo of the Milky Way generate a prompt flux of gamma rays that is given by:
\begin{equation}
\frac{dN_{\gamma}}{dE_{\gamma}} (E_{\gamma}, \Omega) = \bigg(\frac{dN_{\gamma}}{dE_{\gamma}}\bigg)  \frac{1}{4\pi \tau_{X} m_{X}} \int_{los} \rho_{X}(l,\Omega) dl.
\label{los}
\end{equation}
Here $\Omega$ is the direction observed, $\tau_{X}$ is the dark matter's lifetime, $m_{X}$ is the dark matter's mass, and the integral is performed over the observed line-of-sight. The function $(dN_{\gamma}/dE_{\gamma})$ is the spectrum of gamma rays produced per decay, which we calculate for a given mass and decay channel using PYTHIA~\cite{Sjostrand:2014zea}, as implemented in micrOMEGAs 5.0.4~\cite{Belanger:2018mqt}. For the distribution of dark matter in the Milky Way, we adopt a standard Navarro-Frenk-White (NFW) profile~\cite{Navarro:1995iw,Navarro:1996gj}:
\begin{equation}
\rho_X(r) = \frac{\rho_0}{(r/r_s) [1+(r/r_s)]^2},
\end{equation}
where $r$ is the distance from the Galactic Center. We adopt a scale radius of $r_s=20$ kpc and set $\rho_0$ such that the local density (at $r=8.25$ kpc) is 0.4 GeV/cm$^3$. We also assume that the dark matter density is restored to its cosmological average beyond a virial radius of 300 kpc. Although the prompt emission from dark matter decay in the halo of the Milky Way is a function of $\Omega$ and thus not strictly isotropic, the line-of-sight integral in Eq.~\ref{los} departs by less than 10\% from the average value within the range of angles that contribute to Fermi's measurement of the IGRB ($|b| > 20^{\circ}$). A component of gamma-ray emission with such a small degree of variation across the high-latitude sky would be morphologically indistinguishable from the overall IGRB.

In addition to the prompt gamma-ray emission, dark matter decays can produce energetic electrons,\footnote{Throughout this paper, we refer to both electrons and positrons simply as electrons.} which generate gamma rays through inverse Compton scattering (ICS). An electron of energy, $E_e$, generates the following spectrum of inverse Compton emission:
\begin{eqnarray}
\frac{dN_{\gamma}}{dE_{\gamma}}(E_{\gamma}, E_e) \propto   \int    \frac{dn}{d\epsilon}(\epsilon) \, \frac{d\sigma_{ICS}}{dE_{\gamma}}(\epsilon,E_{\gamma},E_{e}) \,   d\epsilon,
\end{eqnarray}
where $dn/d\epsilon$ is the spectrum of the target radiation and the differential cross section for inverse Compton scattering is given by~\cite{aharonian1981}:
\begin{eqnarray}
\frac{d\sigma_{ICS}}{dE_{\gamma}}(\epsilon,E_{\gamma},E_{e})&=&\frac{3\sigma_{T}m_{e}^{2}}{4\epsilon E_{e}^{2}}\,
\bigg[ 1+\bigg( \frac{z^{2}}{2(1-z)}\bigg)+\bigg(\frac{z}{\beta (1-z)}\bigg)-\bigg(\frac{2z^{2}}{\beta^{2}(1-z)}\bigg) \nonumber \\
&-&\bigg(\frac{z^{3}}{2\beta (1-z)^{2}}\bigg)-\bigg(\frac{2z}{\beta (1-z)}\bigg)\ln\bigg(\frac{\beta (1-z)}{z}\bigg) \bigg],
\end{eqnarray}
where $z \equiv E_{\gamma}/E_{e}$, $\beta \equiv 4\epsilon E_{e}/m_{e}^{2}$ and $\sigma_T$ is the Thomson cross section.

For each electron that is injected into the halo through the decay of a dark matter particle, we calculate the total spectrum of inverse Compton emission that is produced. We then use a variation of Eq.~\ref{los}, in which the prompt gamma-ray spectrum per decay is replaced by the spectrum of inverse Compton emission per decay. To determine the injected spectrum of electrons, we again utilize micrOMEGAs~\cite{Belanger:2018mqt}. Our calculation furthermore takes into account the fact that a fraction of each electron's energy will be lost through its interactions with the Galactic magnetic field. For a randomly oriented magnetic field of uniform strength, the synchrotron energy loss rate is given by:
\begin{equation}
\frac{dE_e}{dt} = \frac{4\sigma_{T}\rho_{B} E_e^2}{3m_e^2 c^3}, 
\label{loss1}
\end{equation}
where $\rho_B = B^2/8 \pi$ is the energy density of the magnetic field. Approximating the spectrum of target radiation as a sum of greybody contributions of temperature, $T_i$, the energy loss rate from inverse Compton scattering can be written as follows~\cite{schlickeiser2010}: 
\begin{equation}
\frac{dE_e}{dt} = \frac{4\sigma_{T}}{3m_e^2 c^3} \sum_i \rho_{i, \rm rad} \, E_e^2 \, \bigg(\frac{\gamma_{i, k}^2}{\gamma_{i, k}^2+\gamma^2}\bigg),
\label{loss2}
\end{equation}
where $\gamma=E_e/m_e$, $\rho_{i, \rm rad}$ is the energy density in the $i$th component of radiation, and $\gamma_{i, k} \equiv 3\sqrt{5}m_{e}c^{2}/8\pi k_{b}T_i$. In addition to the cosmic microwave background, we use the model described in Refs.~\cite{Moskalenko2006,Zhang2006,Porter2005} to characterize the spectral and spatial distribution of starlight and infrared radiation in the halo of the Milky Way. We further assume that the magnetic field energy density roughly traces the density of radiation, $\rho_B = 0.2\times(\rho_{\rm star} +\rho_{\rm IR})$.

While undergoing inverse Compton and synchrotron processes, cosmic-ray electrons also undergo spatial diffusion, and thus move throughout the halo. In principle, this could lead to an angular distribution of inverse Compton emission that is distinct from that of the prompt component. From Eqns.~\ref{loss1} and~\ref{loss2} it follows that an electron will lose an order one fraction of its energy over a timescale given by $t_{\rm loss} \simeq 10^{13} \, {\rm s} \, \times ({\rm TeV}/E_e) \, [( \rho_B+\rho_{\rm rad})/({\rm eV}/{\rm cm}^3)]^{-1}$, during which it will typically diffuse a distance $L_{\rm dif} \sim \sqrt{D \,t_{\rm loss}}$. Adopting a value of the diffusion constant consistent with local cosmic-ray measurements, $D(E_e) \simeq 4\times 10^{28}$ cm$^2/$s $\times (E_e/{\rm GeV})^{0.3}$~\cite{Strong:2007nh,Pato:2010ih,Hooper:2017tkg}, this yields an expected diffusion length of $L_{\rm dif} \sim 0.6$ kpc $\times  ({\rm TeV}/E_e)^{0.3} \, [(\rho_B+\rho_{\rm rad})/({\rm eV}/{\rm cm}^3)]^{-0.5}$. Given that the dark matter density varies only slightly on such length scales, we conclude that we can safely neglect the effects of diffusion in our calculations.

\subsection{The Cosmological Contribution}

Dark matter decays that take place beyond the boundaries of the Milky Way also contribute to the IRGB. Over cosmological distances, however, gamma rays are much more likely to be attenuated via pair production, thereby initiating electromagnetic cascades. Neglecting attenuation for the moment, the spectrum of gamma rays per area per time per solid angle from decaying dark matter is given by:
\begin{equation}
\frac{dN_{\gamma}}{dE_{\gamma}}(E_{\gamma}) = \frac{c}{4\pi} \int \frac{\rho_{\rm X}(z) \, dz}{H(z) (1+z)^3 \, \tau_{\rm X} m_{\rm X}}  \,\bigg(\frac{dN_{\gamma}}{dE'}\bigg)_{E' = E_{\gamma}(1+z)},  
\end{equation}
where $H(z) = H_0 [\Omega_M (1+z)^3 + \Omega_{\Lambda}]^{0.5}$ is the expansion rate of the universe in terms of the cosmological parameters $\Omega_M=0.31$, $\Omega_{\Lambda} =0.69$ and $H_0=67.7$ km/s~\cite{Aghanim:2018eyx}. The average dark matter density evolves as $\rho_{\rm X}(z)= \rho_0 (1+z)^3$, where $\rho_0$ is the current cosmologically averaged density. The quantity $dN_{\gamma}/dE'$ is the gamma-ray spectrum produced per decay, after accounting for the effects of cosmological redshift.

High-energy gamma rays are significantly attenuated through their scattering with infrared, optical and microwave radiation fields~\cite{Murase:2011yw,Murase:2012xs,Murase:2011cy,Murase:2012df,Berezinsky:2016feh}. The inverse mean free path of these interactions is given by:
\begin{eqnarray}
l^{-1} (E_{\gamma},z)&=&  \int  \sigma_{\gamma\gamma}(E_{\gamma},\epsilon) \, \frac{dn}{d\epsilon}(\epsilon,z) \, d\epsilon, \nonumber
\end{eqnarray}
where $\sigma_{\gamma\gamma}$ is the total pair-production cross section and $dn(\epsilon,z)/d\epsilon$ is the differential number density of target photons at redshift, $z$. The cross section for pair-production is well approximated by the following expression~\cite{aharonian1983}:
\begin{eqnarray}
\sigma_{\gamma\gamma}(s)&=&\frac{3\sigma_{T}}{2s^{2}}\bigg[\left(s-1+\frac{1}{2s}-\frac{\ln s}{2}+\ln2\right) \ln (\sqrt{s} +\sqrt{s-1})\\
&+&\frac{\left(\ln s\right)^{2}}{8}-\frac{\left(\ln\left(\sqrt{s}+\sqrt{s-1}\right)\right)^{2}}{2}+\frac{\ln2 \, \ln s}{2}- \sqrt{s^2-s} \bigg], \nonumber
\end{eqnarray}
where $s=E_{\gamma} \epsilon/m_{e}^{2}$. The spectrum of electrons generated from these interactions is given by~\cite{aharonian1983}:
\begin{equation}
\frac{dN_e}{dE_{e}}(E_{e}) \propto \iint\frac{dN_{\gamma}}{dE_{\gamma}}(E_{\gamma}) \,\frac{dn}{d\epsilon}(\epsilon)\,\frac{d\sigma_{\gamma\gamma}}{dE_{e}}(\epsilon,E_{\gamma},E_{e})    \,d\epsilon \, dE_{\gamma},
\end{equation}
where $dN_{\gamma}/dE_{\gamma}$ is the injected spectrum of gamma rays and $d\sigma_{\gamma \gamma}/dE_e$ is the differential pair-production cross section, given by:
\begin{eqnarray}
\frac{d\sigma_{\gamma\gamma}}{dE_{e}}(\epsilon,E_{\gamma},E_{e})&=&\frac{3\sigma_{T}m_{e}^{4}}{32\epsilon^{2}E_{\gamma}^{3}} \, \bigg[ \frac{4E_{\gamma}^{2}}{(E_{\gamma}-E_{e}) E_{e}} \, 
\ln \bigg(\frac{4\epsilon E_{e}(E_{\gamma}-E_{e})}{m_{e}^{2}E_{\gamma}}\bigg) -
\frac{8\epsilon E_{\gamma}}{m_{e}^{2}} \nonumber \\
&+& \bigg(\frac{2E_{\gamma}^{2} (2\epsilon E_{\gamma}-m_{e}^{2})}{(E_{\gamma}-E_{e})E_{e}m_{e}^{2}}\bigg)
-\bigg(1-\frac{m_{e}^{2}}{\epsilon E_{\gamma}}\bigg) \, 
\frac{E_{\gamma}^{4}}{(E_{\gamma}-E_{e})^{2}E_{e}^{2}}\bigg) \bigg].
\end{eqnarray}
In order to calculate the total spectrum of the resulting electromagnetic cascade, we repeat the above procedure iteratively, evolving each electron until its energy falls below 0.1 GeV.  In our calculations, we include both the cosmic microwave background and the infrared and optical background model of Ref.~\cite{Dominguez:2010bv}. For a publicly available version of the code used to perform these calculations, we direct the reader to Ref.~\cite{Blanco:2018bbf}.

\section{Results}
\label{results}

\begin{figure}[t]
\includegraphics[scale=0.47]{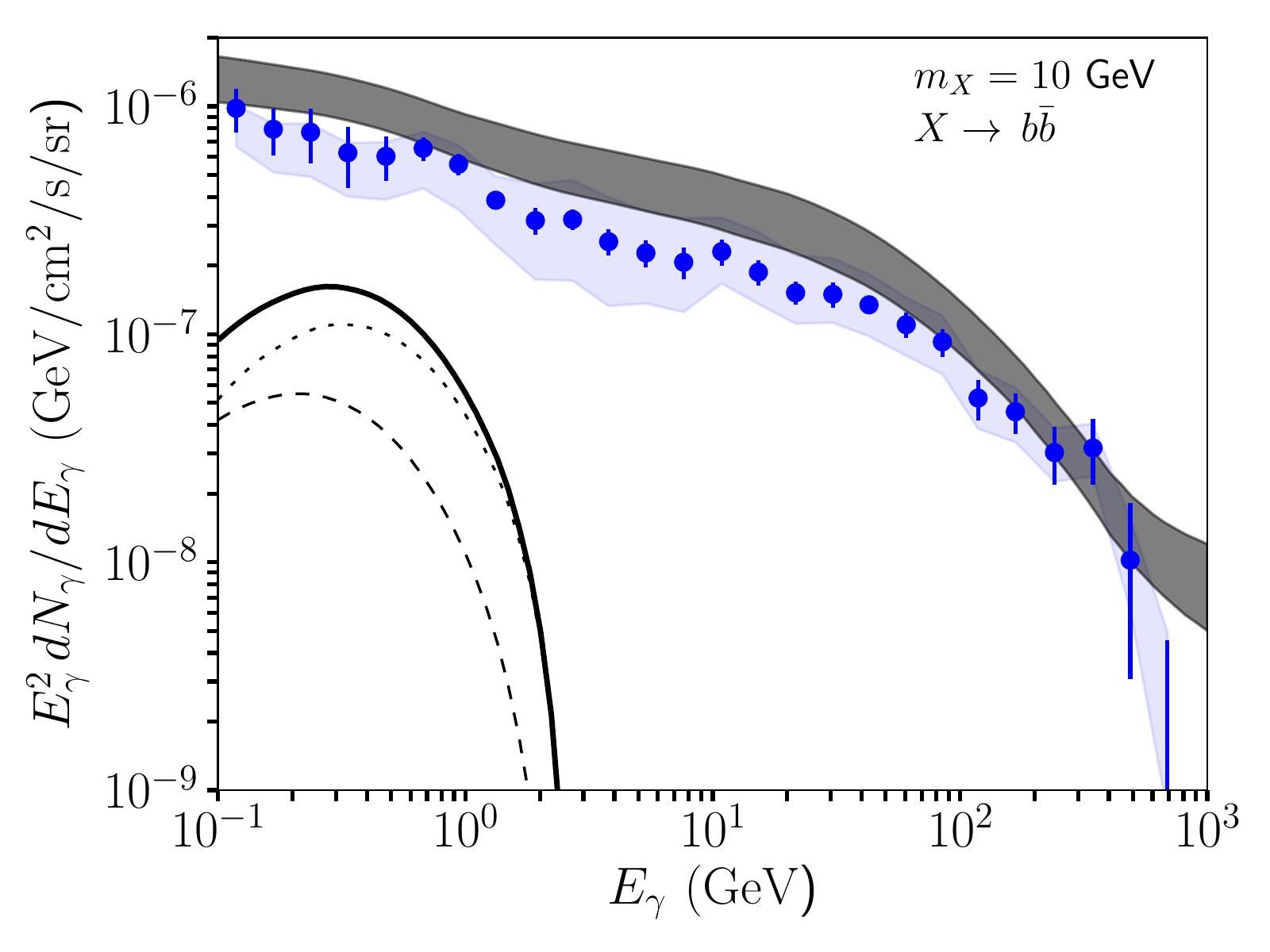} 
\includegraphics[scale=0.47]{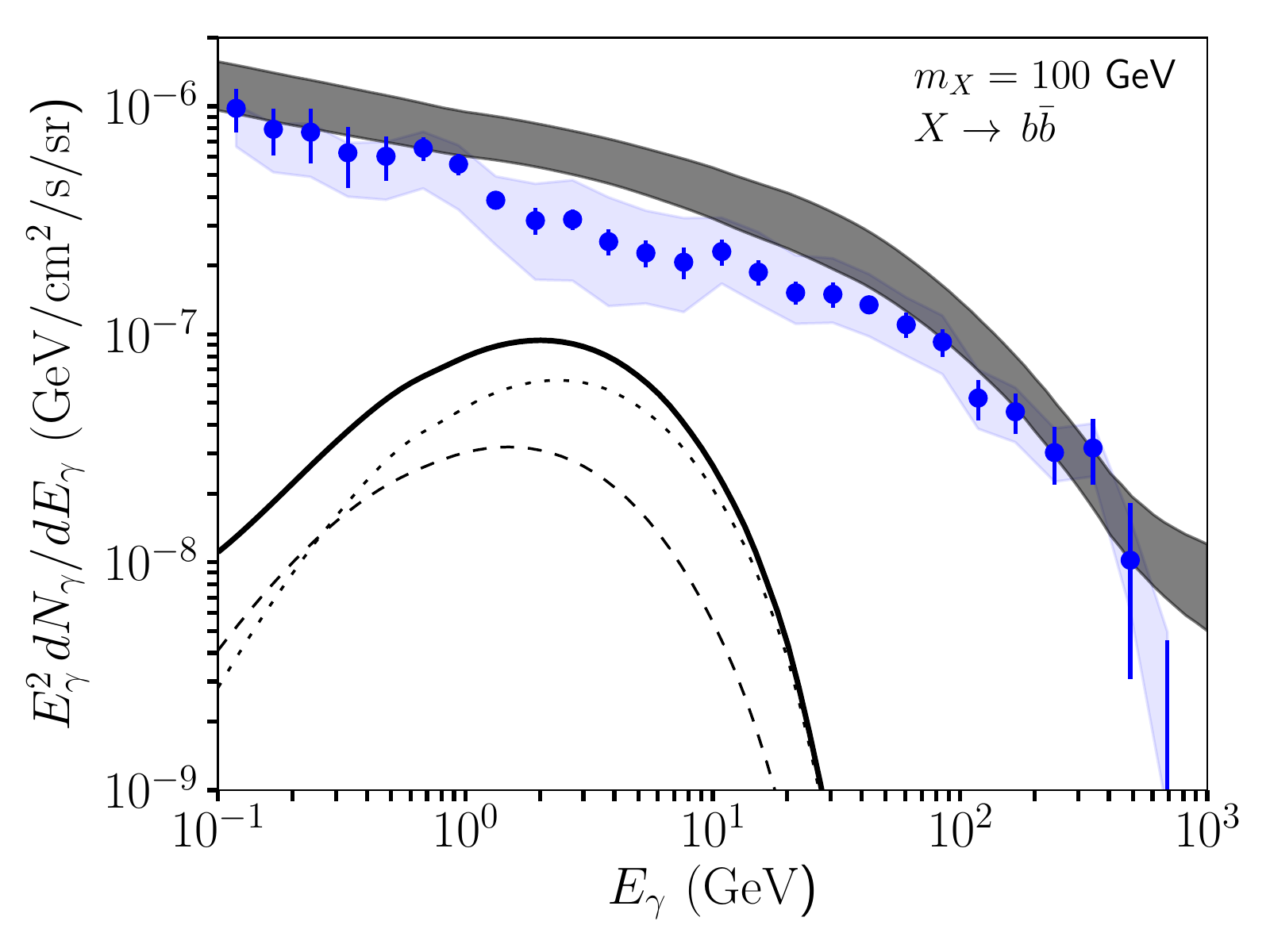} \\
\includegraphics[scale=0.47]{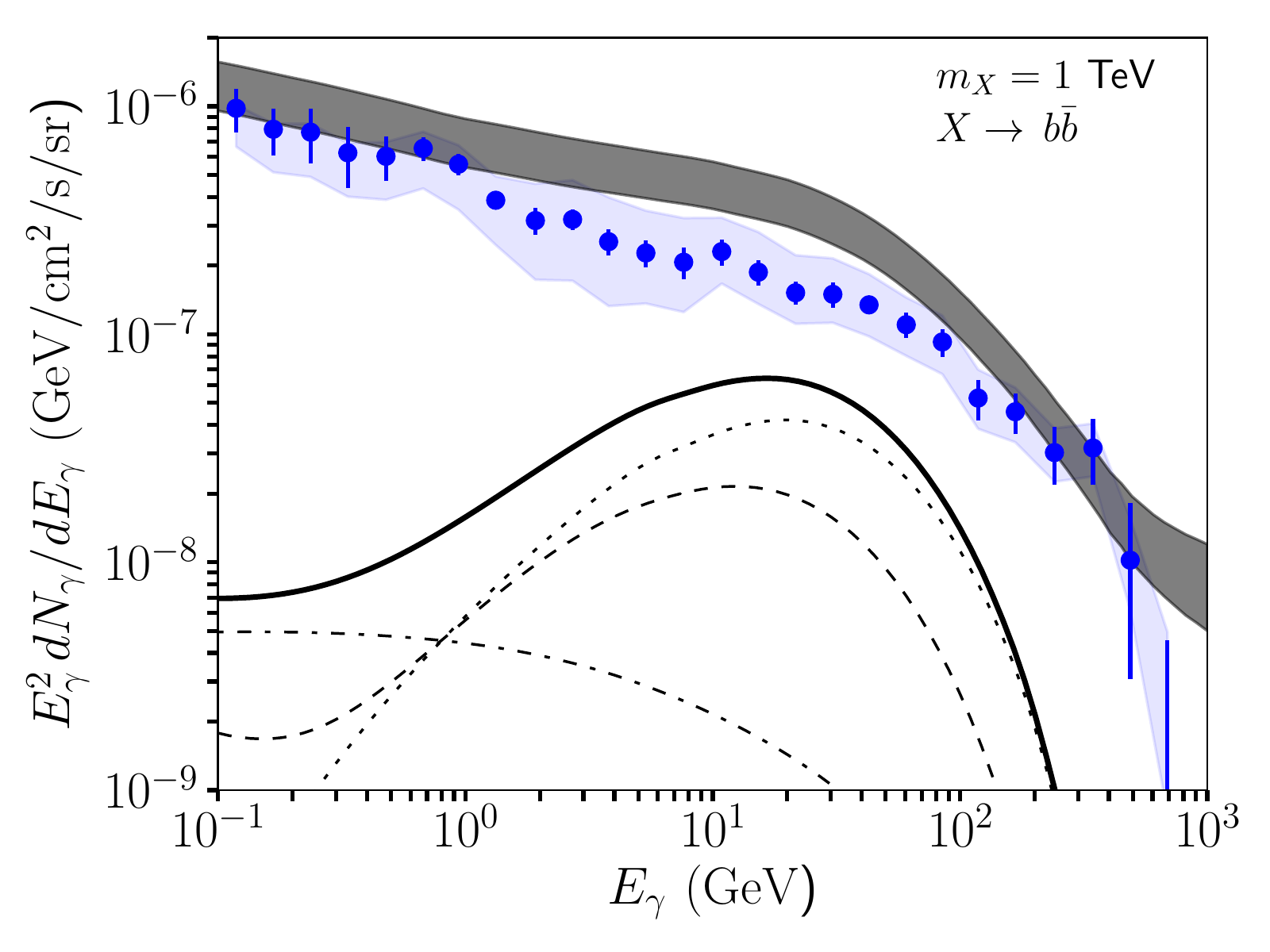} 
\includegraphics[scale=0.47]{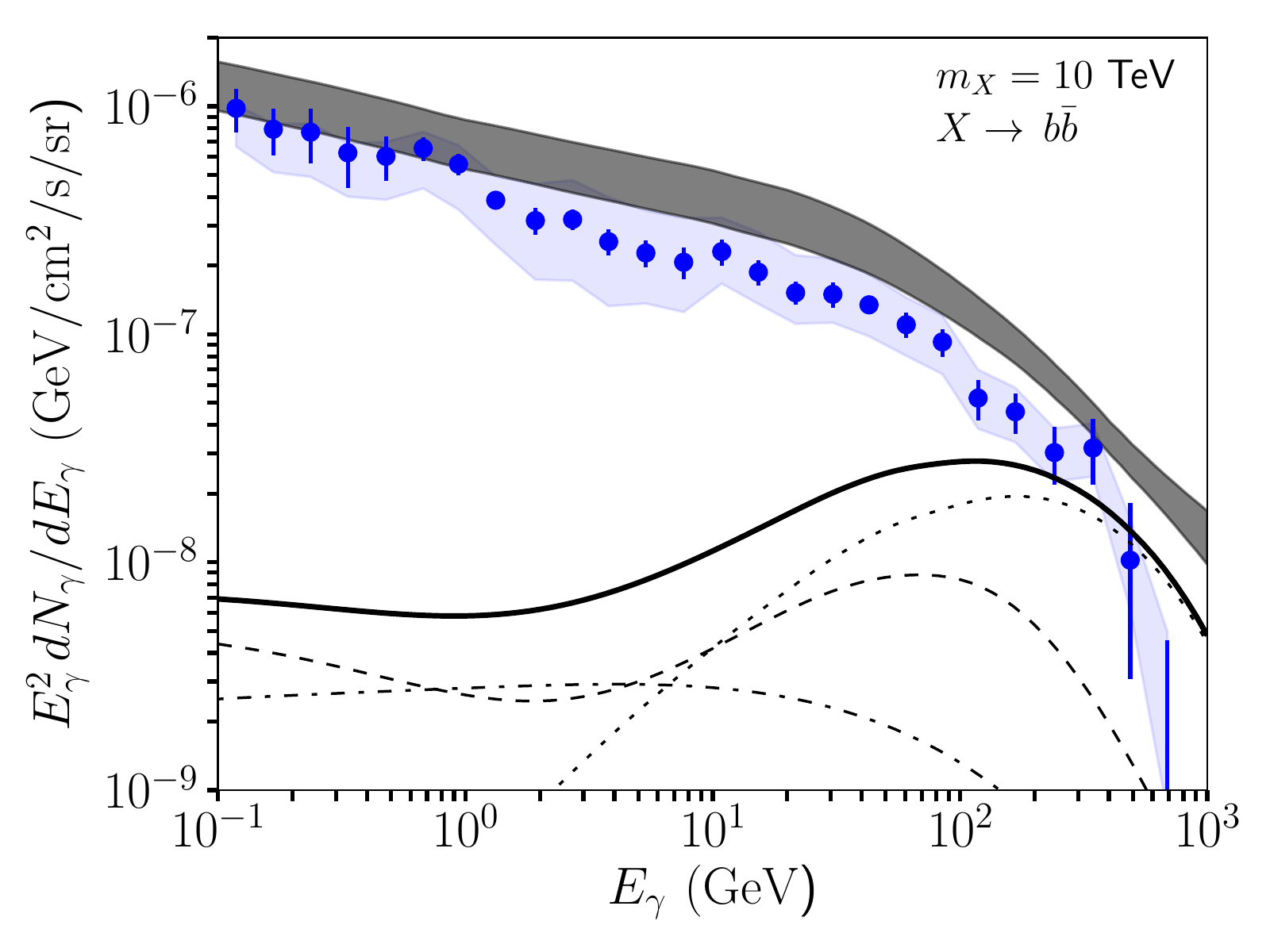} \\
\includegraphics[scale=0.47]{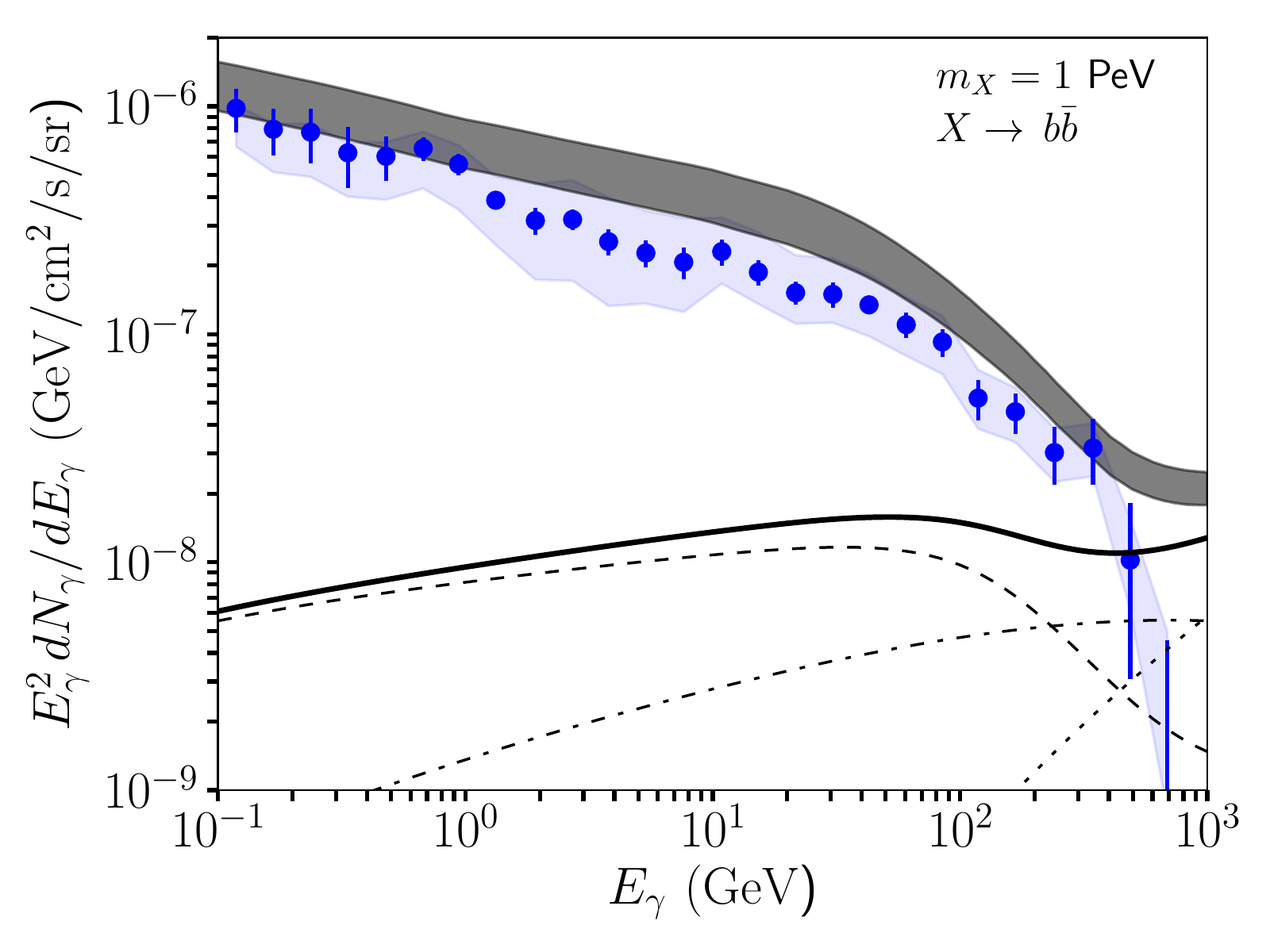} 
\includegraphics[scale=0.47]{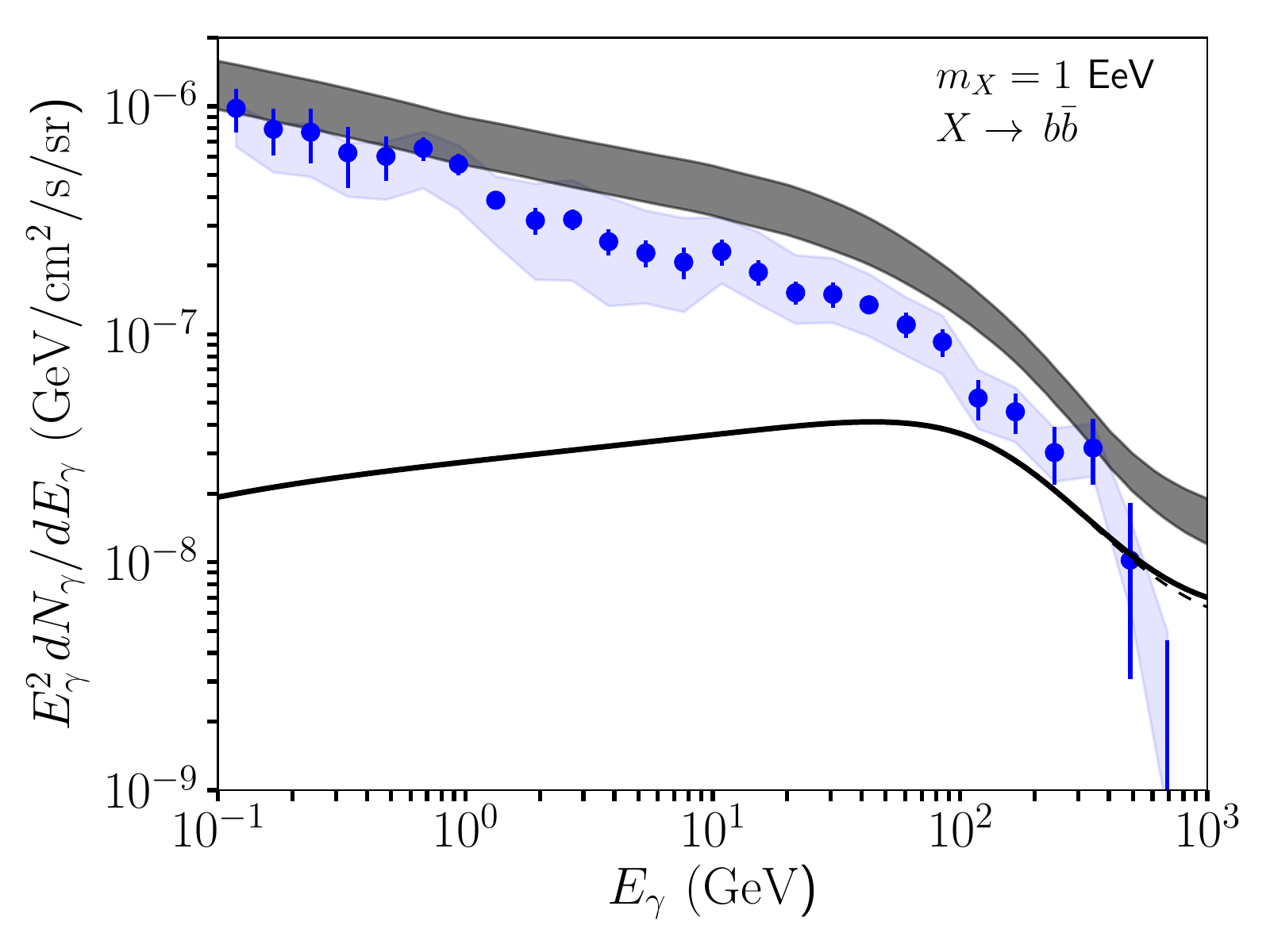}
\caption{The contribution to the isotropic gamma-ray background (IGRB) from decaying dark matter, for the case of decays to $b\bar{b}$. In each frame, the contributions from the Galactic Halo are shown as dotted and dot-dashed lines, representing the emission from direct production and from inverse Compton scattering, respectively. The dashed lines represent the cosmological contribution, including electromagnetic cascades. The solid lines denote the sum of these components. In each frame, the normalization (and corresponding dark matter decay rate) has been set to the maximum value allowed by out fit (at the 95\% confidence level); see Fig.~\ref{limits}. Note that in the $m_X=1$ EeV case, the spectrum is almost entirely dominated by the cosmological cascade.}
\label{bb}
\vspace{0.5cm}
\end{figure}

\begin{figure}[t]
\vspace{2.5cm}
\includegraphics[scale=0.47]{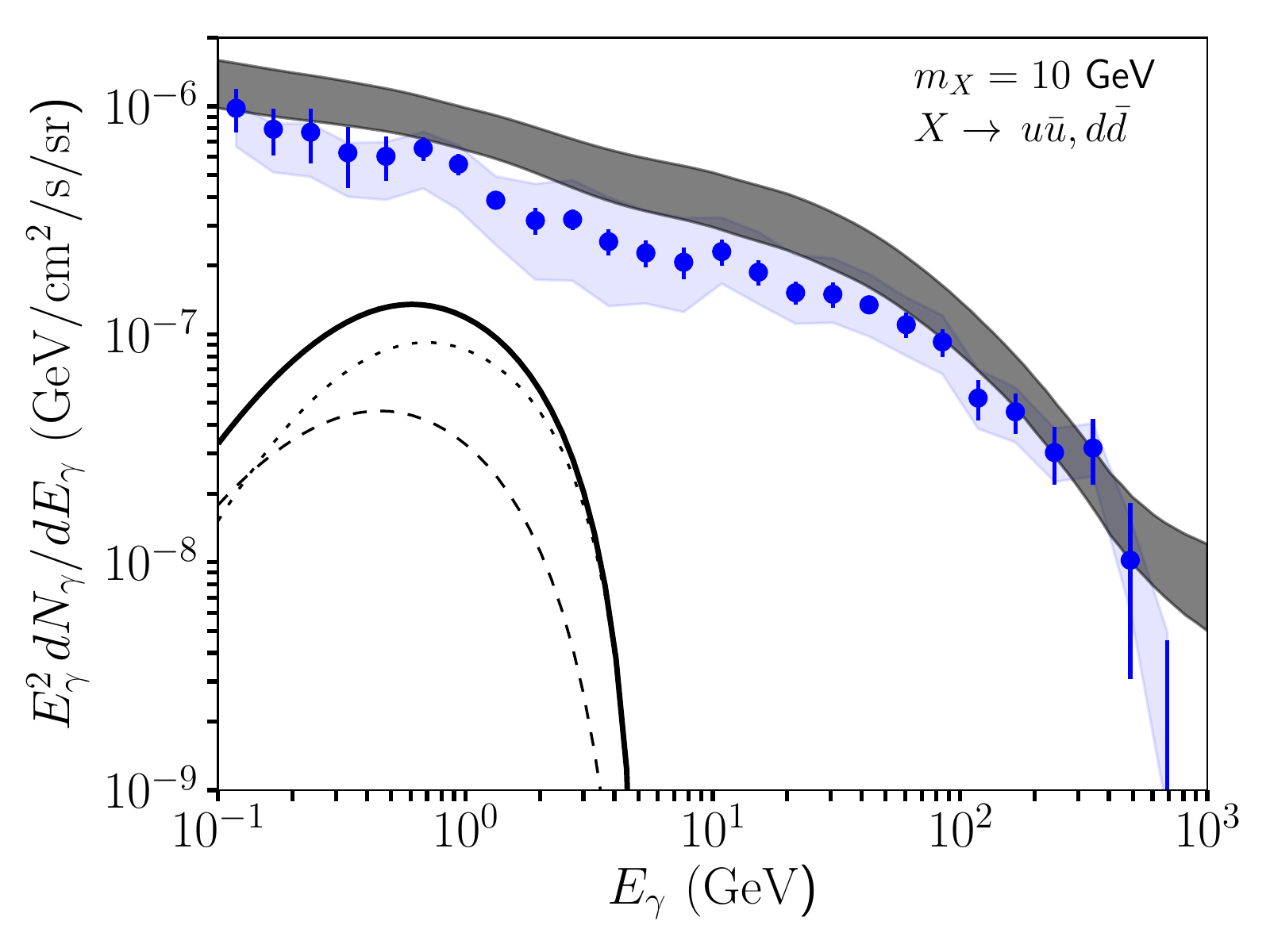} 
\includegraphics[scale=0.47]{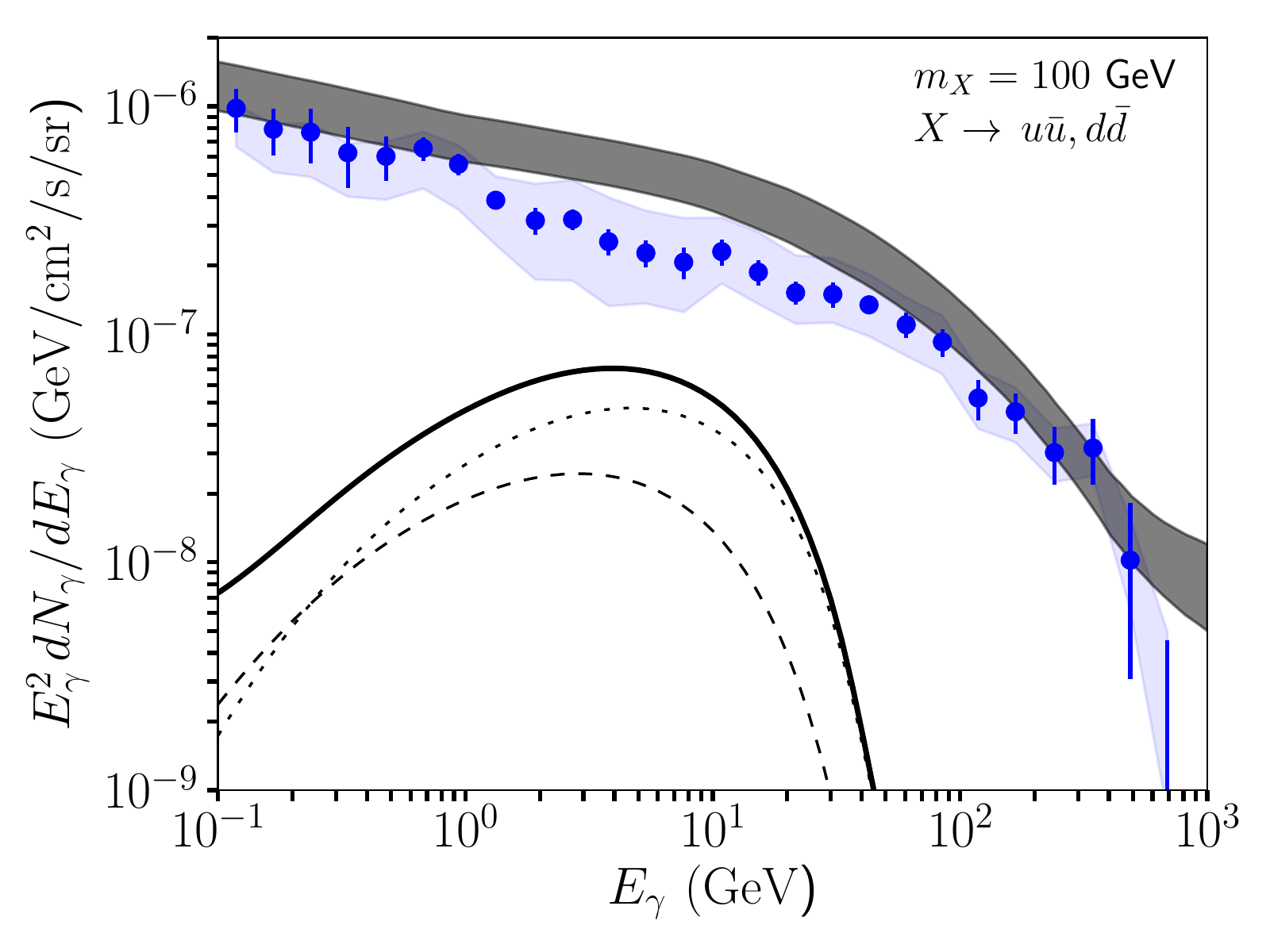} \\
\includegraphics[scale=0.47]{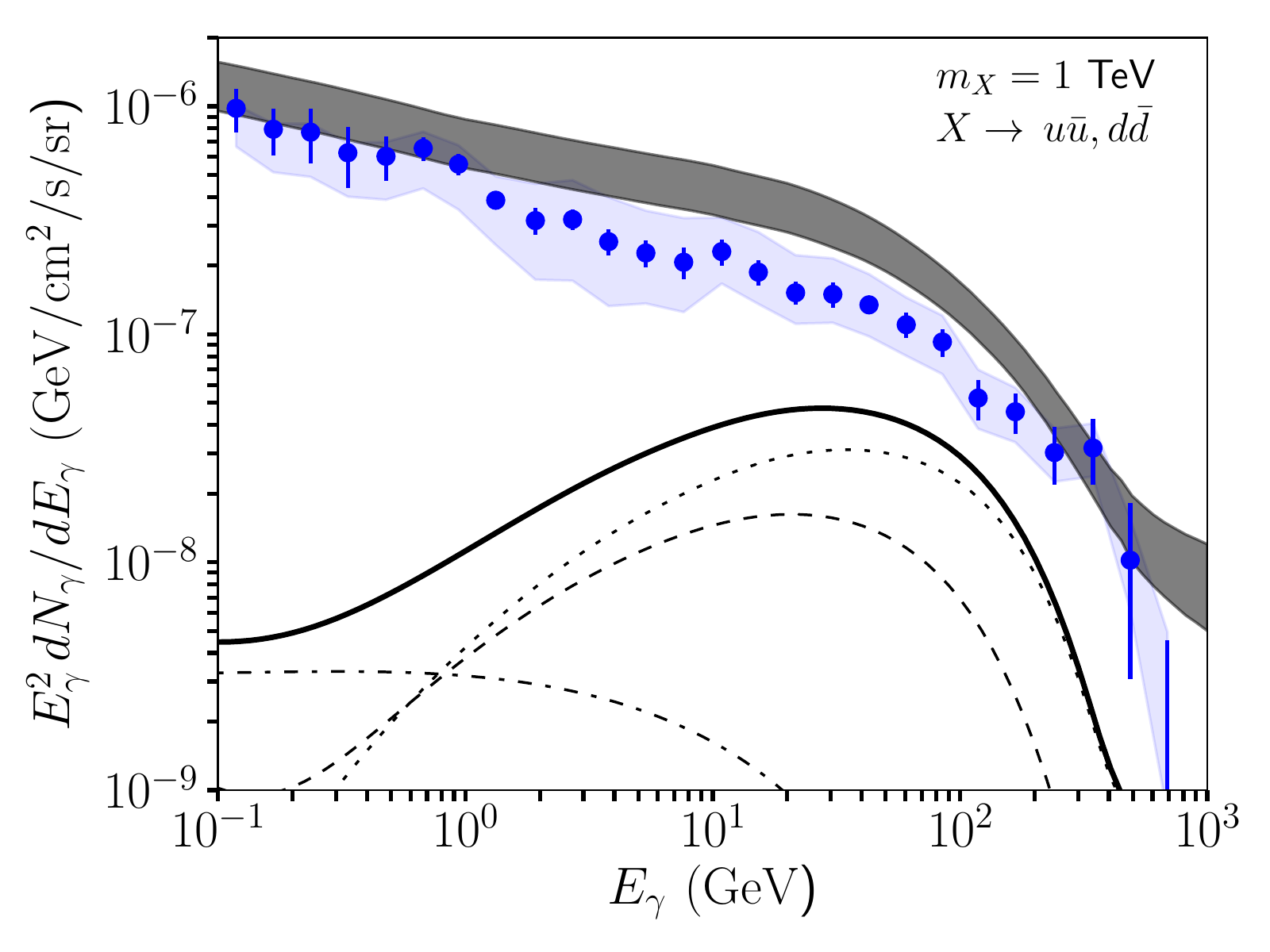} 
\includegraphics[scale=0.47]{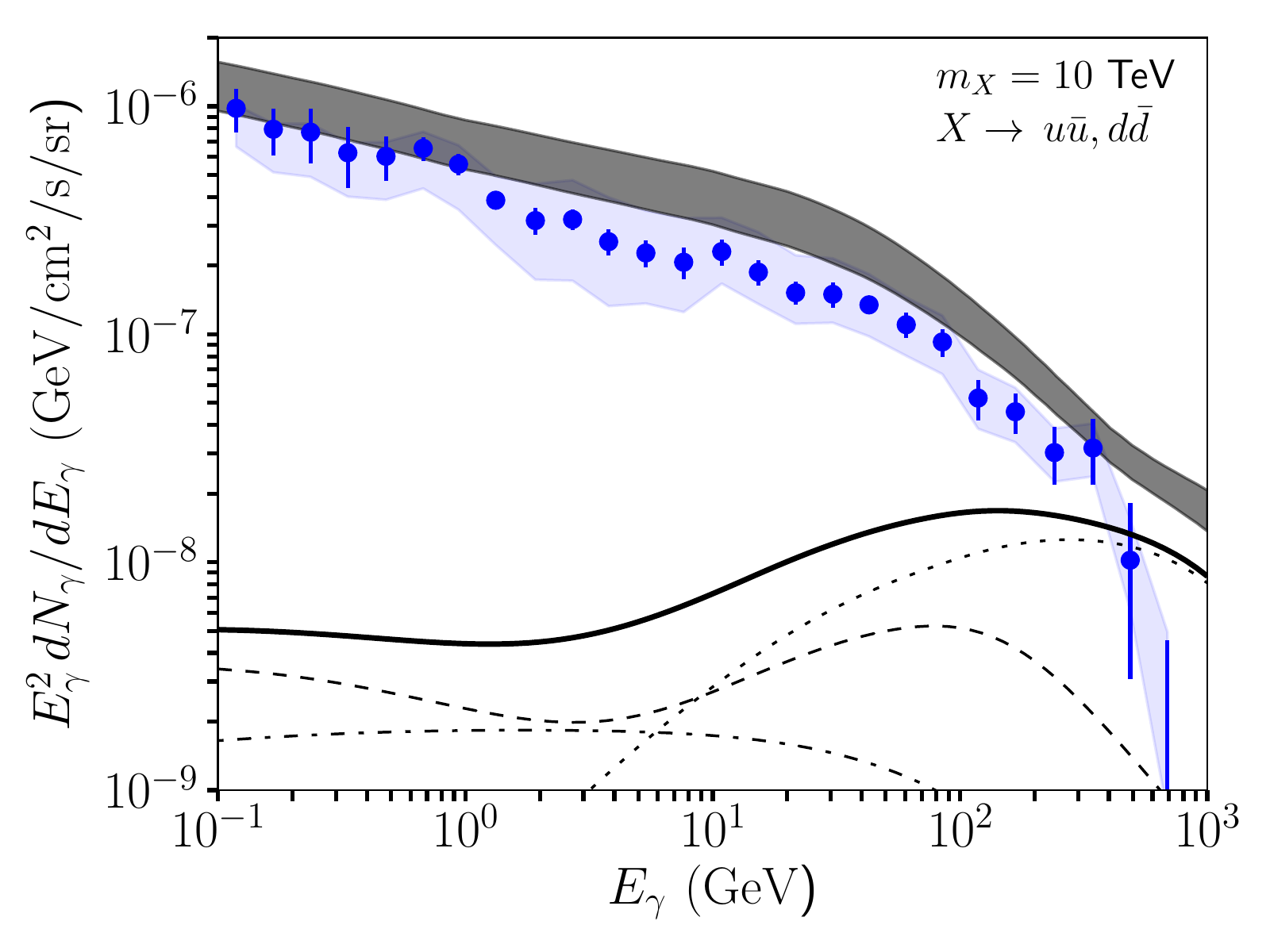} \\
\includegraphics[scale=0.47]{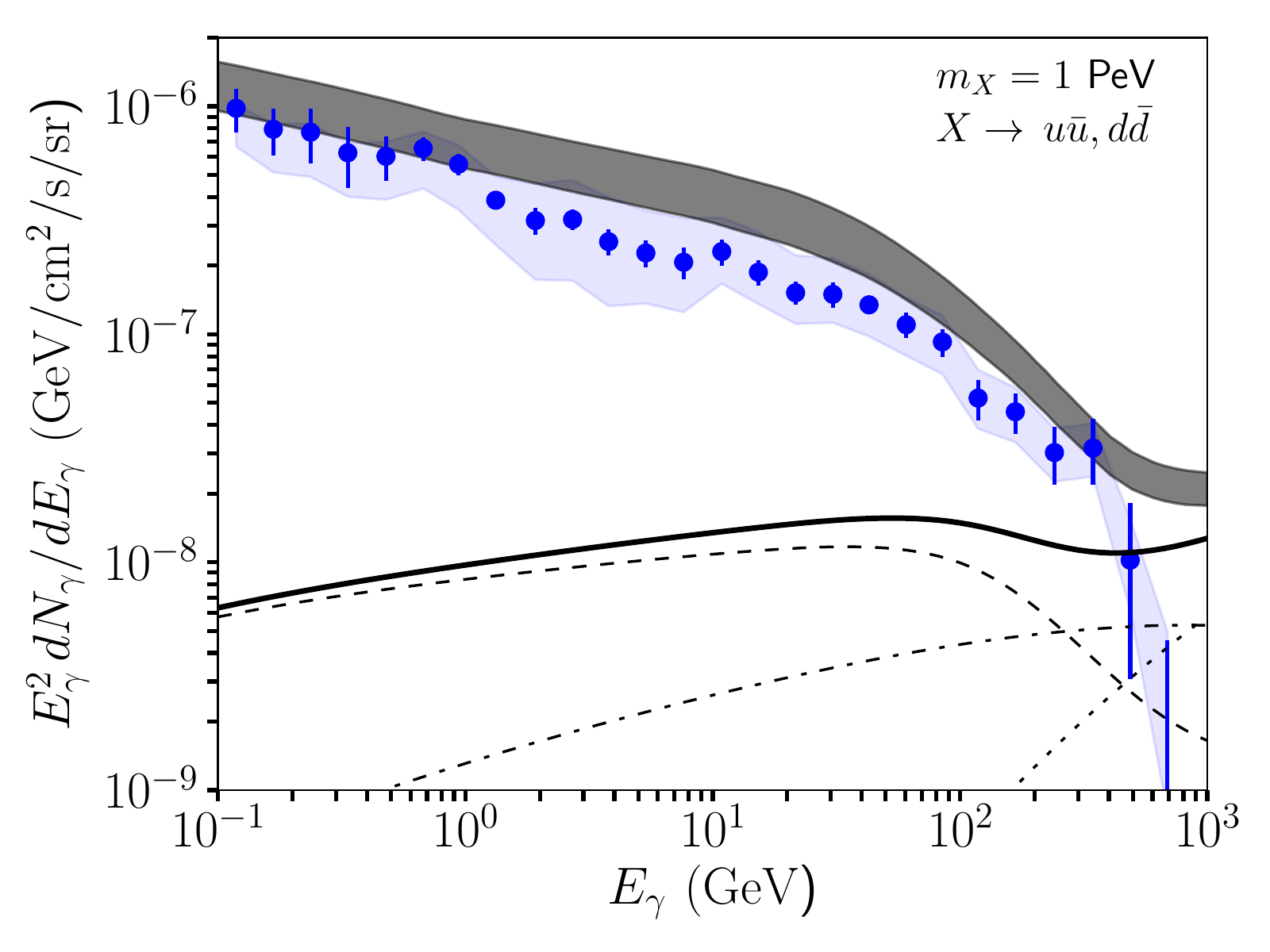} 
\includegraphics[scale=0.47]{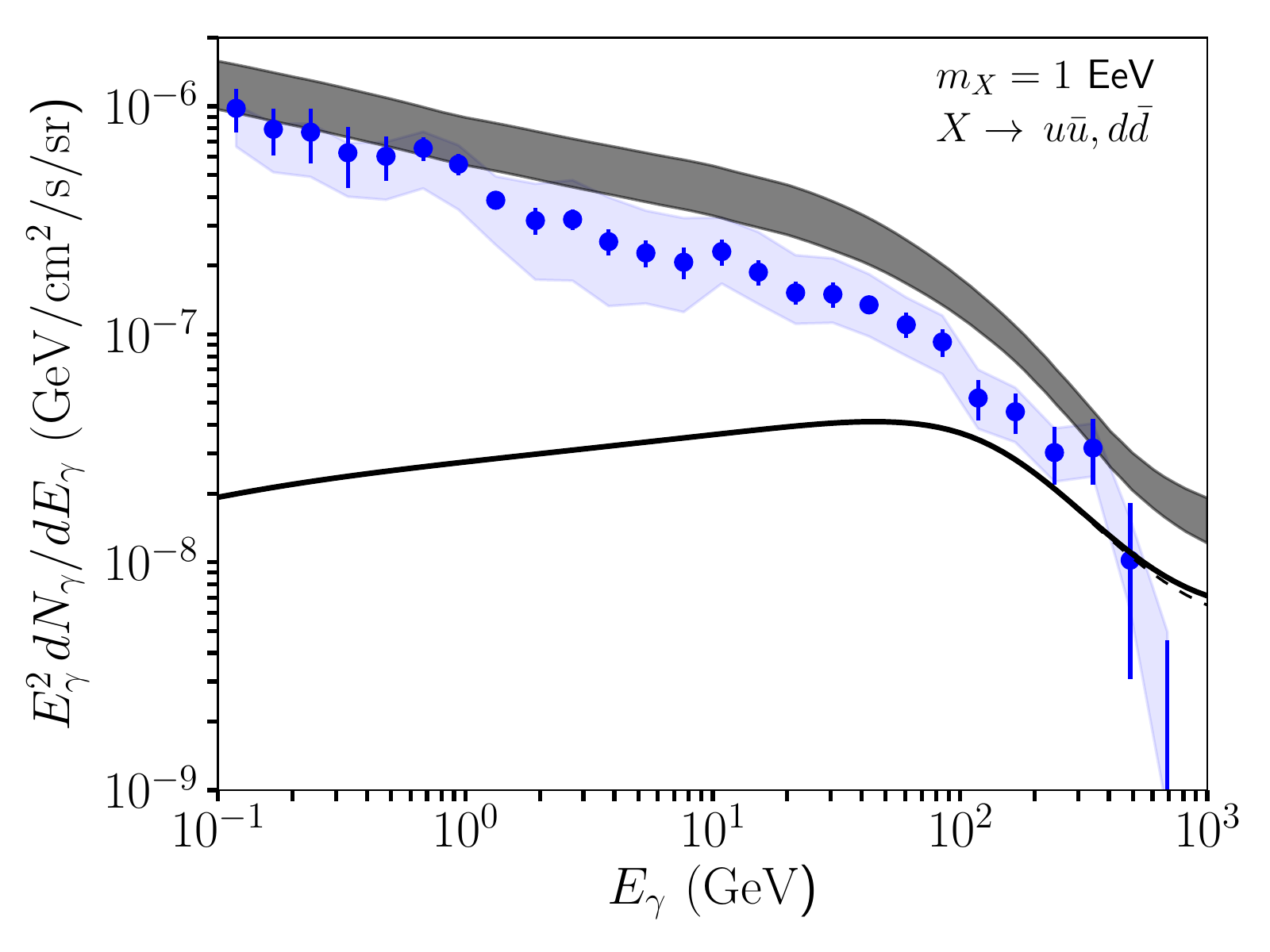}
\caption{As in Fig.~\ref{bb}, but for the case of decays to light quarks.}
\label{uu}
\vspace{1.0cm}
\end{figure}

\begin{figure}[t]
\includegraphics[scale=0.47]{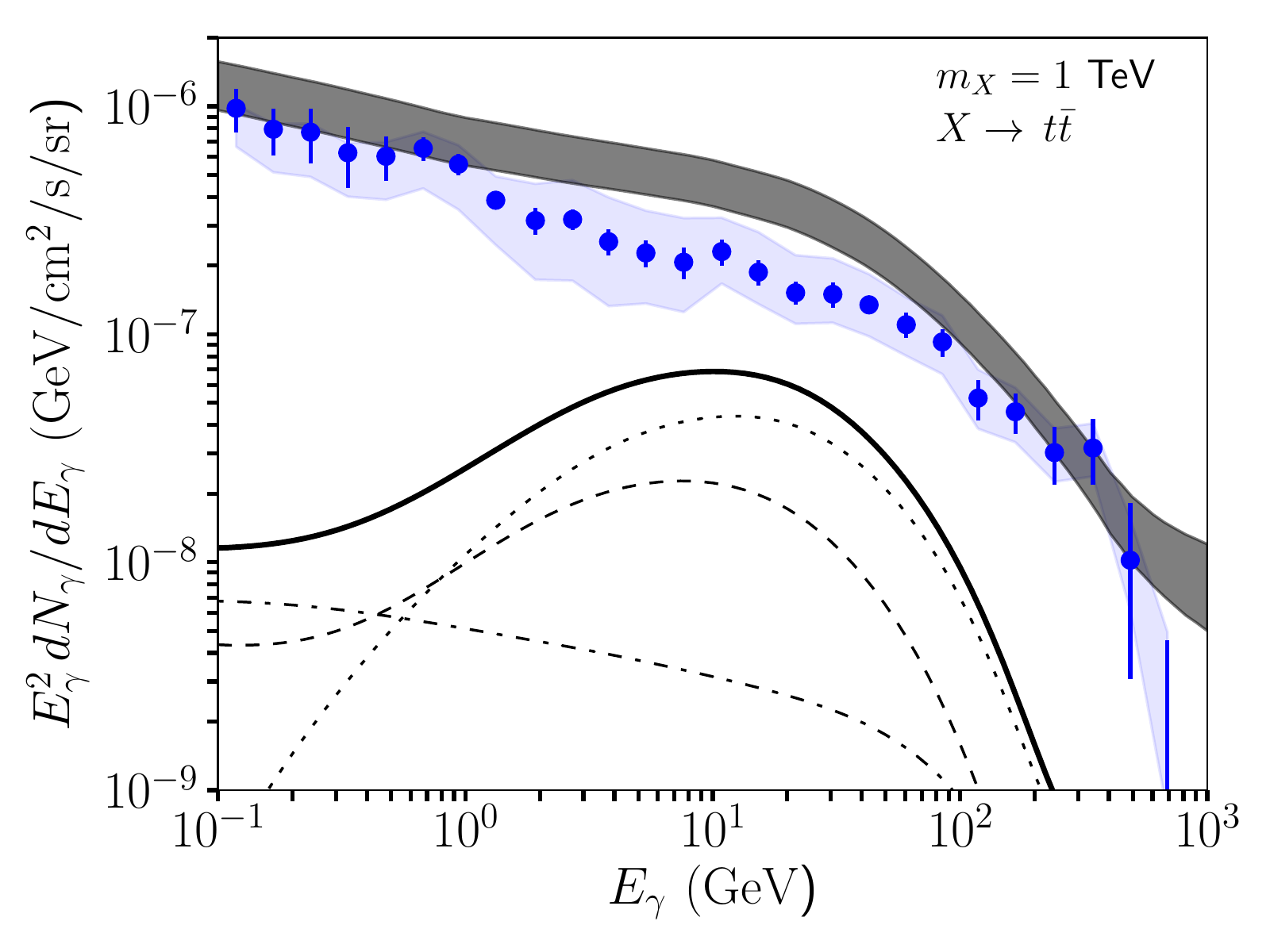} 
\includegraphics[scale=0.47]{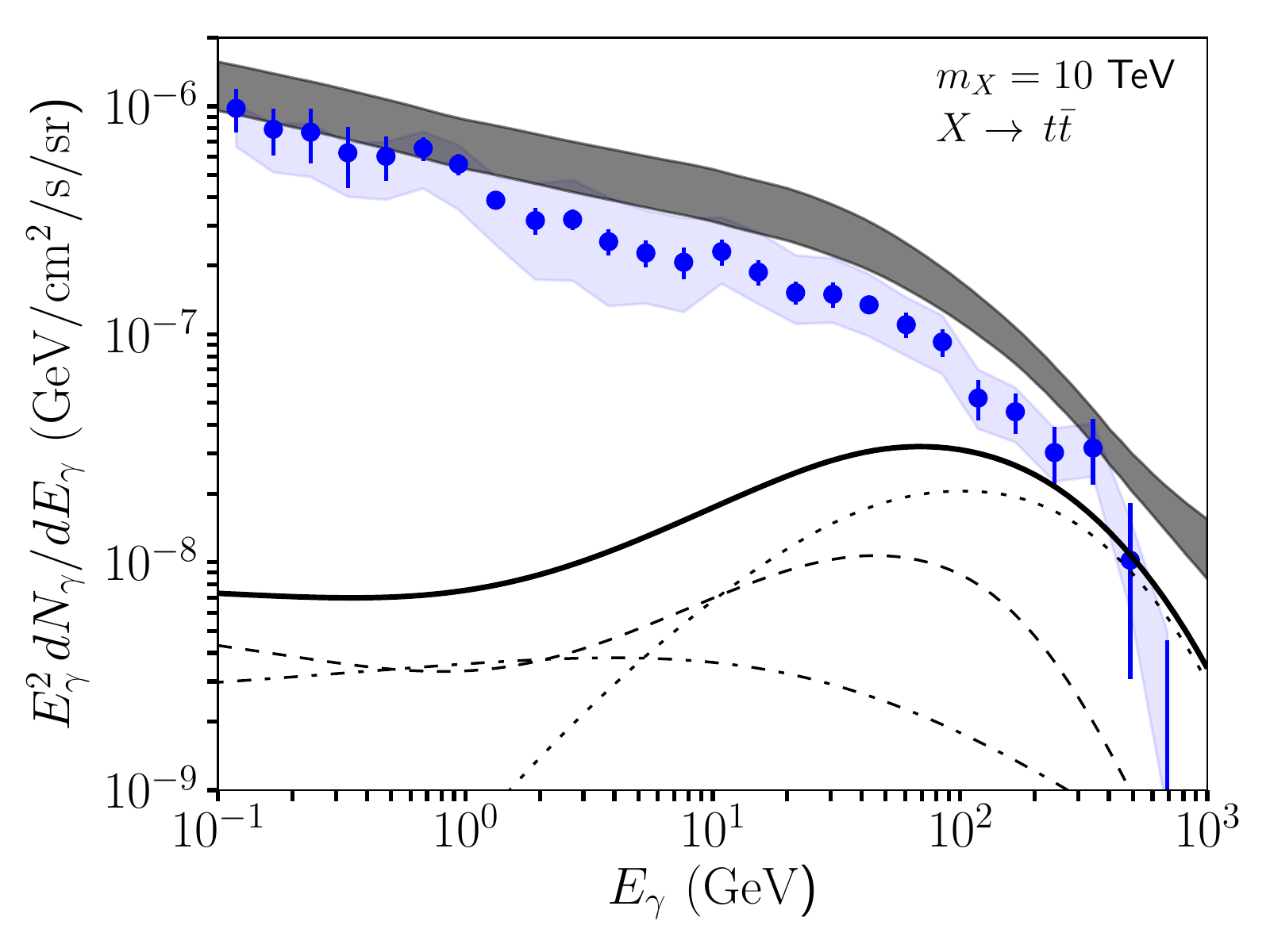} \\
\includegraphics[scale=0.47]{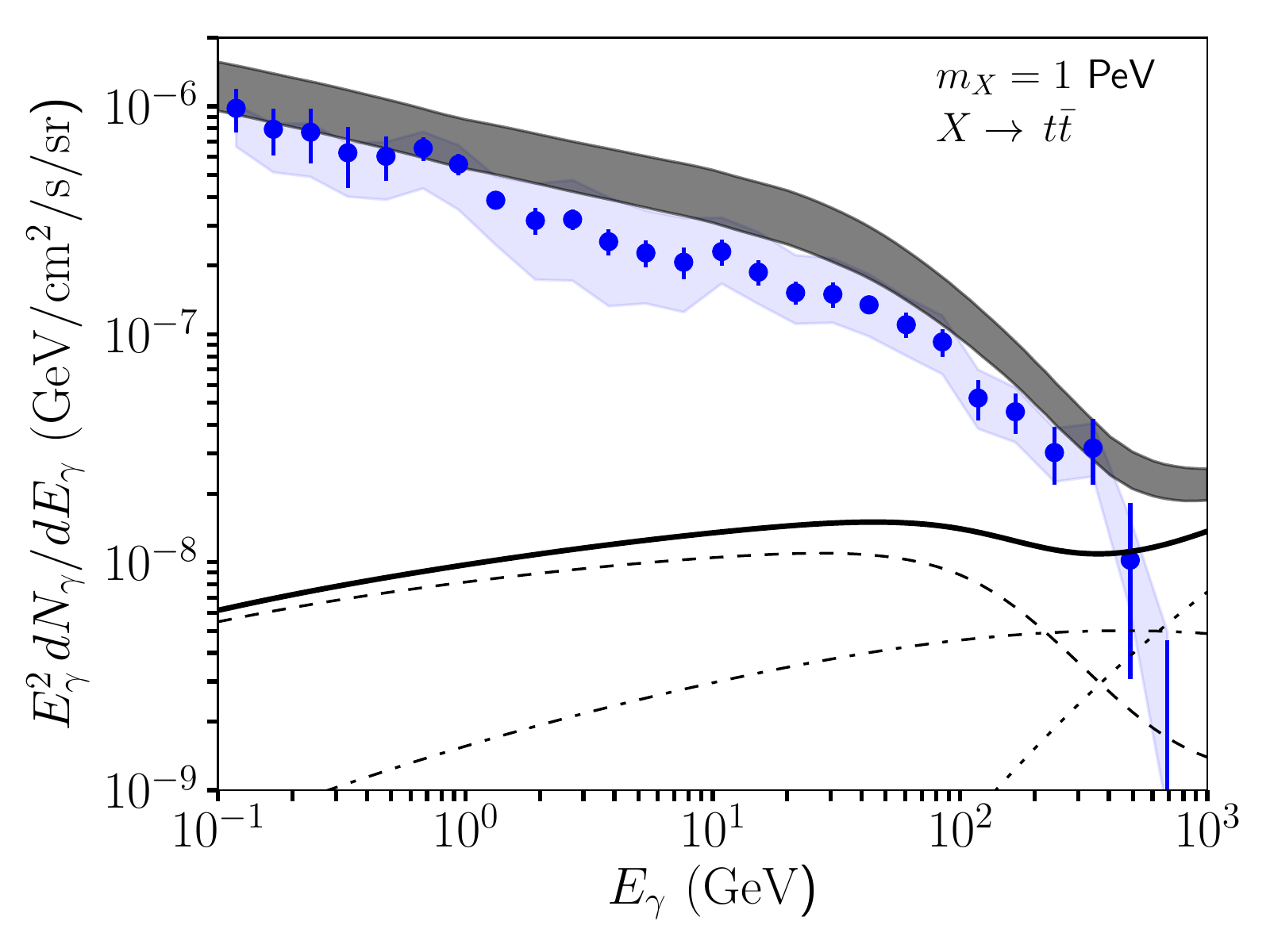} 
\includegraphics[scale=0.47]{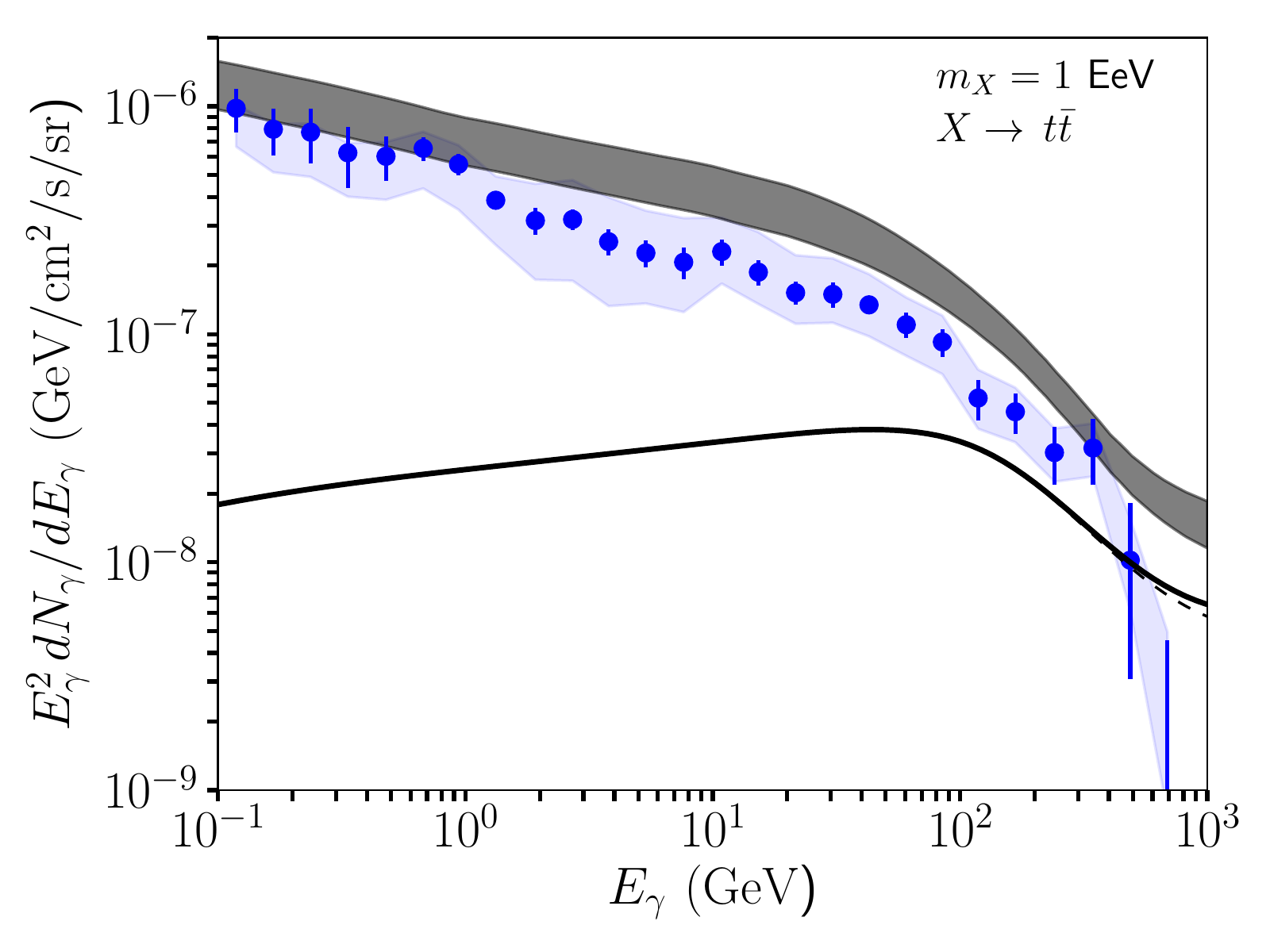}
\caption{As in Fig.~\ref{bb}, but for the case of decays to $t\bar{t}$.}
\label{tt}
\end{figure}

\begin{figure}[t]
\vspace{2.5cm}
\includegraphics[scale=0.47]{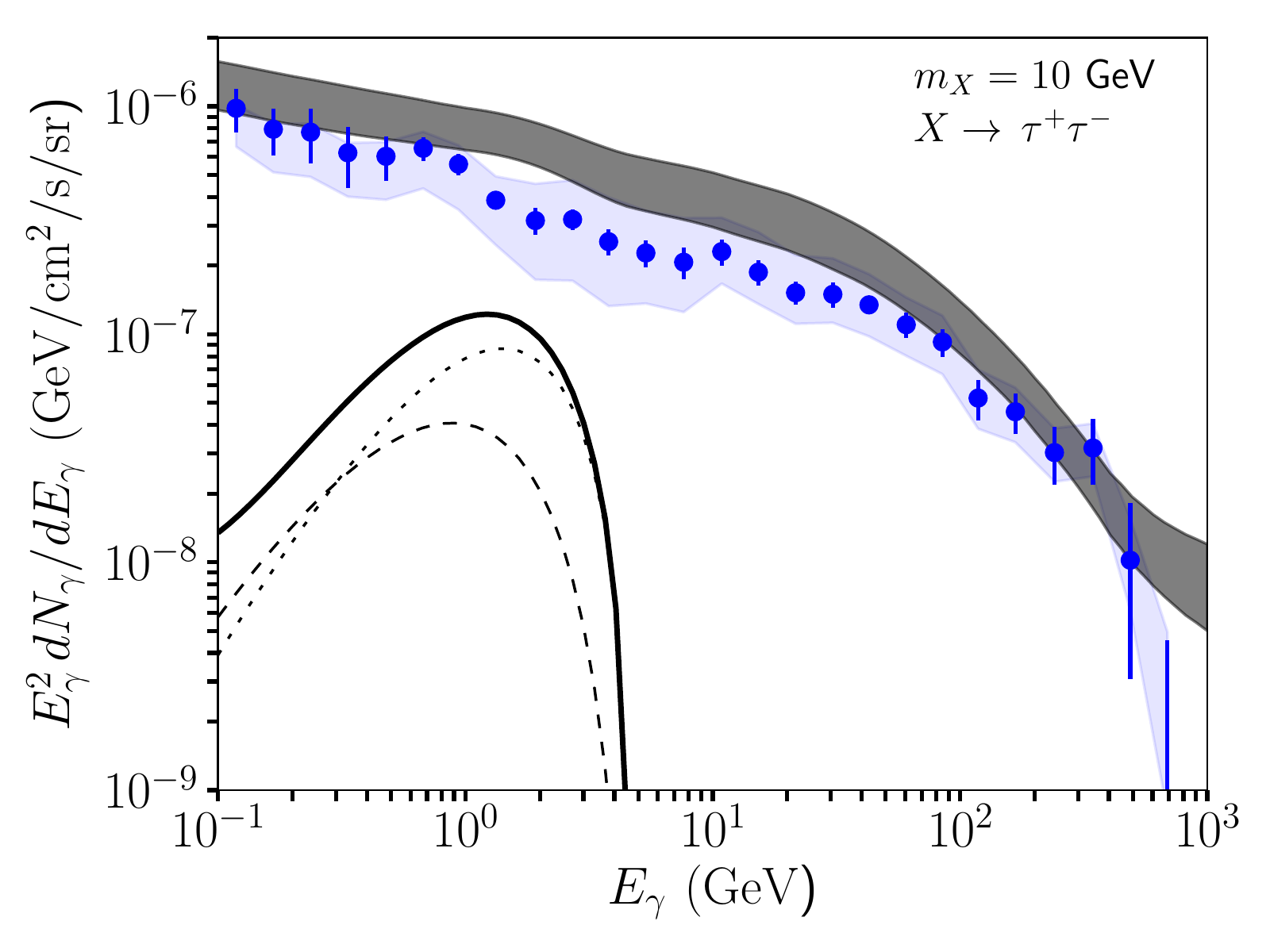} 
\includegraphics[scale=0.47]{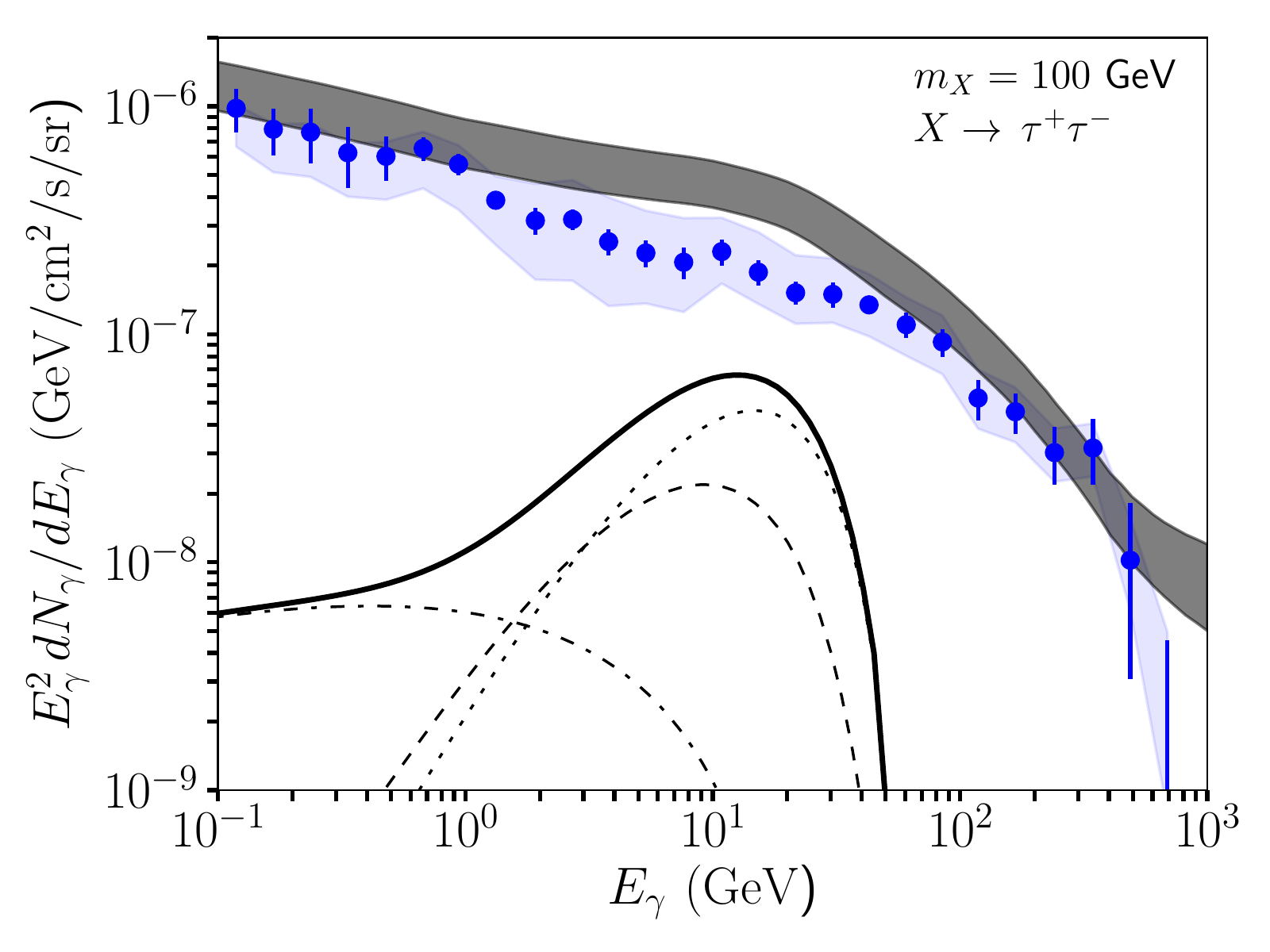} \\
\includegraphics[scale=0.47]{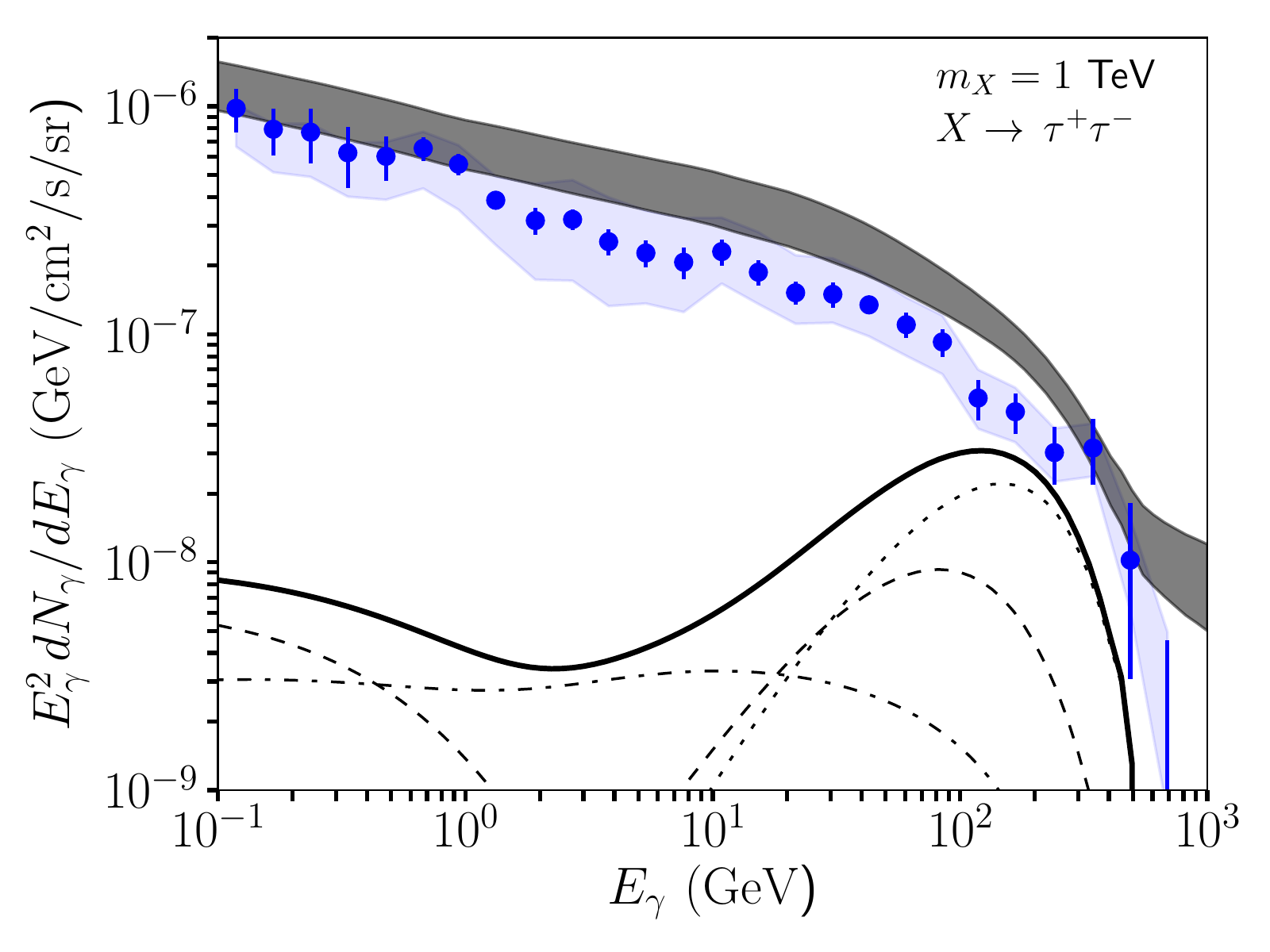} 
\includegraphics[scale=0.47]{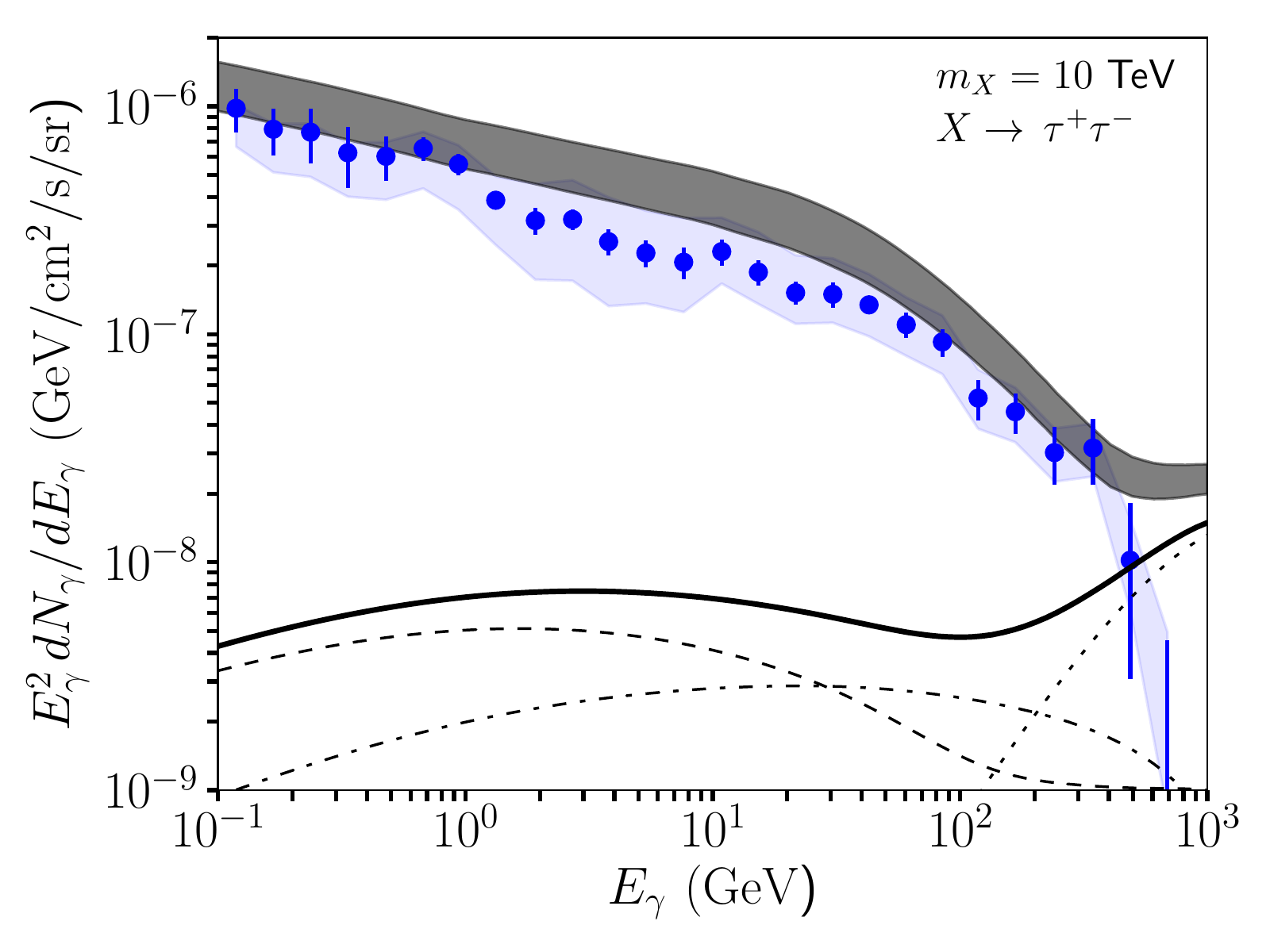} \\
\includegraphics[scale=0.47]{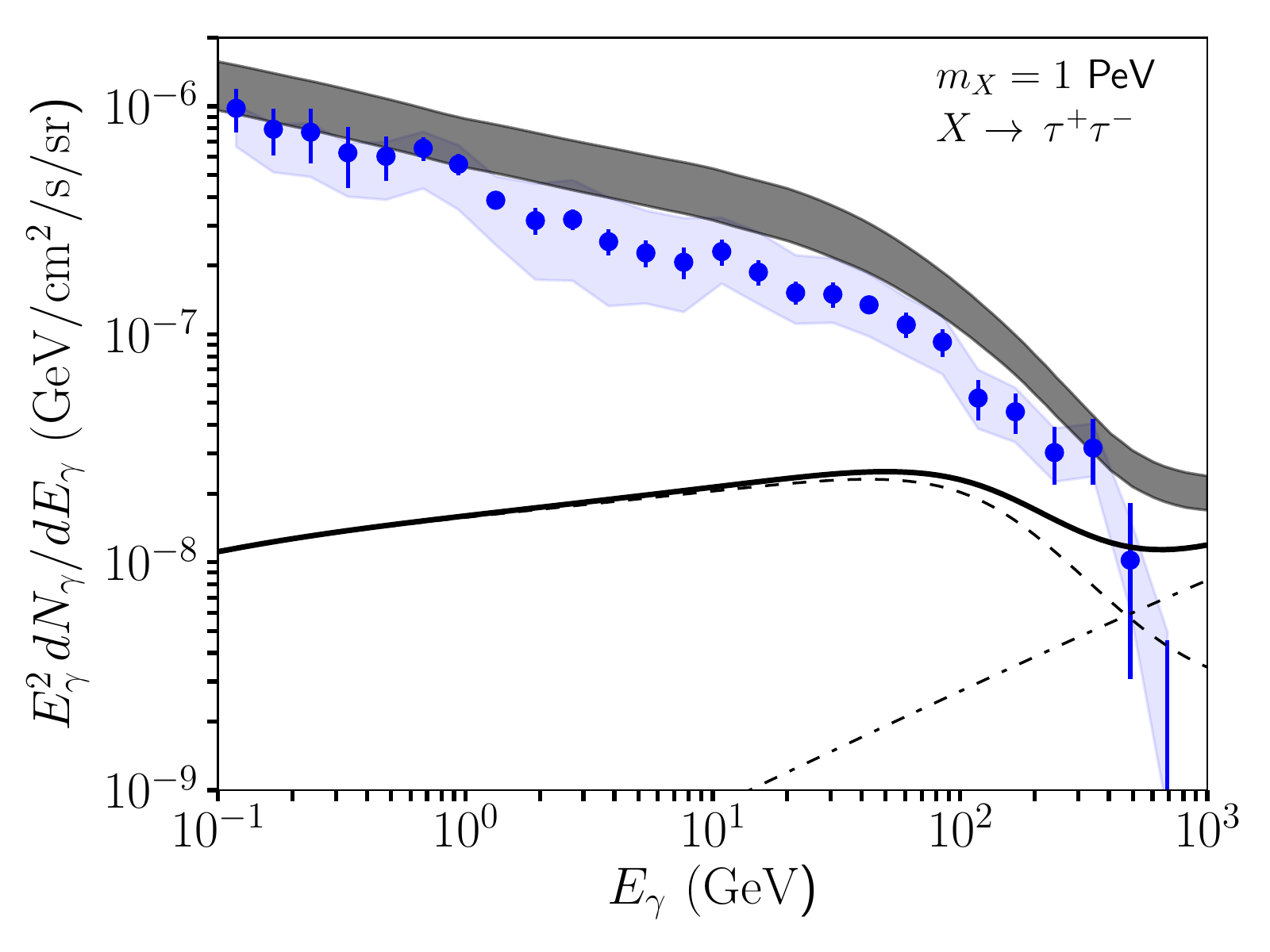} 
\includegraphics[scale=0.47]{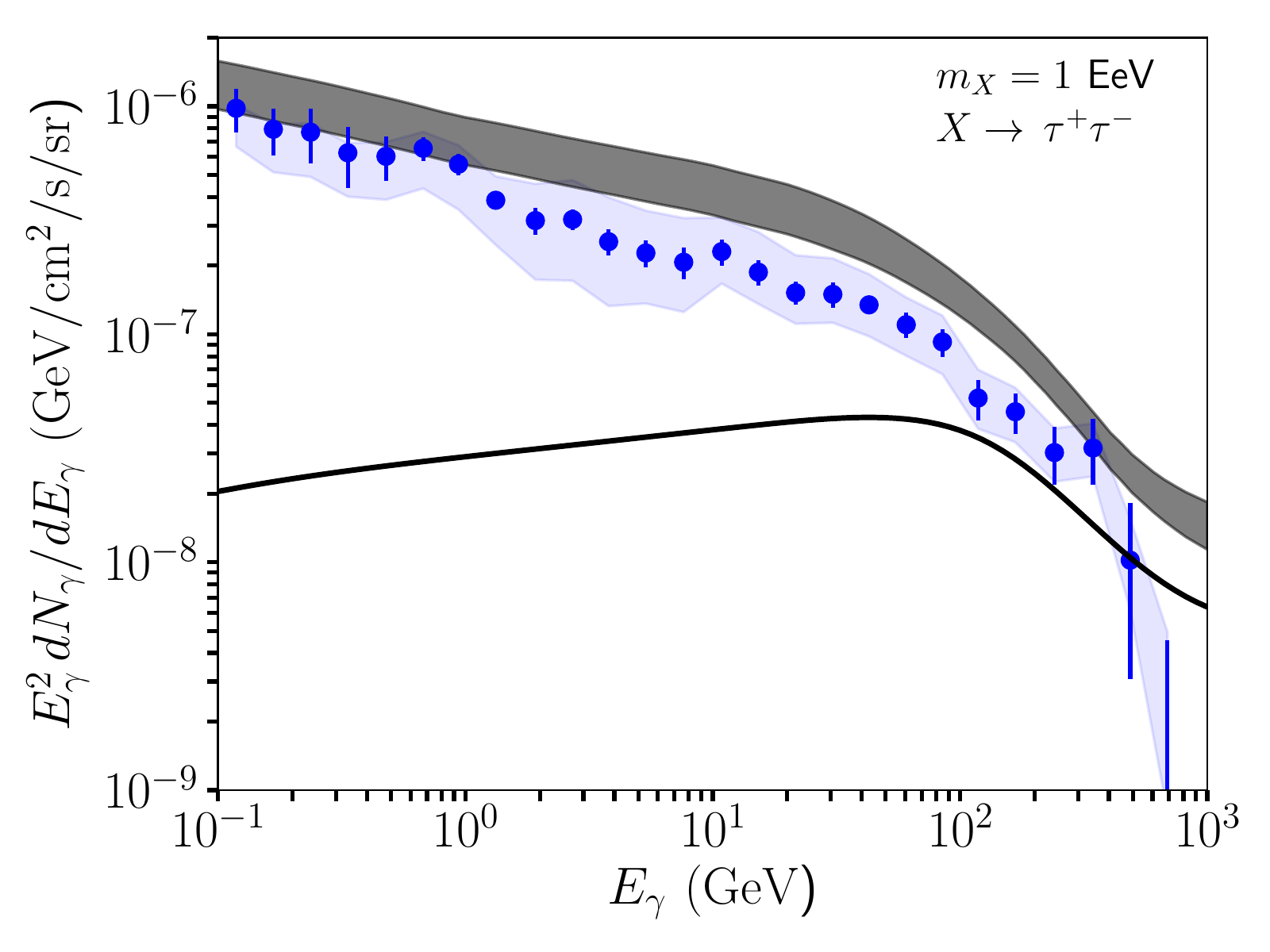}
\caption{As in Fig.~\ref{bb}, but for the case of decays to $\tau^+ \tau^-$.}
\label{tautau}
\vspace{1.0cm}
\end{figure}

\begin{figure}[t]
\vspace{2.5cm}
\includegraphics[scale=0.47]{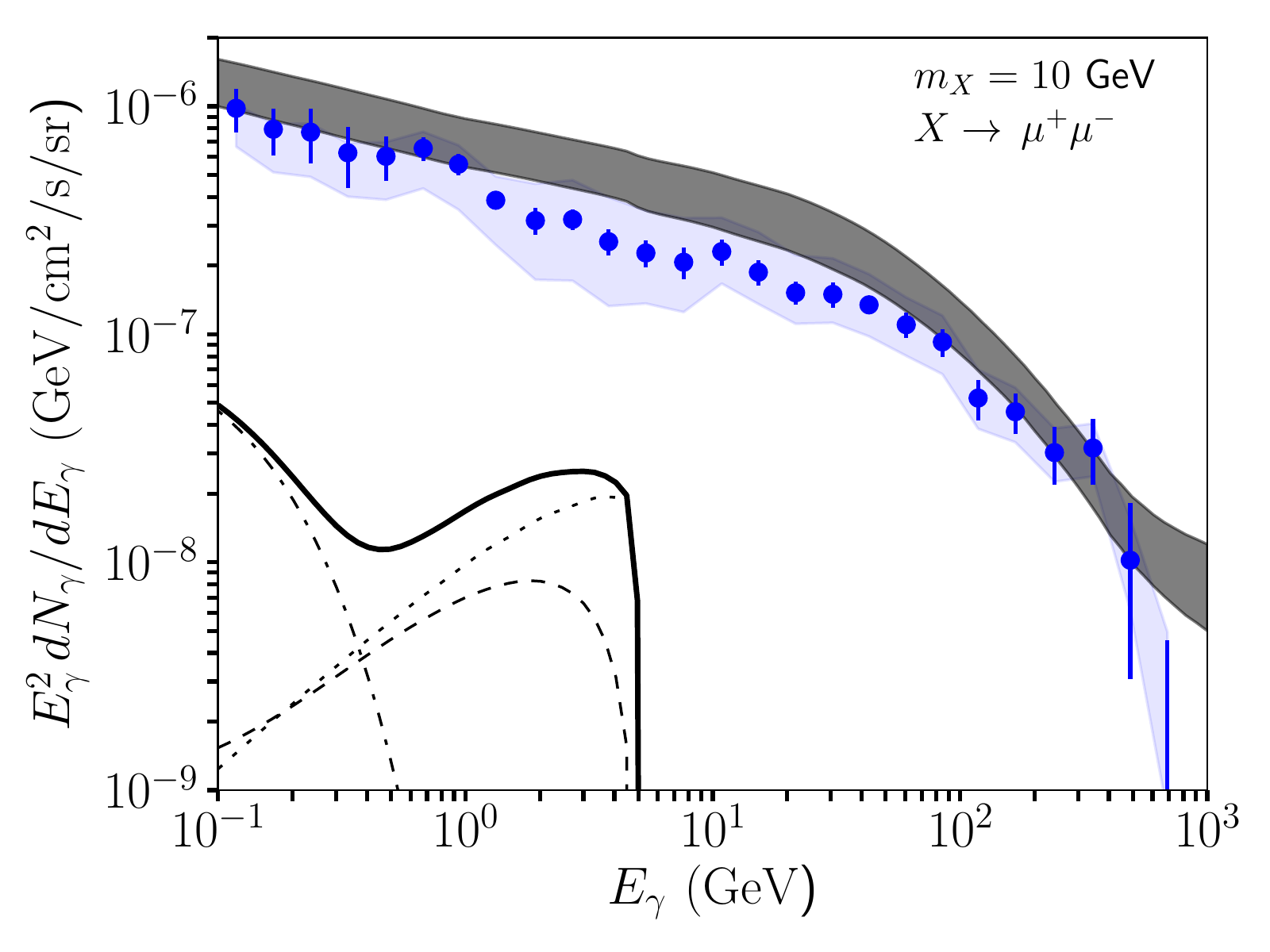} 
\includegraphics[scale=0.47]{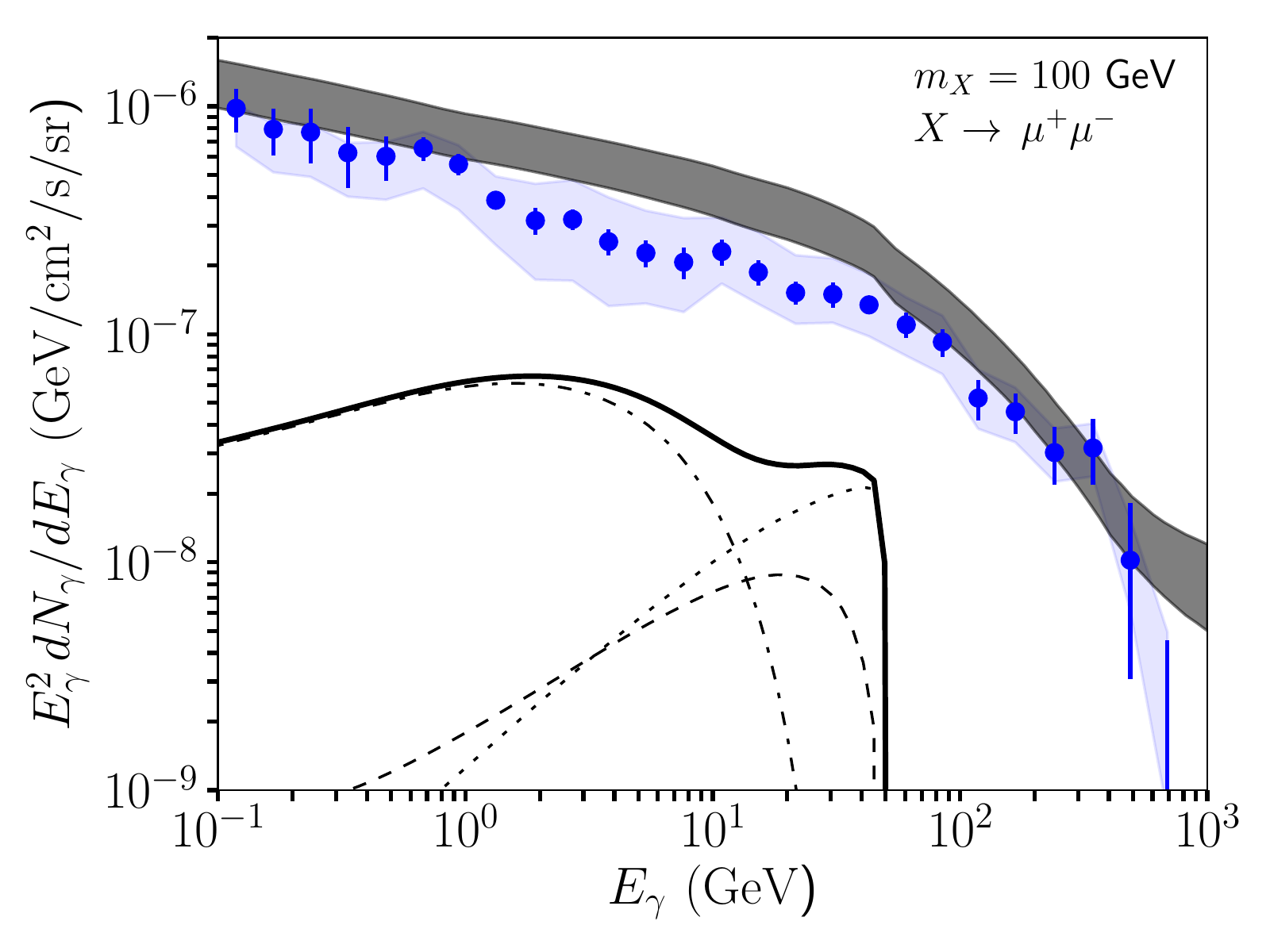} \\
\includegraphics[scale=0.47]{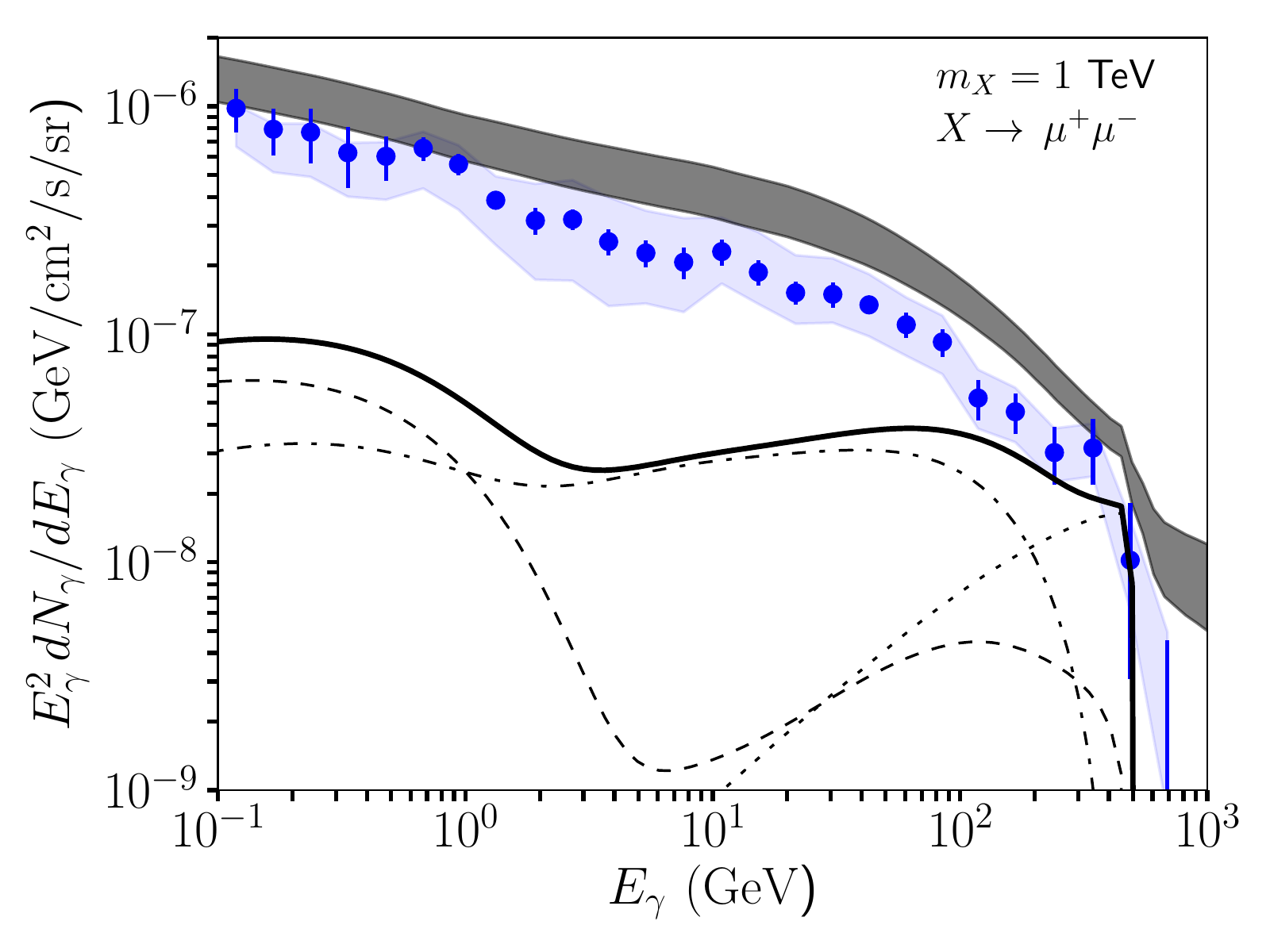} 
\includegraphics[scale=0.47]{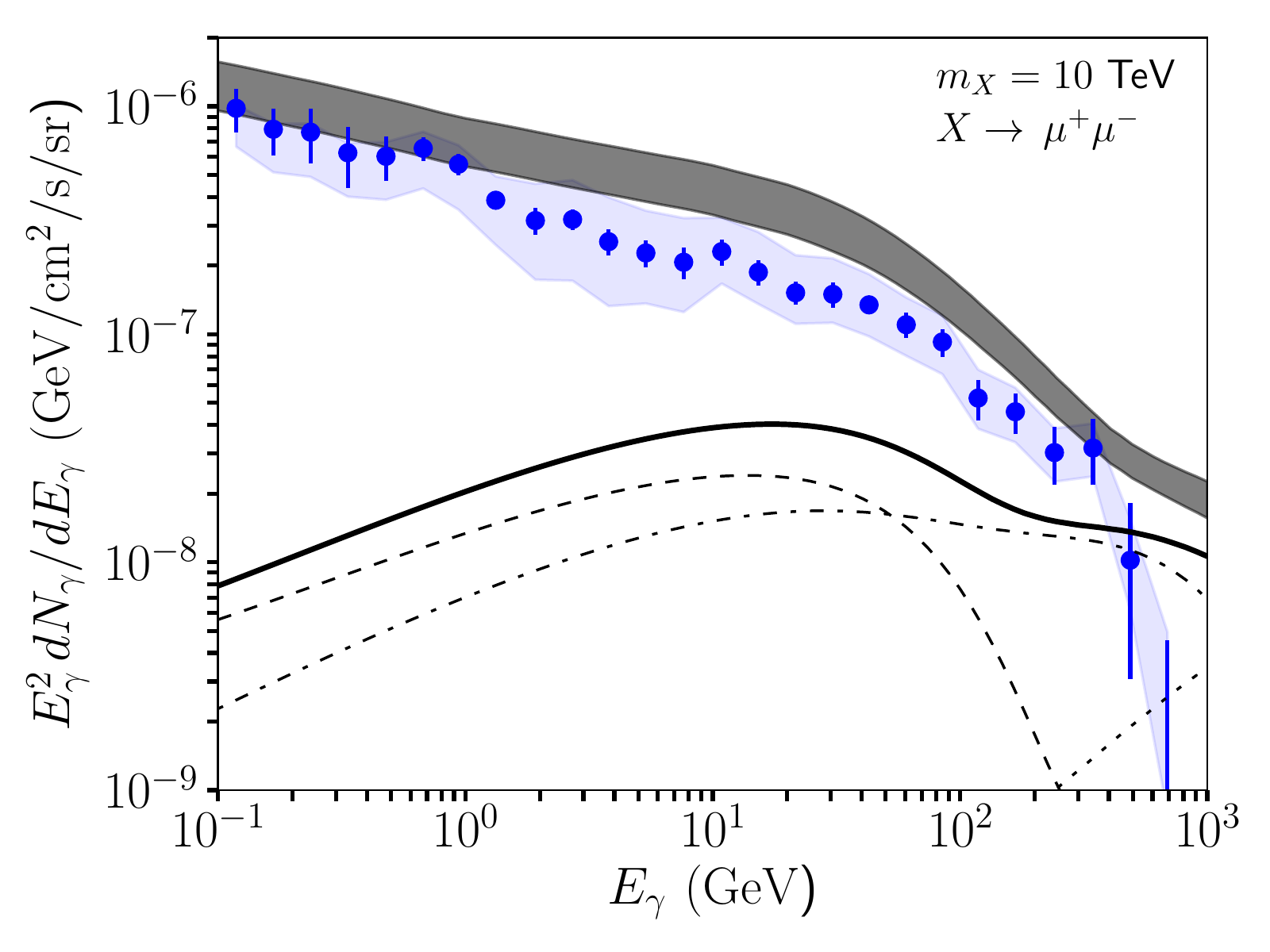} \\
\includegraphics[scale=0.47]{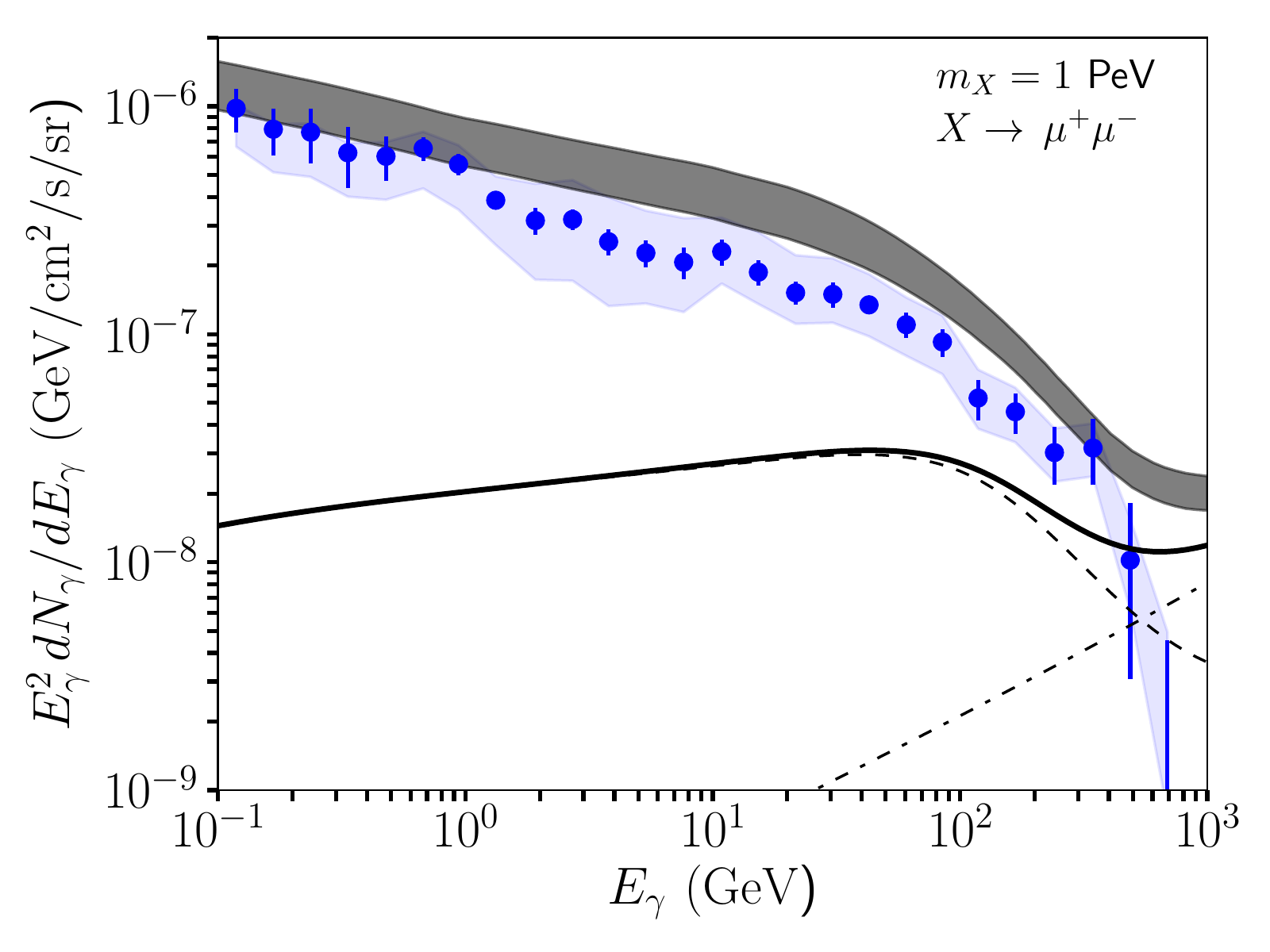} 
\includegraphics[scale=0.47]{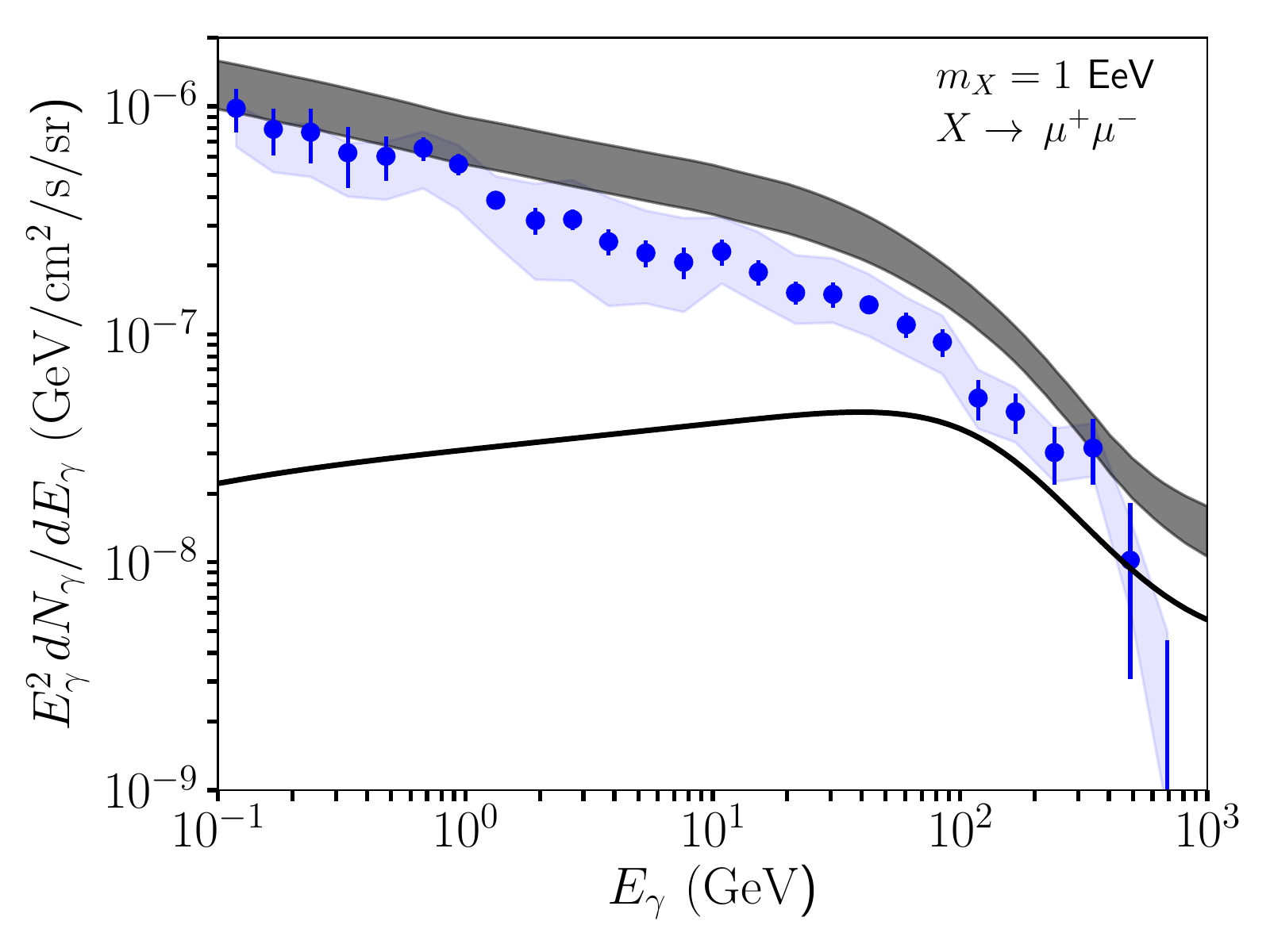}
\caption{As in Fig.~\ref{bb}, but for the case of decays to $\mu^+ \mu^-$.}
\label{mumu}
\vspace{1.0cm}
\end{figure}

\begin{figure}[t]
\vspace{2.5cm}
\includegraphics[scale=0.47]{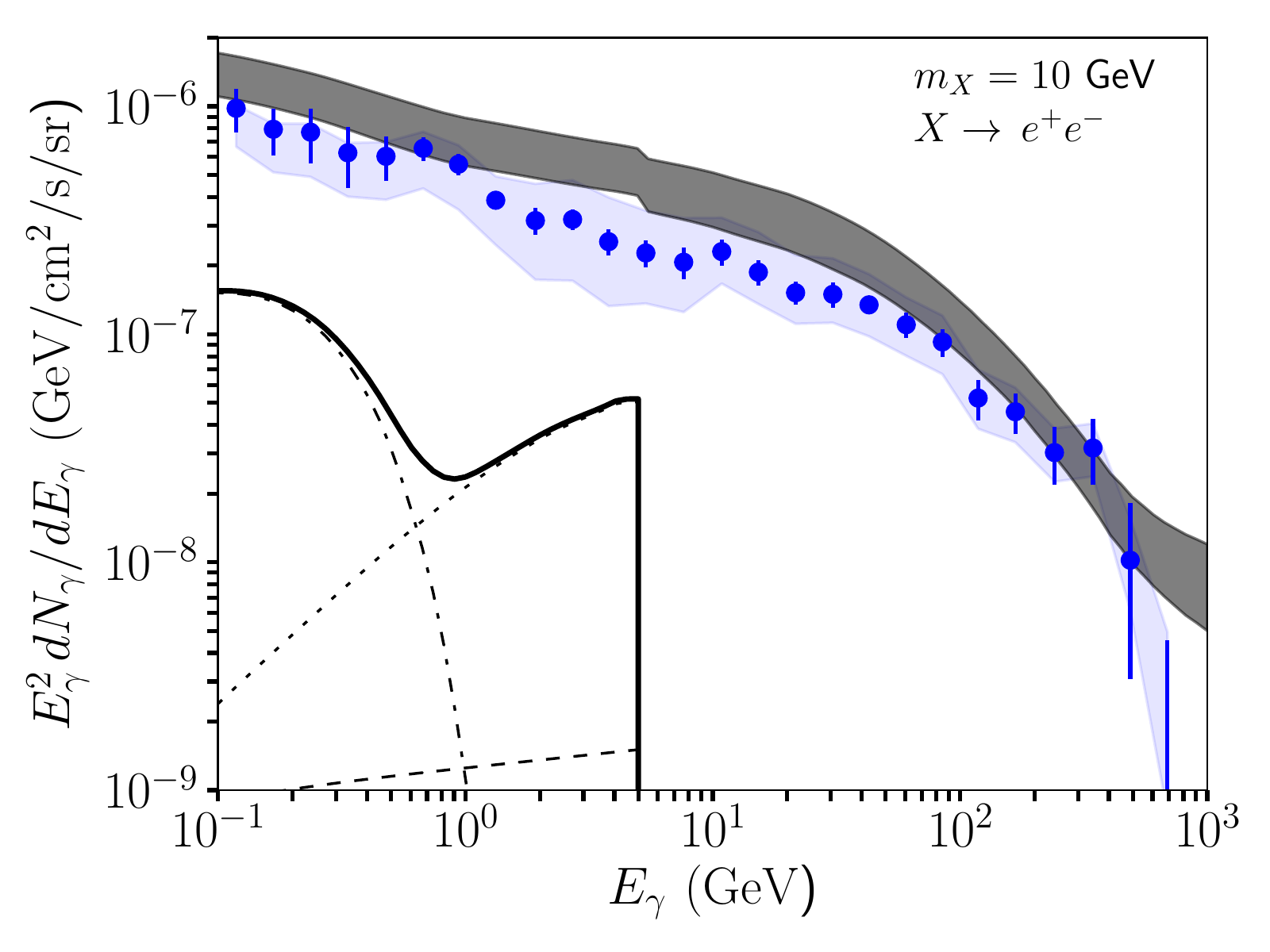} 
\includegraphics[scale=0.47]{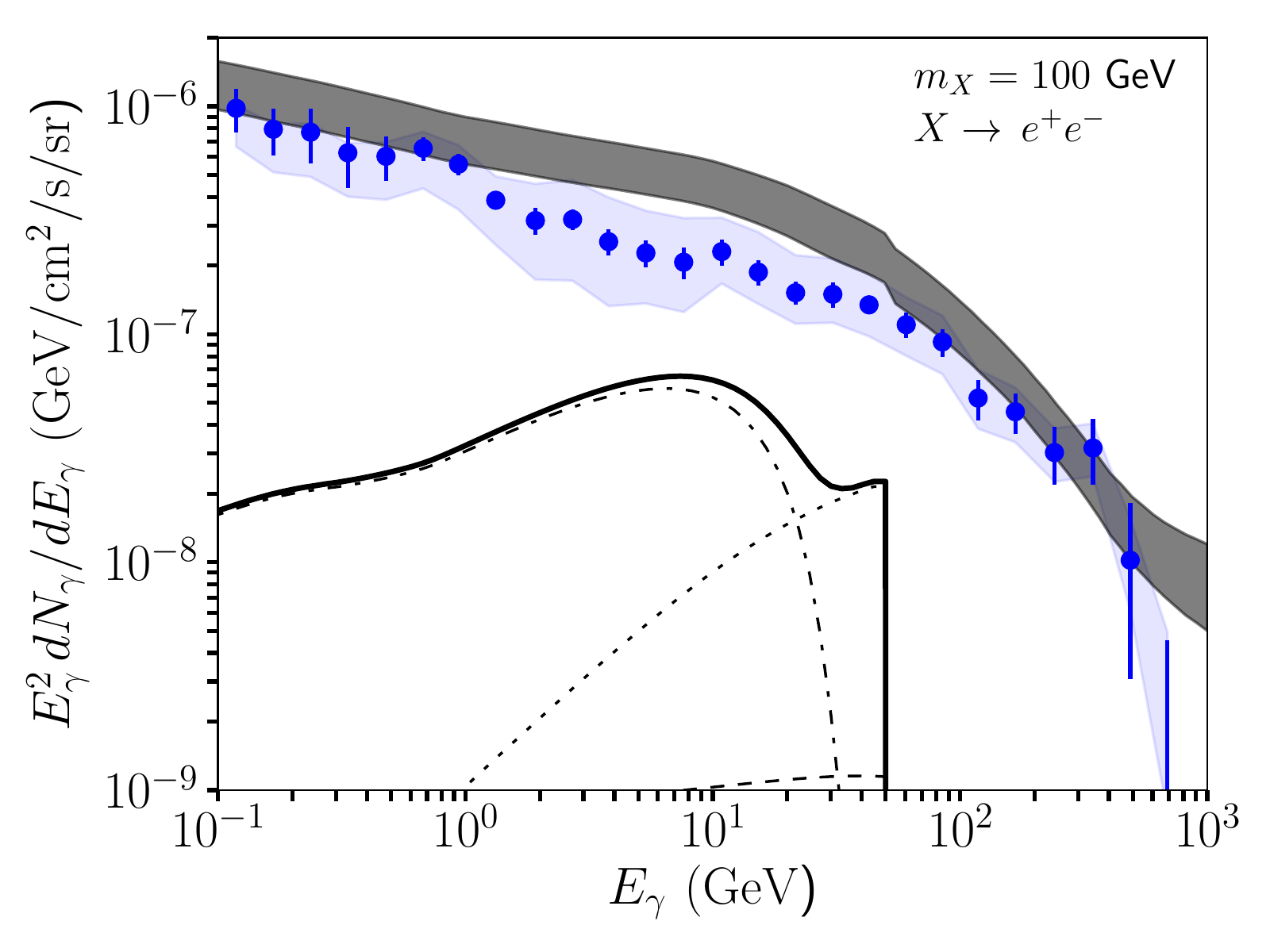} \\
\includegraphics[scale=0.47]{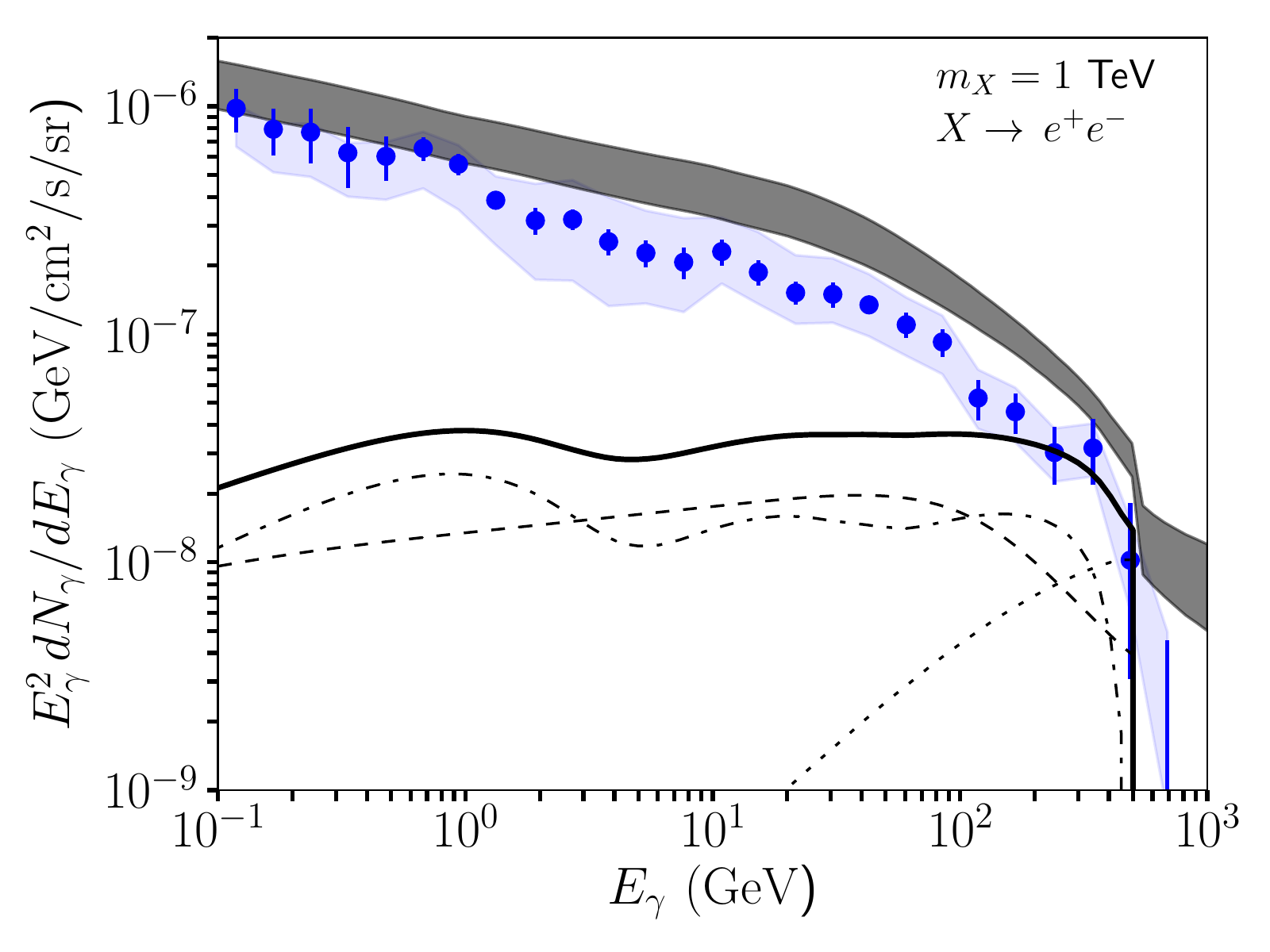} 
\includegraphics[scale=0.47]{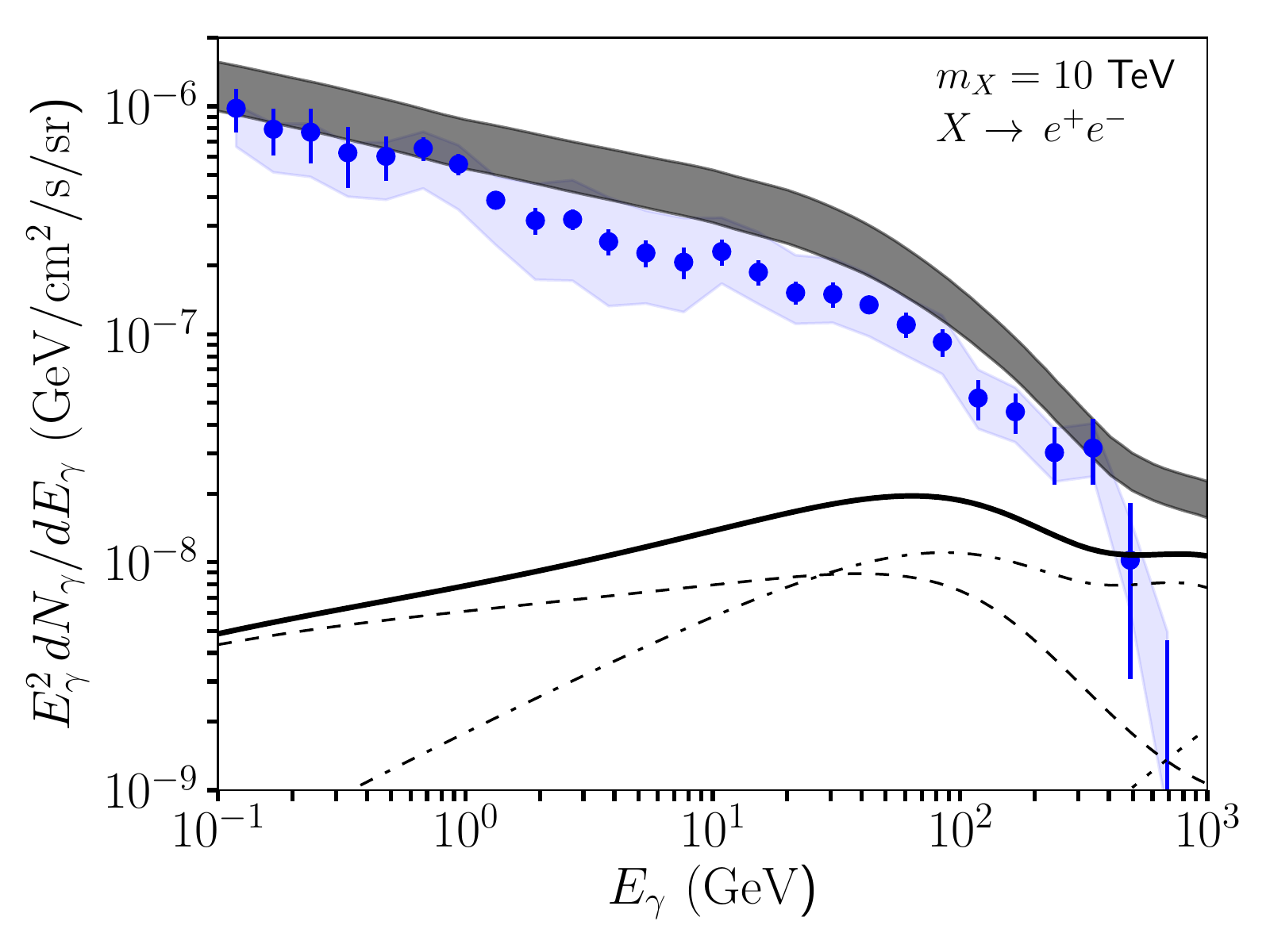} \\
\includegraphics[scale=0.47]{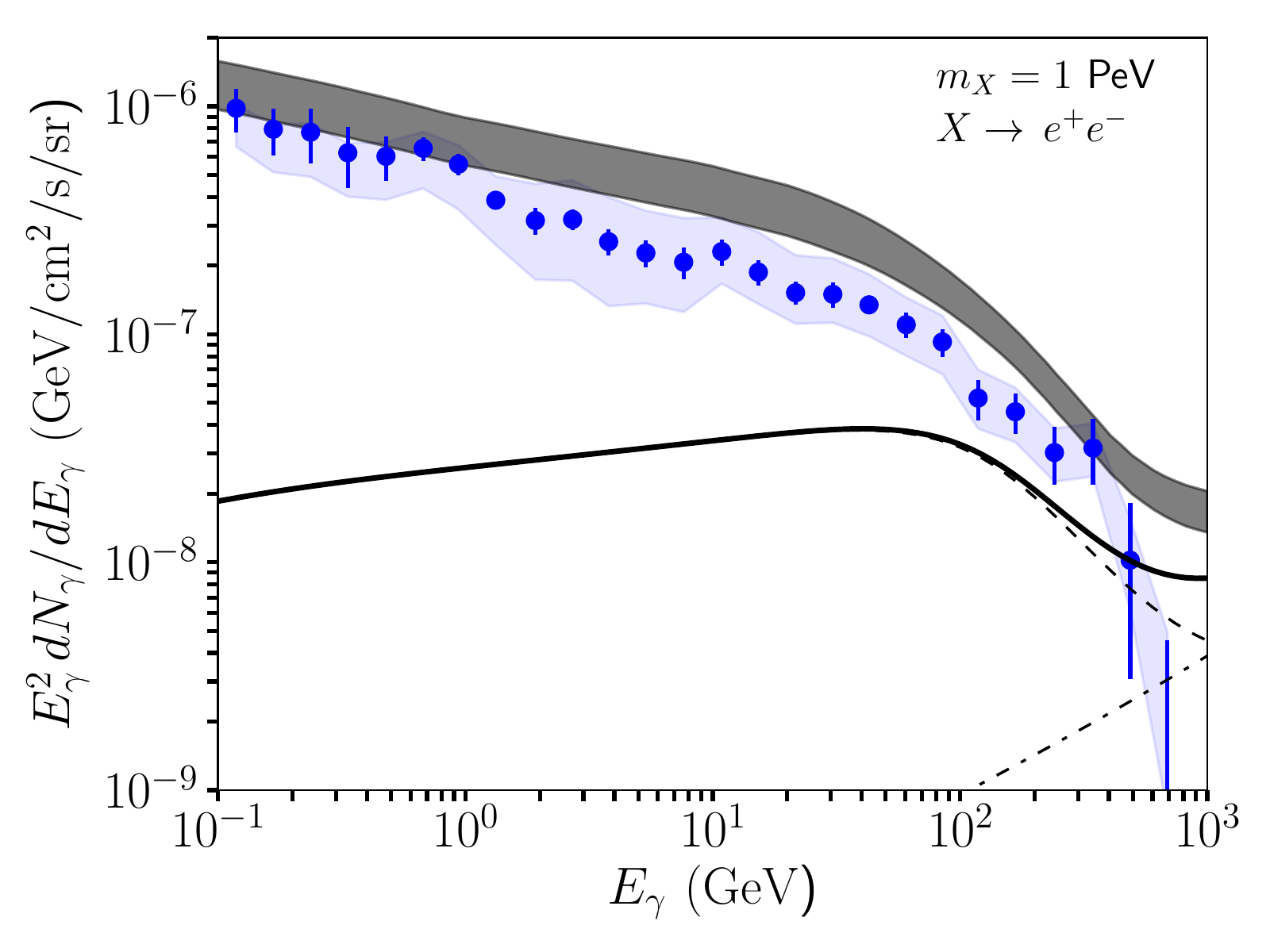} 
\includegraphics[scale=0.47]{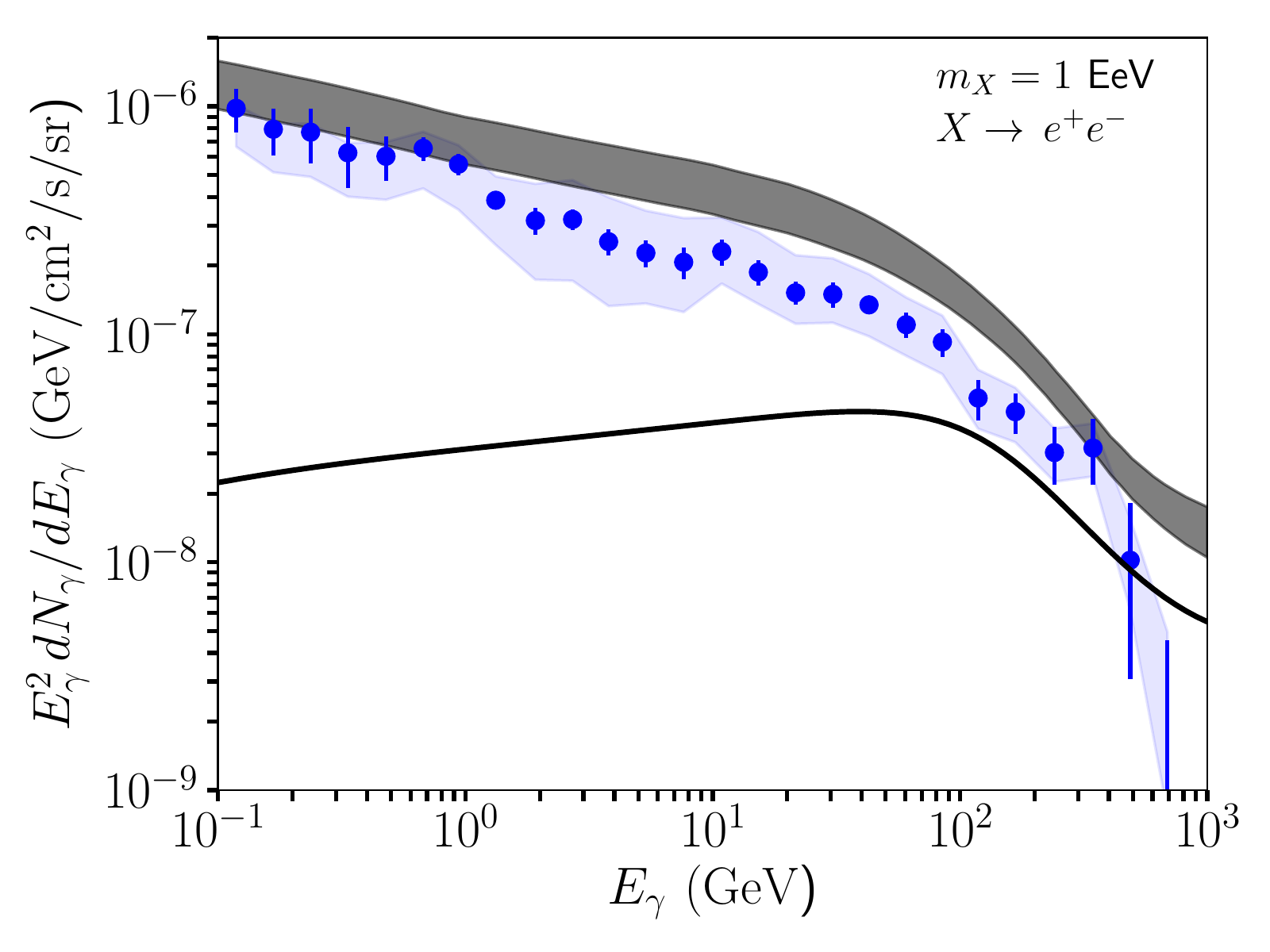}
\caption{As in Fig.~\ref{bb}, but for the case of decays to $e^+ e^-$.}
\label{ee}
\vspace{1.0cm}
\end{figure}

\begin{figure}[t]
\includegraphics[scale=0.43]{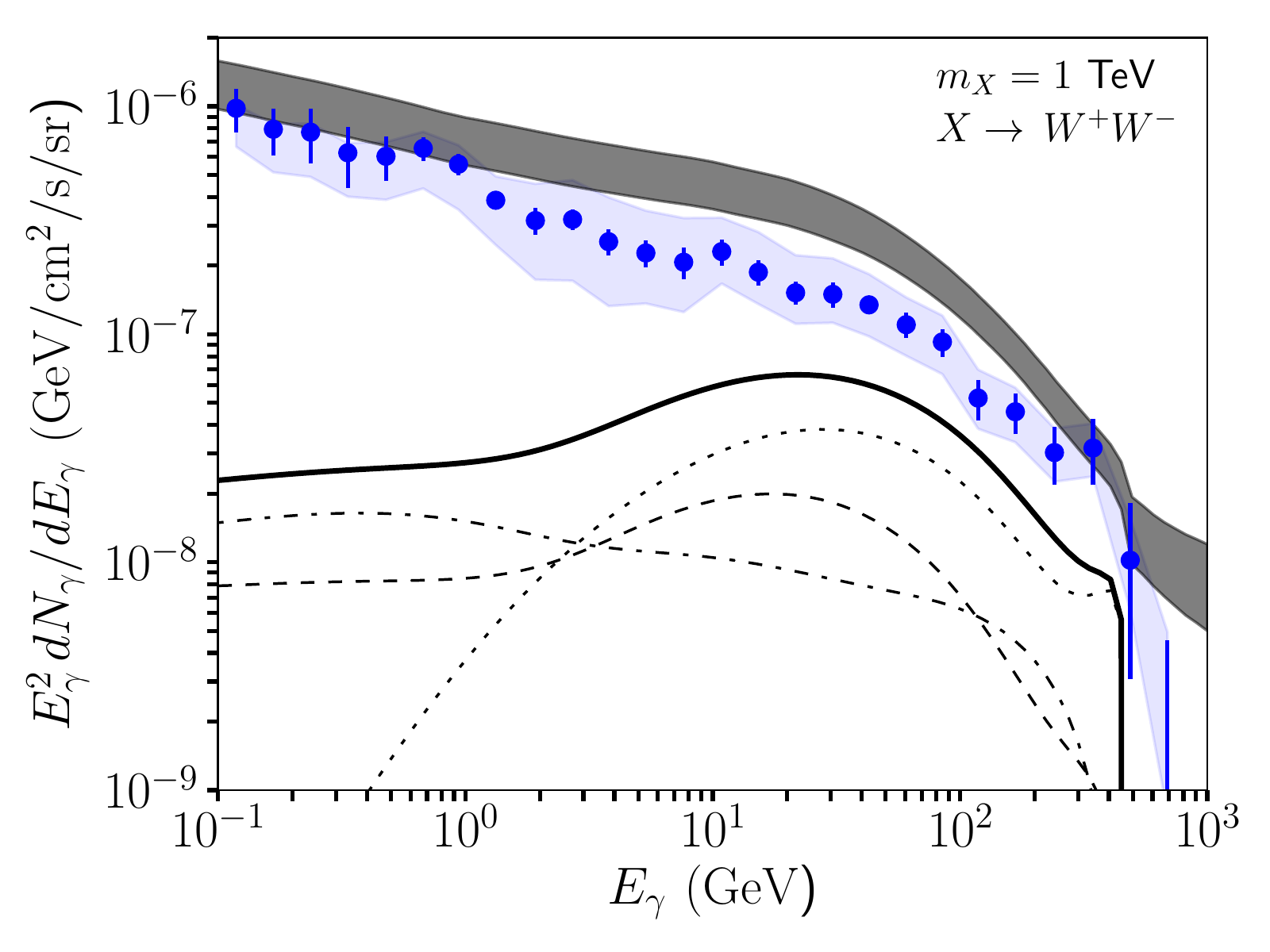} 
\includegraphics[scale=0.43]{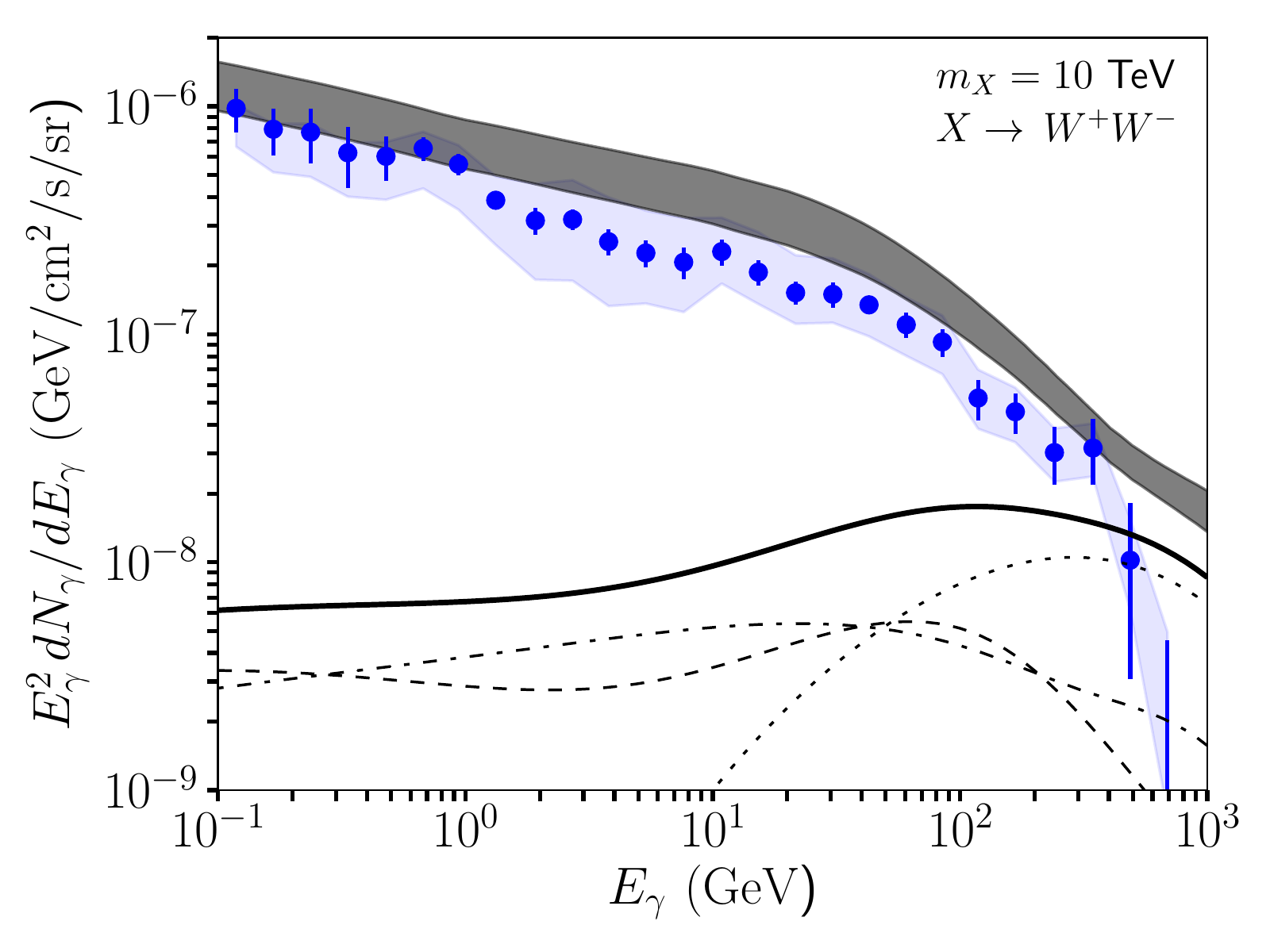} \\
\includegraphics[scale=0.43]{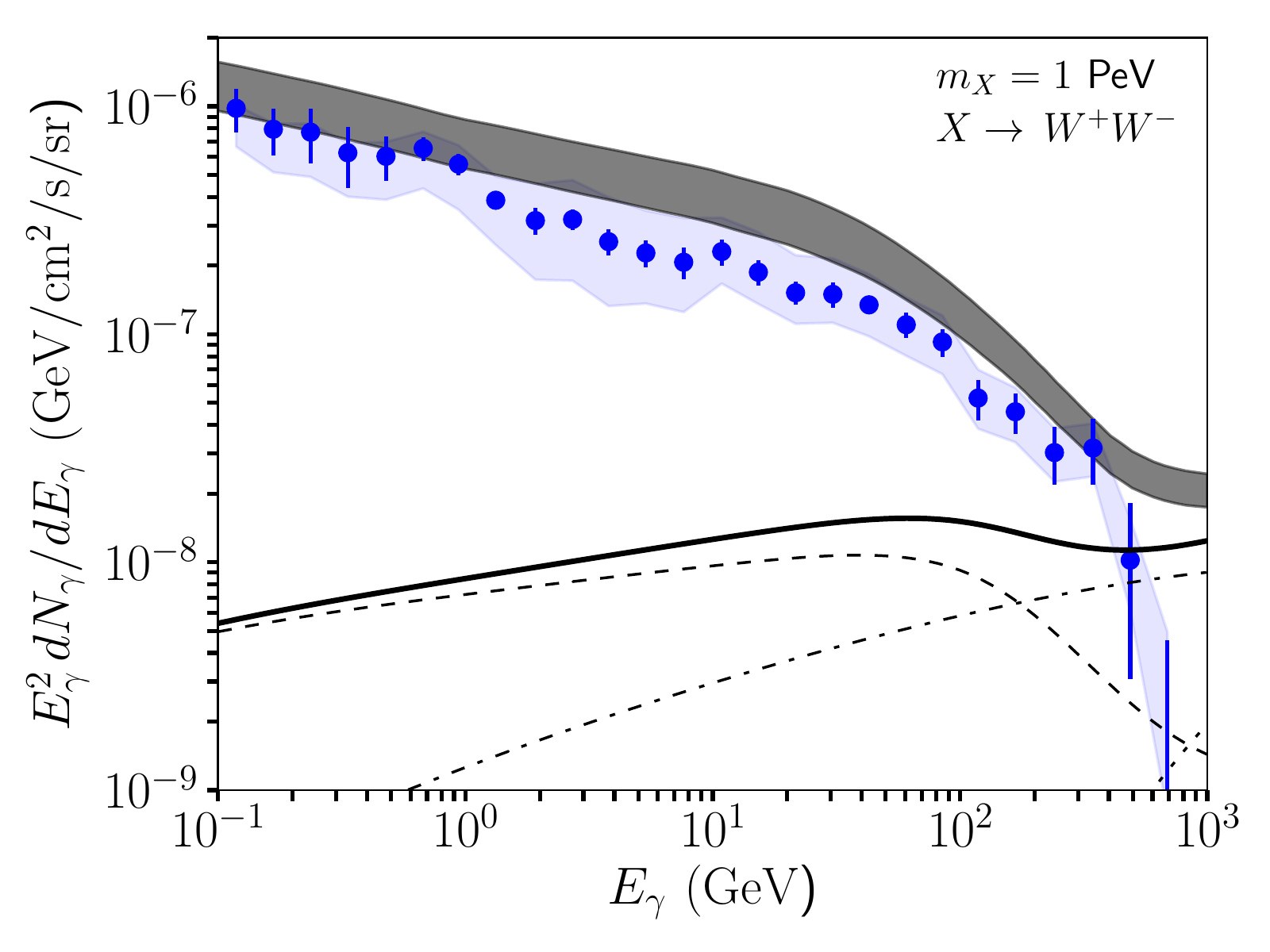} 
\includegraphics[scale=0.43]{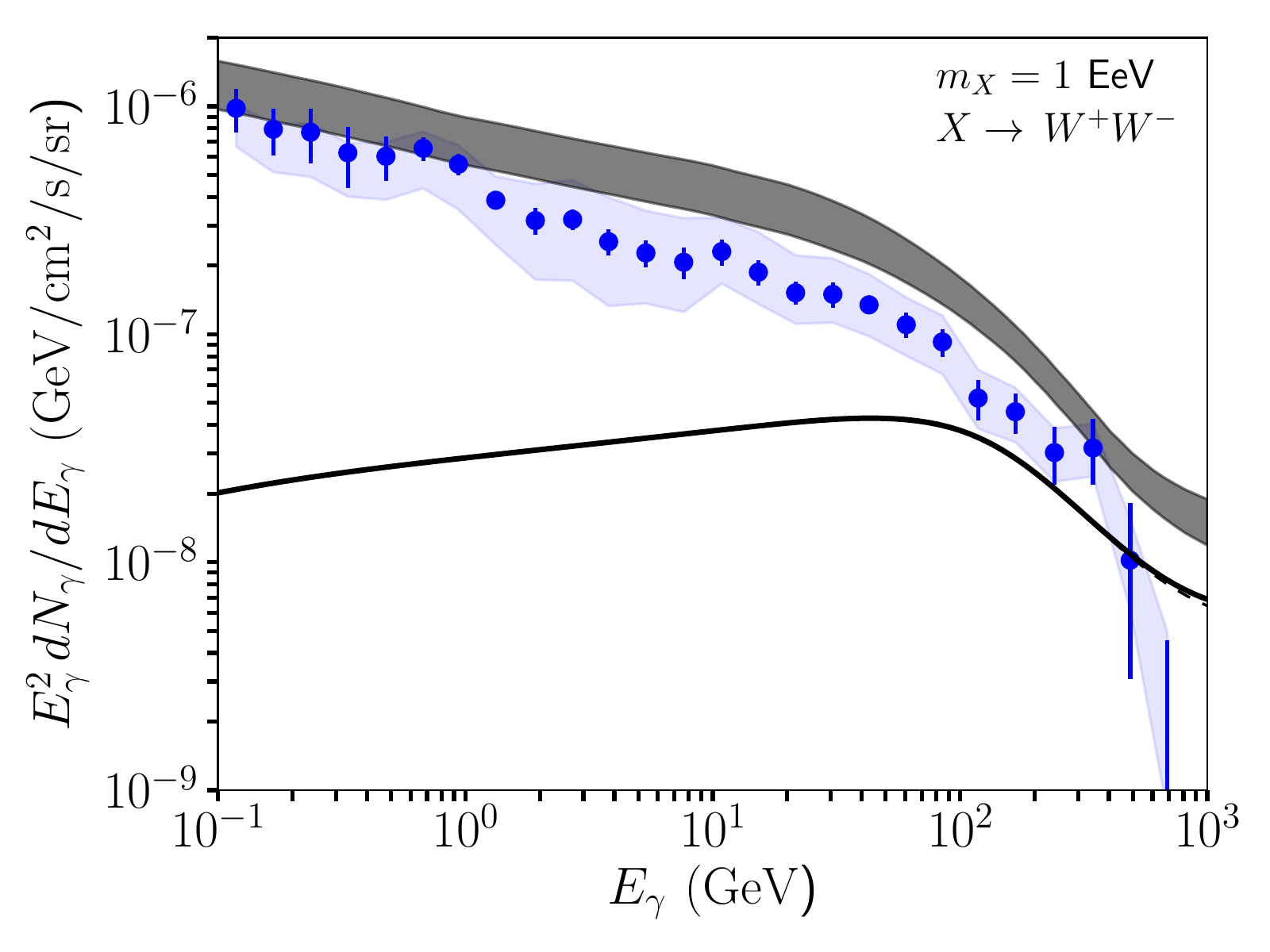} \\
\includegraphics[scale=0.43]{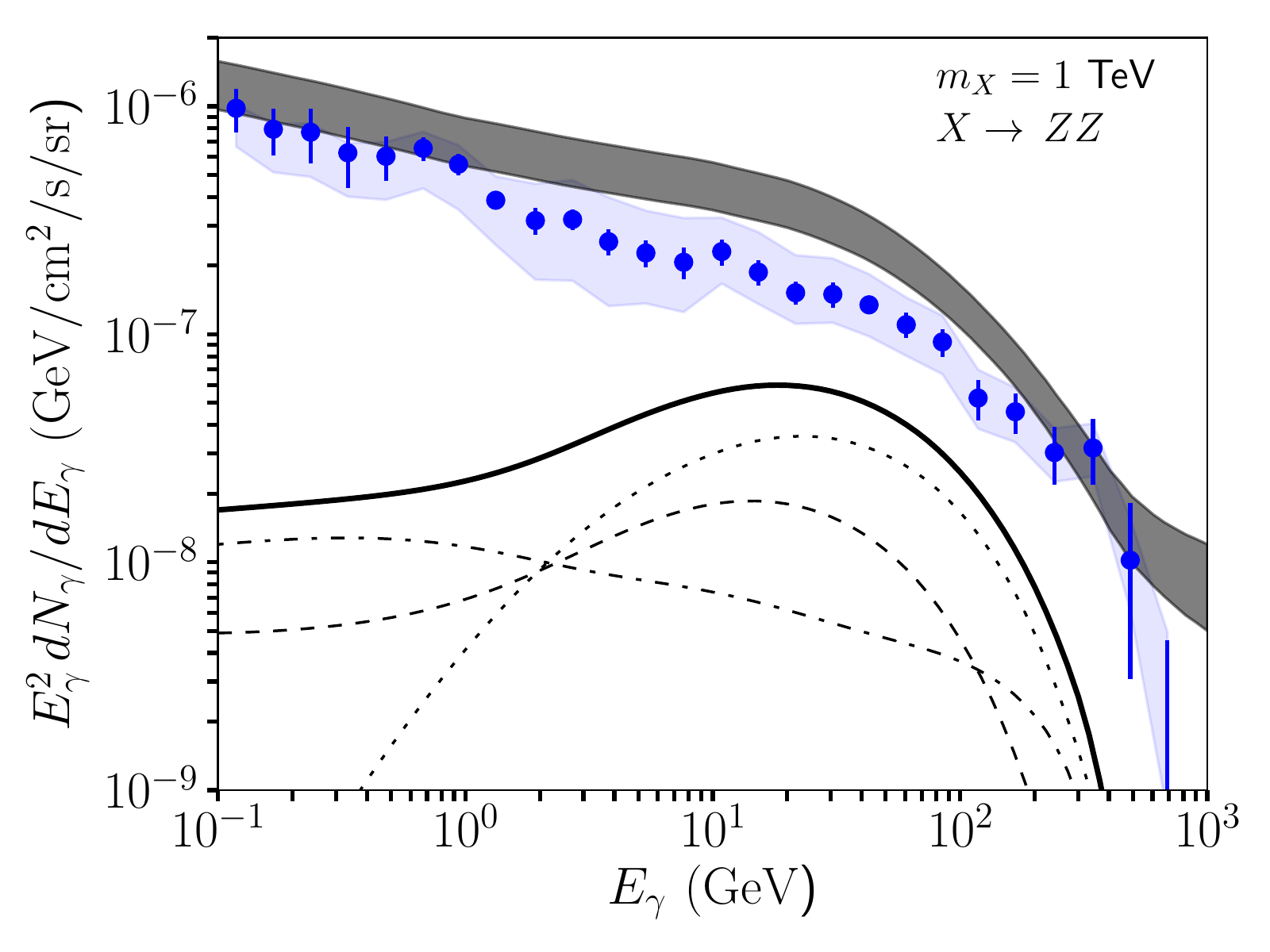} 
\includegraphics[scale=0.43]{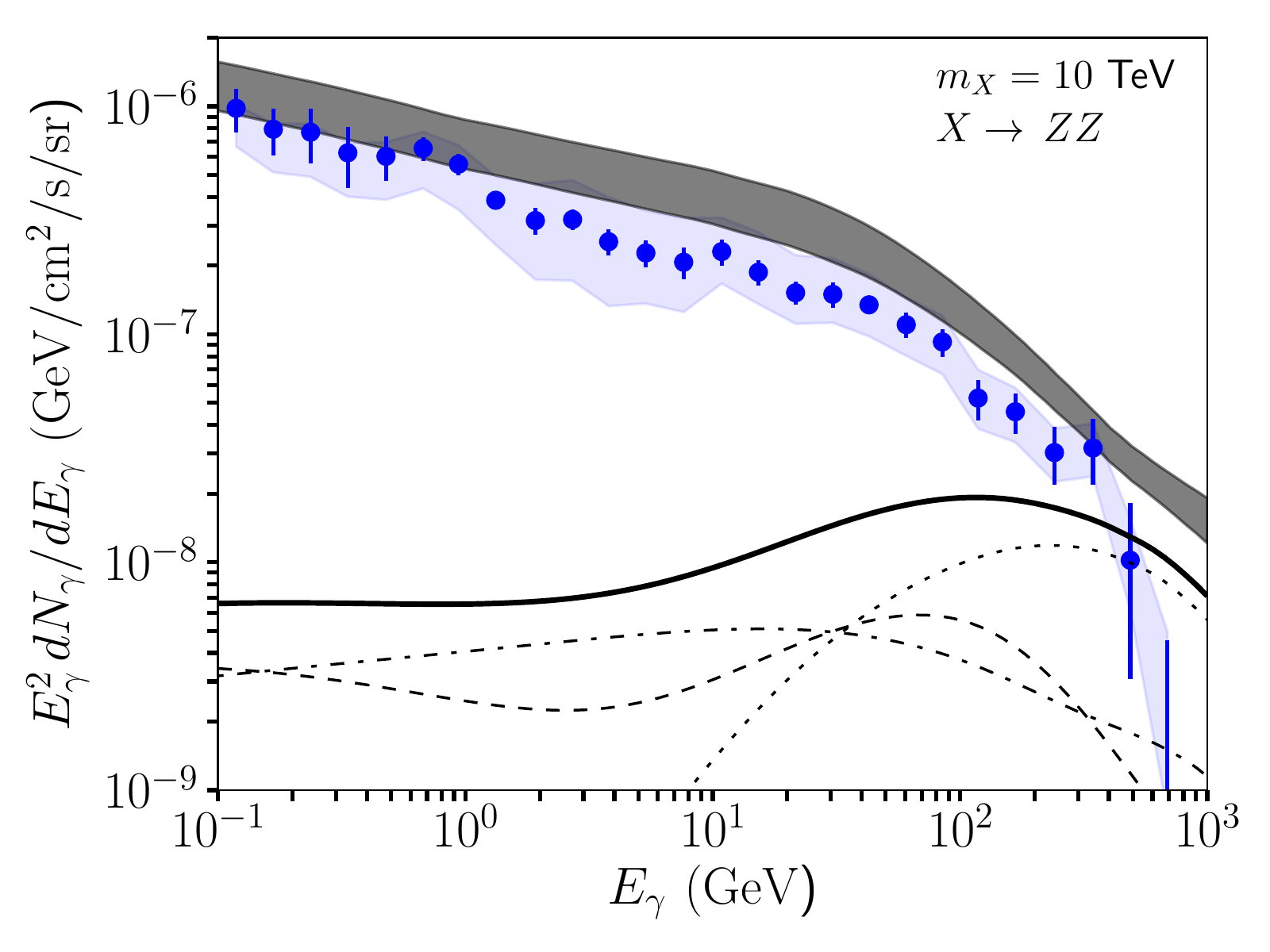} \\
\includegraphics[scale=0.43]{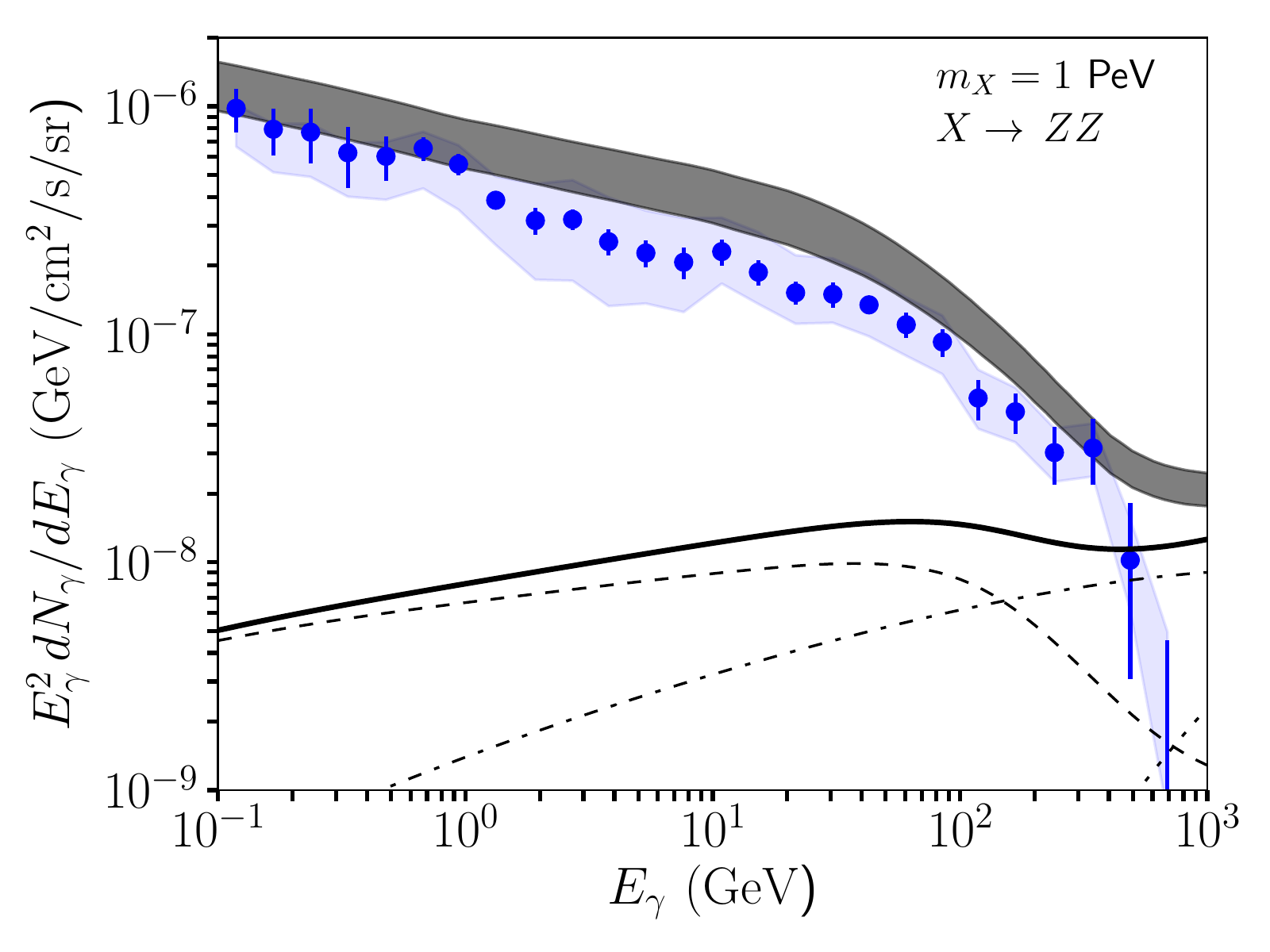} 
\includegraphics[scale=0.43]{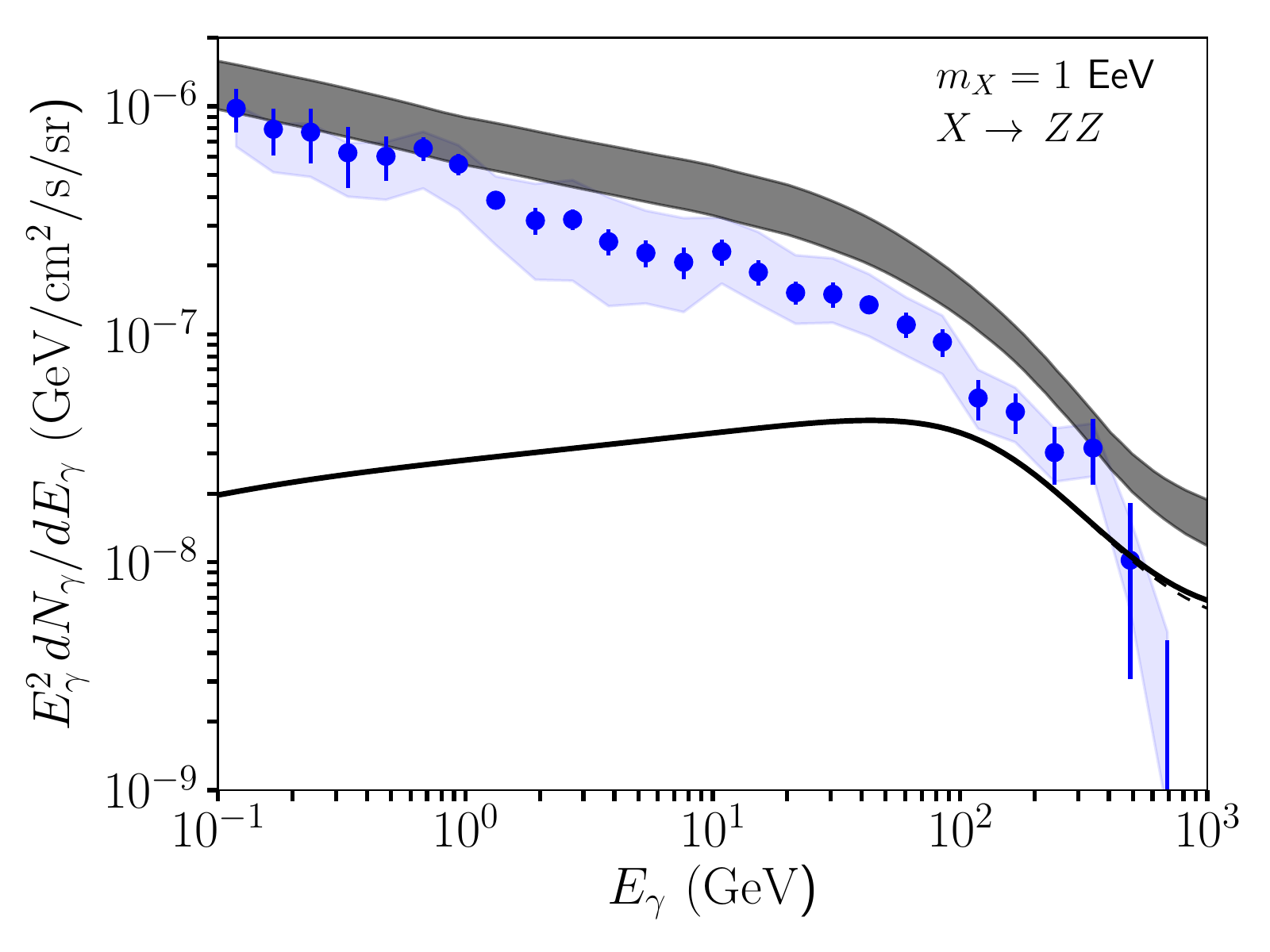}
\caption{As in Fig.~\ref{bb}, but for the case of decays to $W^+W^-$ or $ZZ$.}
\label{WWZZ}
\end{figure}

\begin{figure}[t]
\includegraphics[scale=0.43]{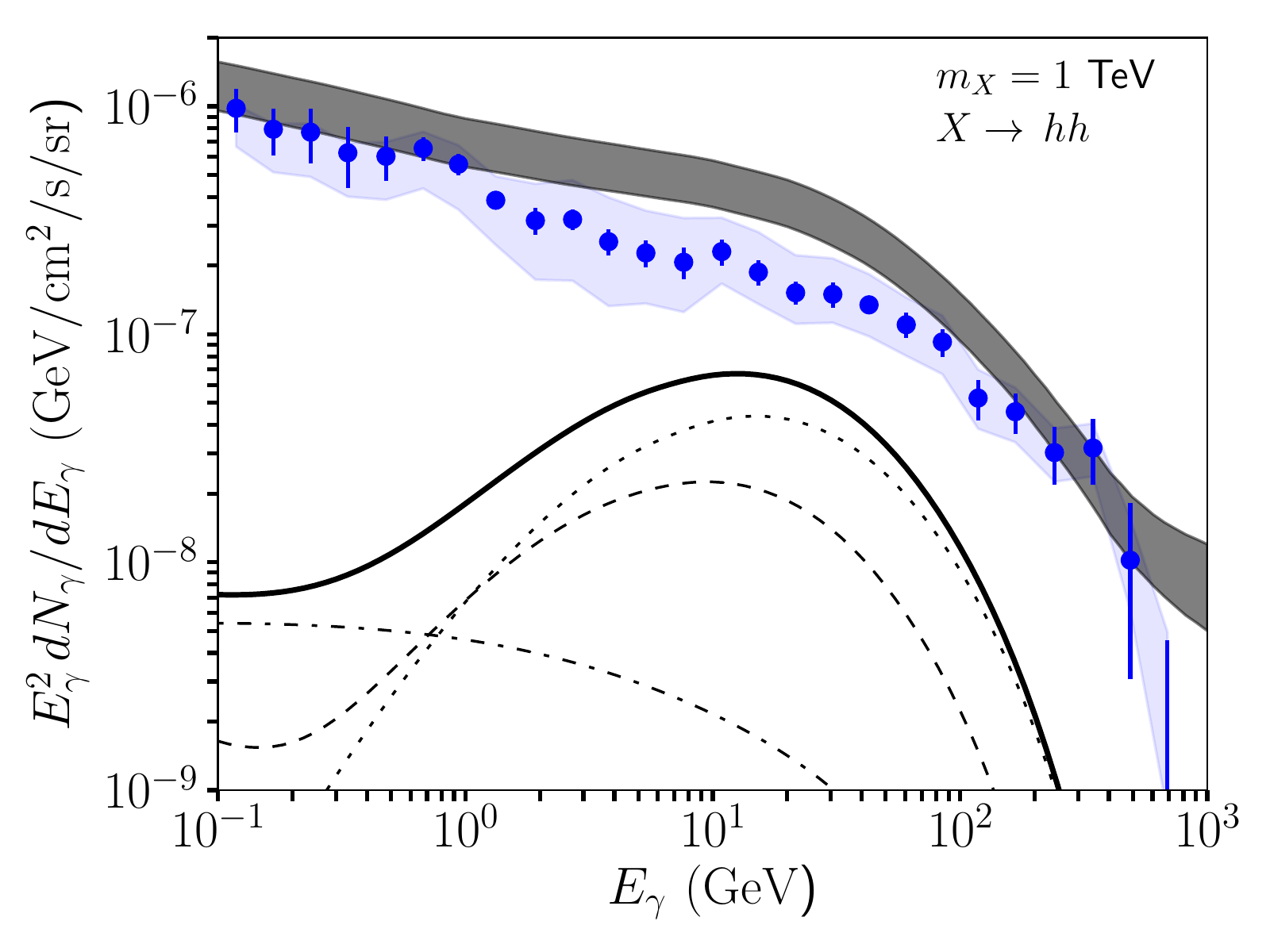} 
\includegraphics[scale=0.43]{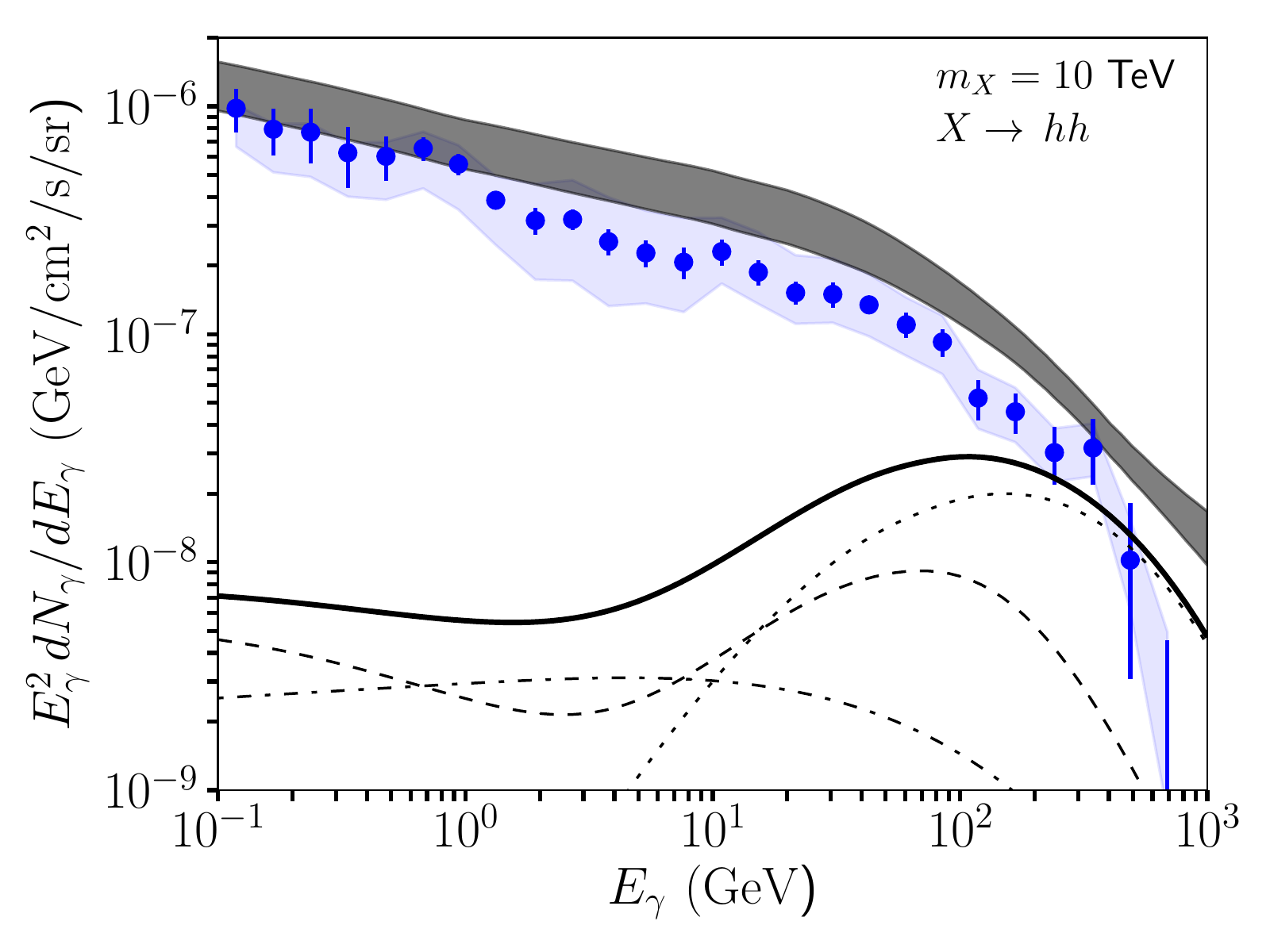} \\
\includegraphics[scale=0.43]{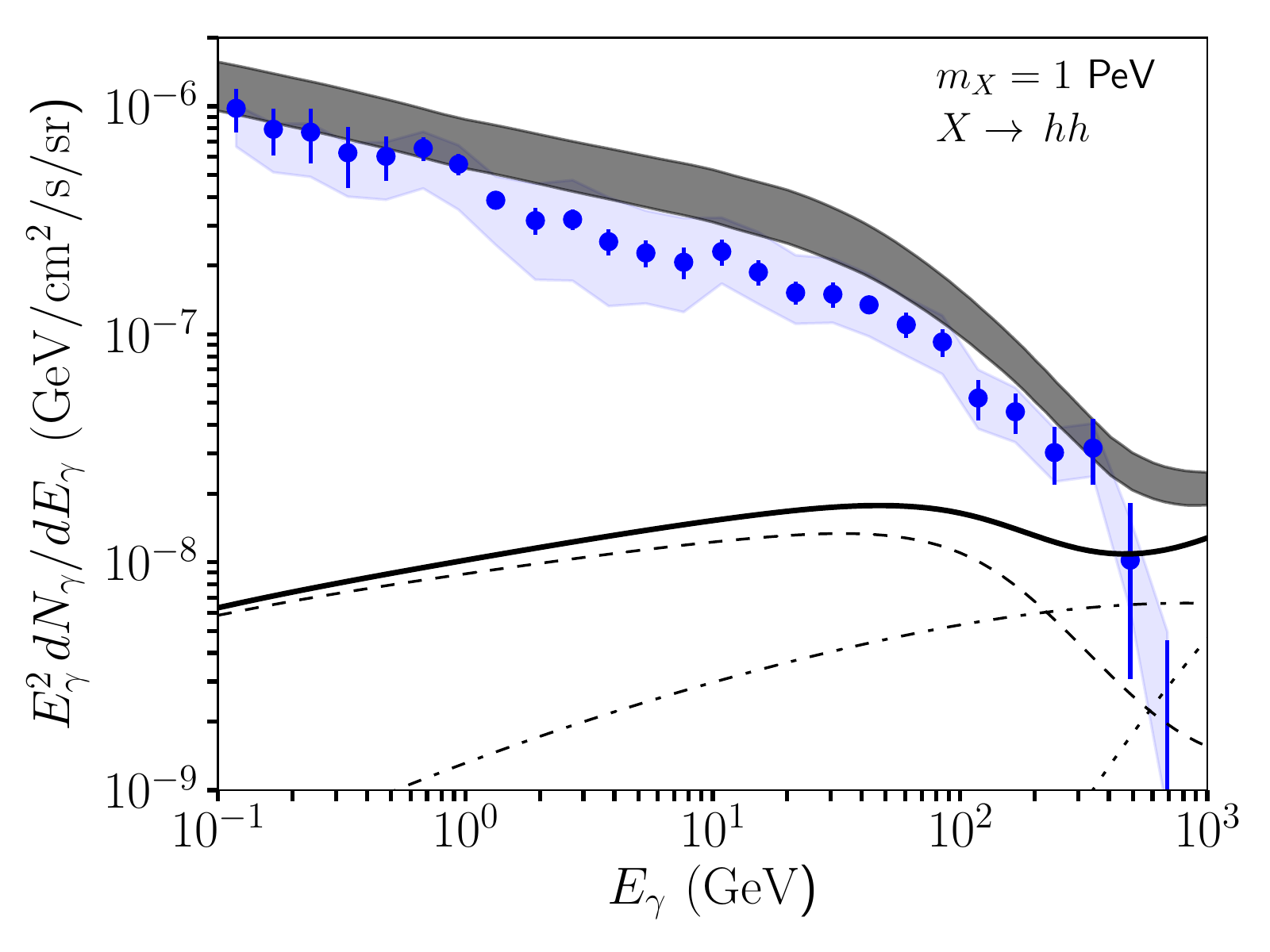} 
\includegraphics[scale=0.43]{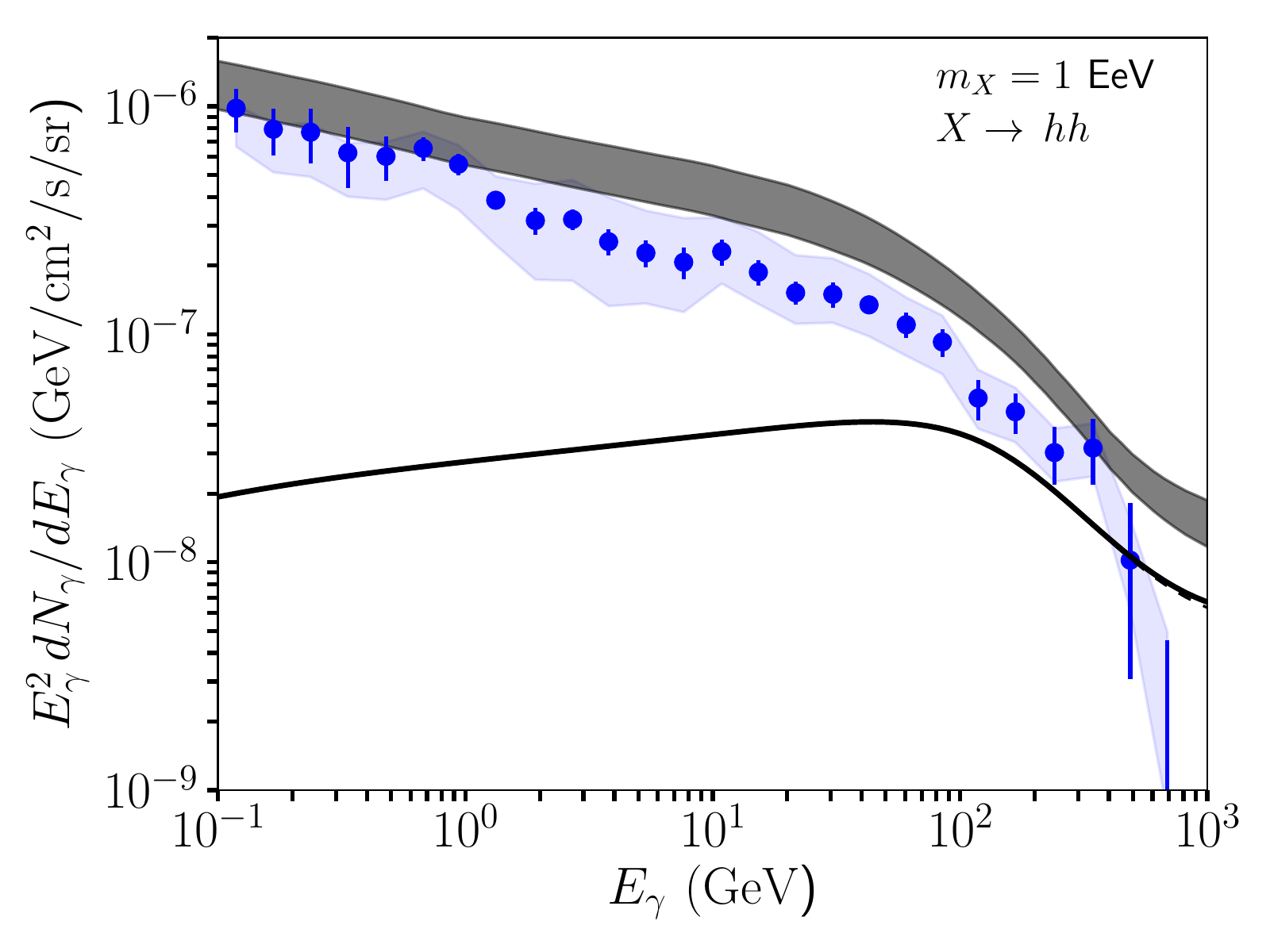} \\
\includegraphics[scale=0.43]{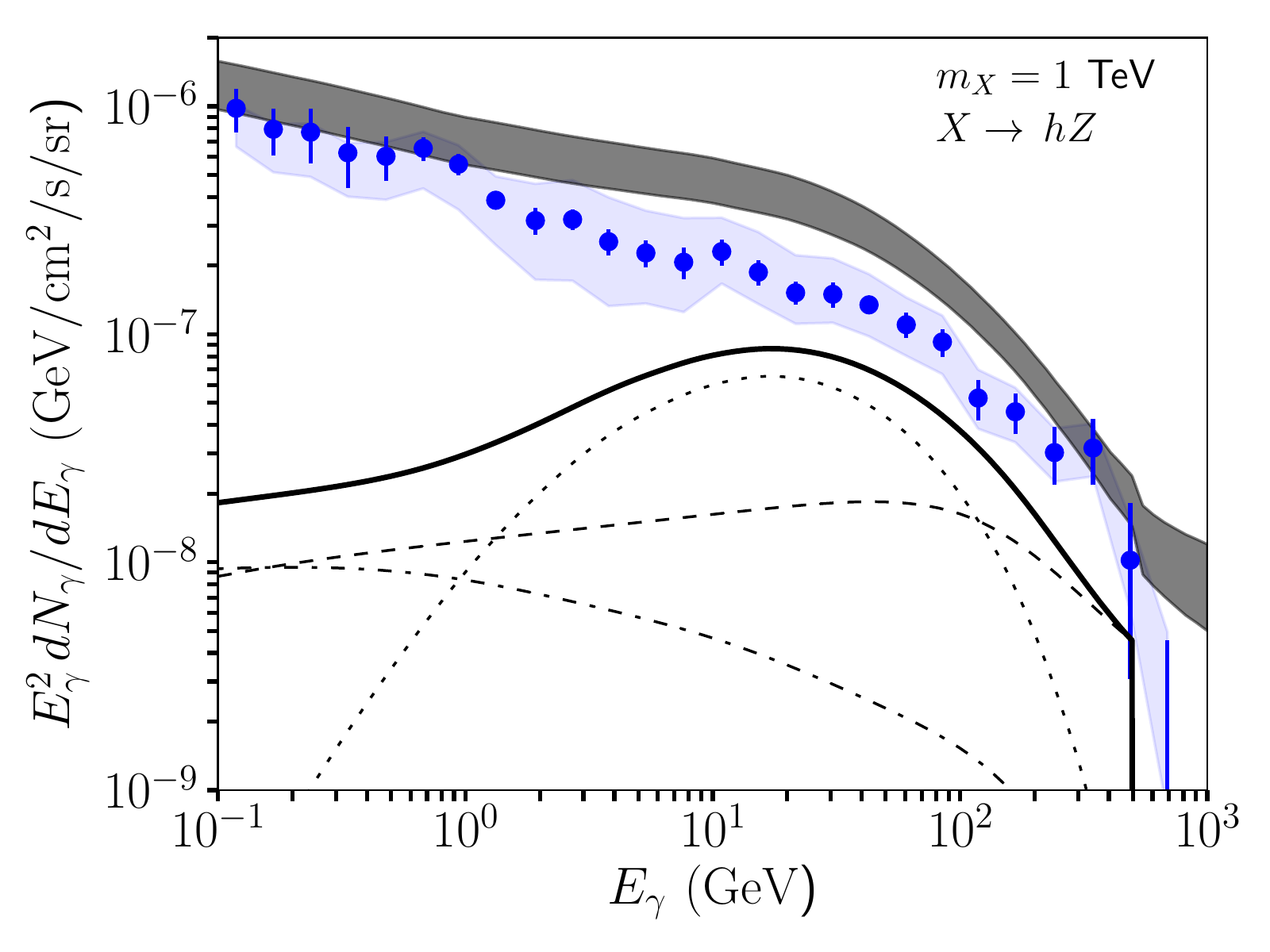} 
\includegraphics[scale=0.43]{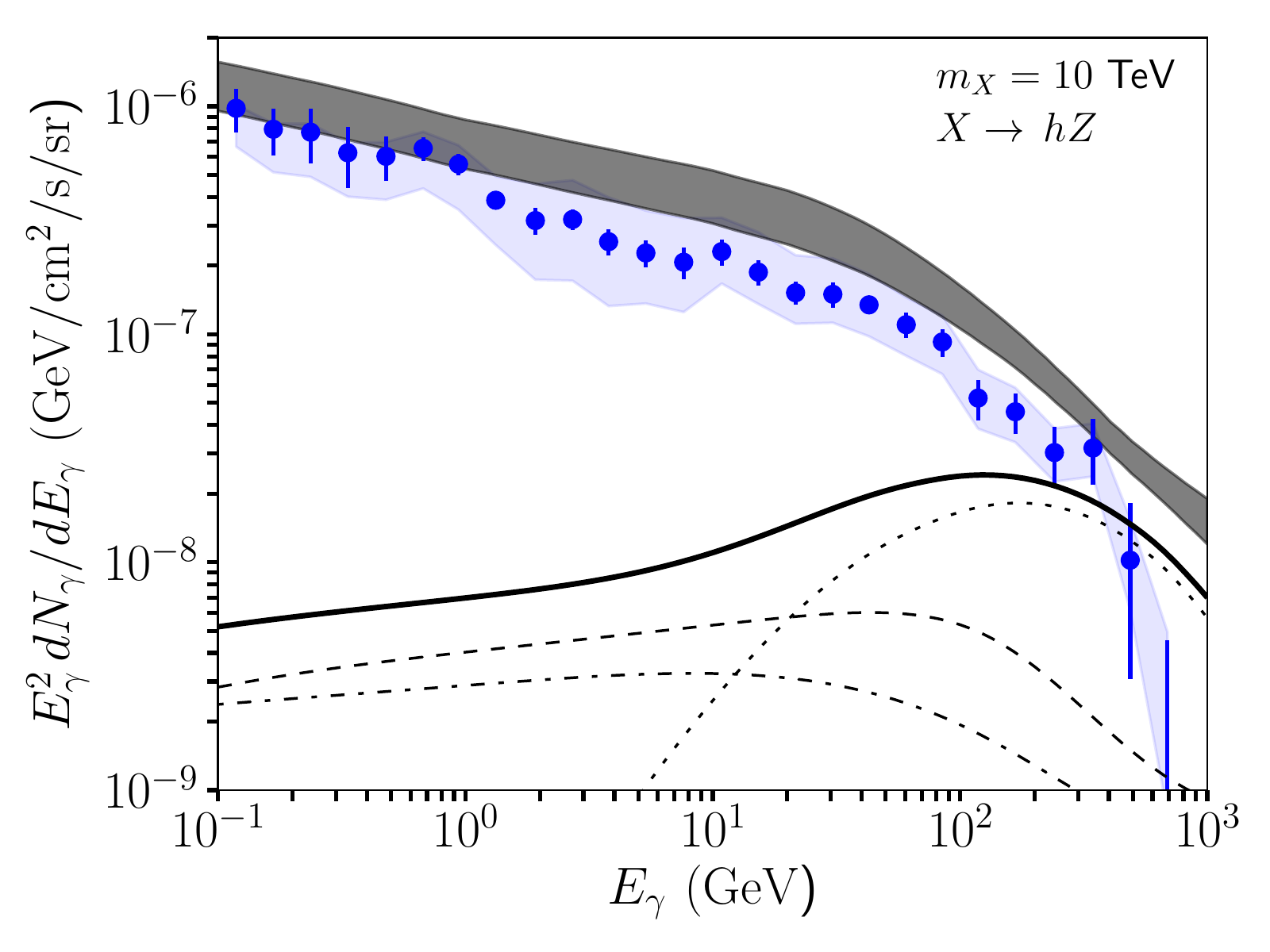} \\
\includegraphics[scale=0.43]{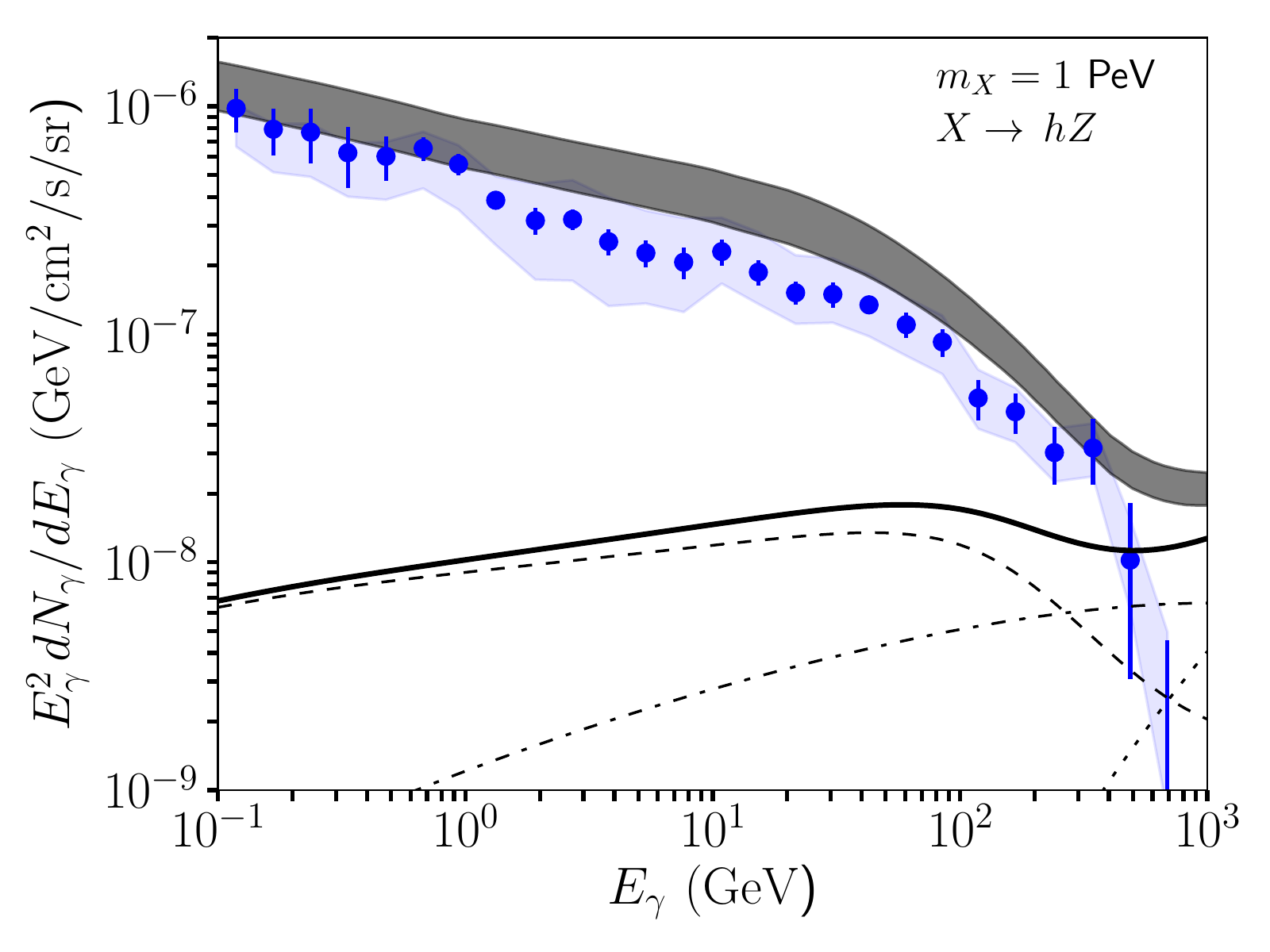} 
\includegraphics[scale=0.43]{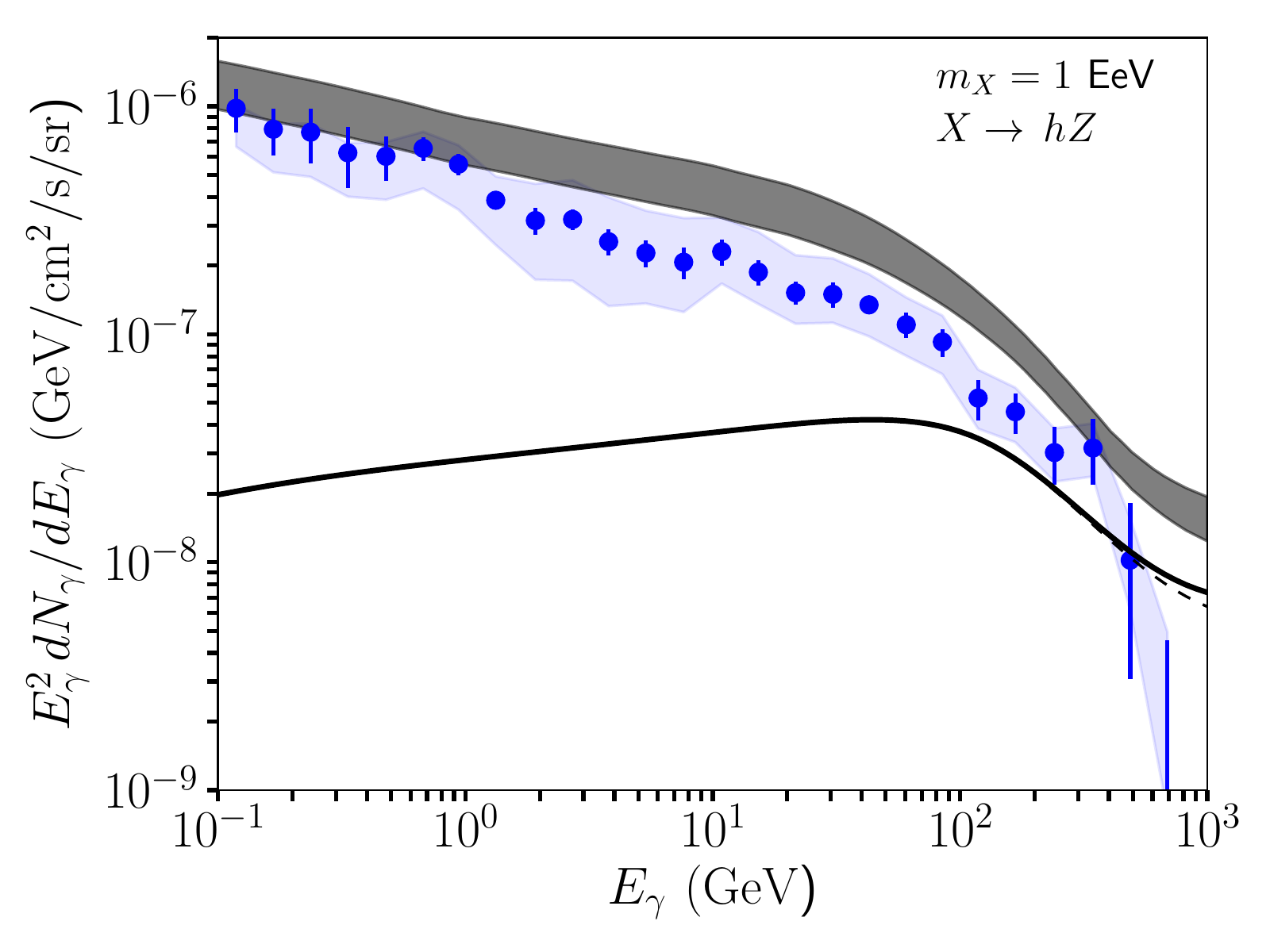}
\caption{As in Fig.~\ref{bb}, but for the case of decays to $hh$ or $hZ$.}
\label{hhhZ}
\end{figure}

\begin{figure}[t]
\includegraphics[scale=0.31]{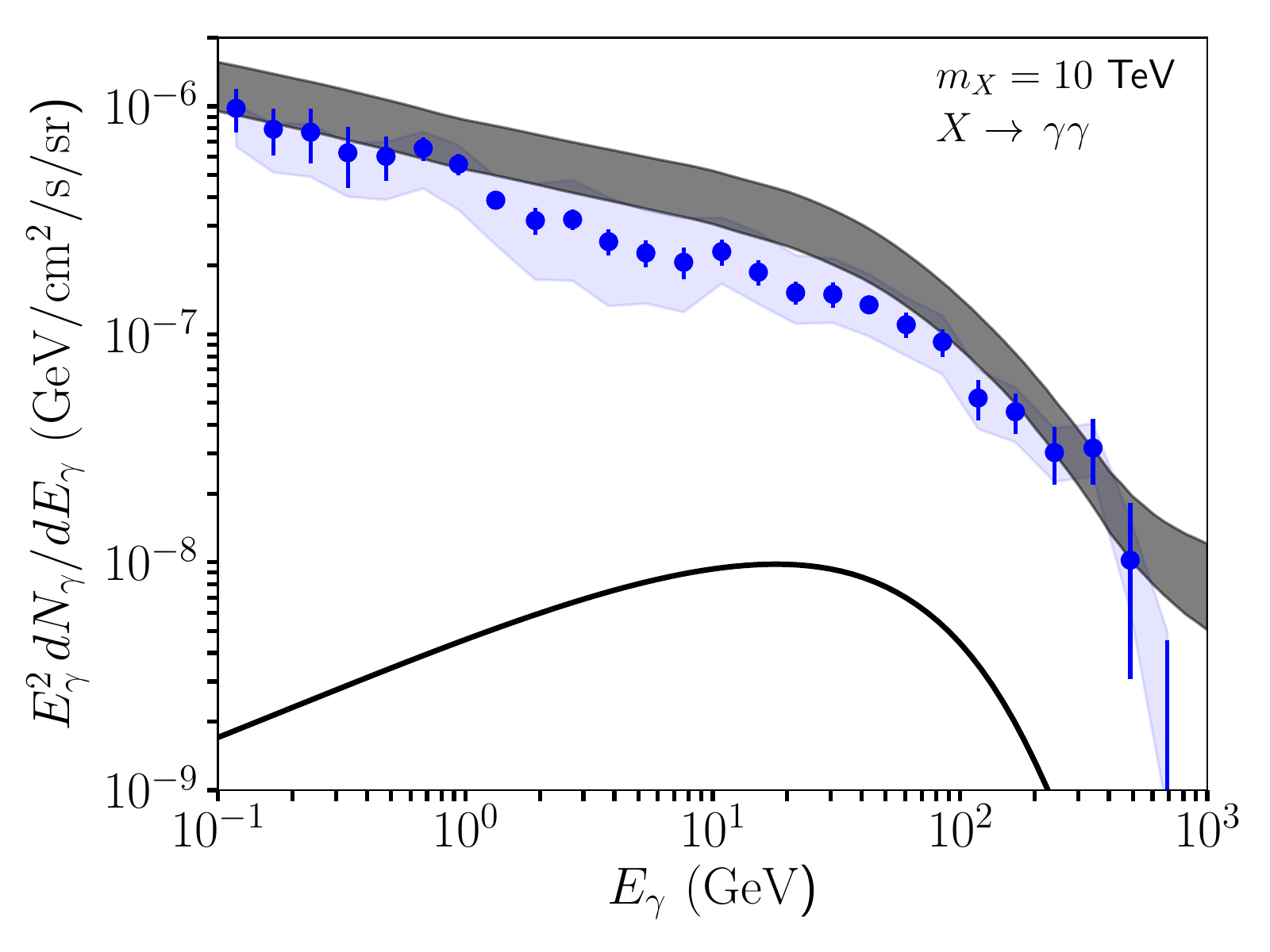} 
\includegraphics[scale=0.31]{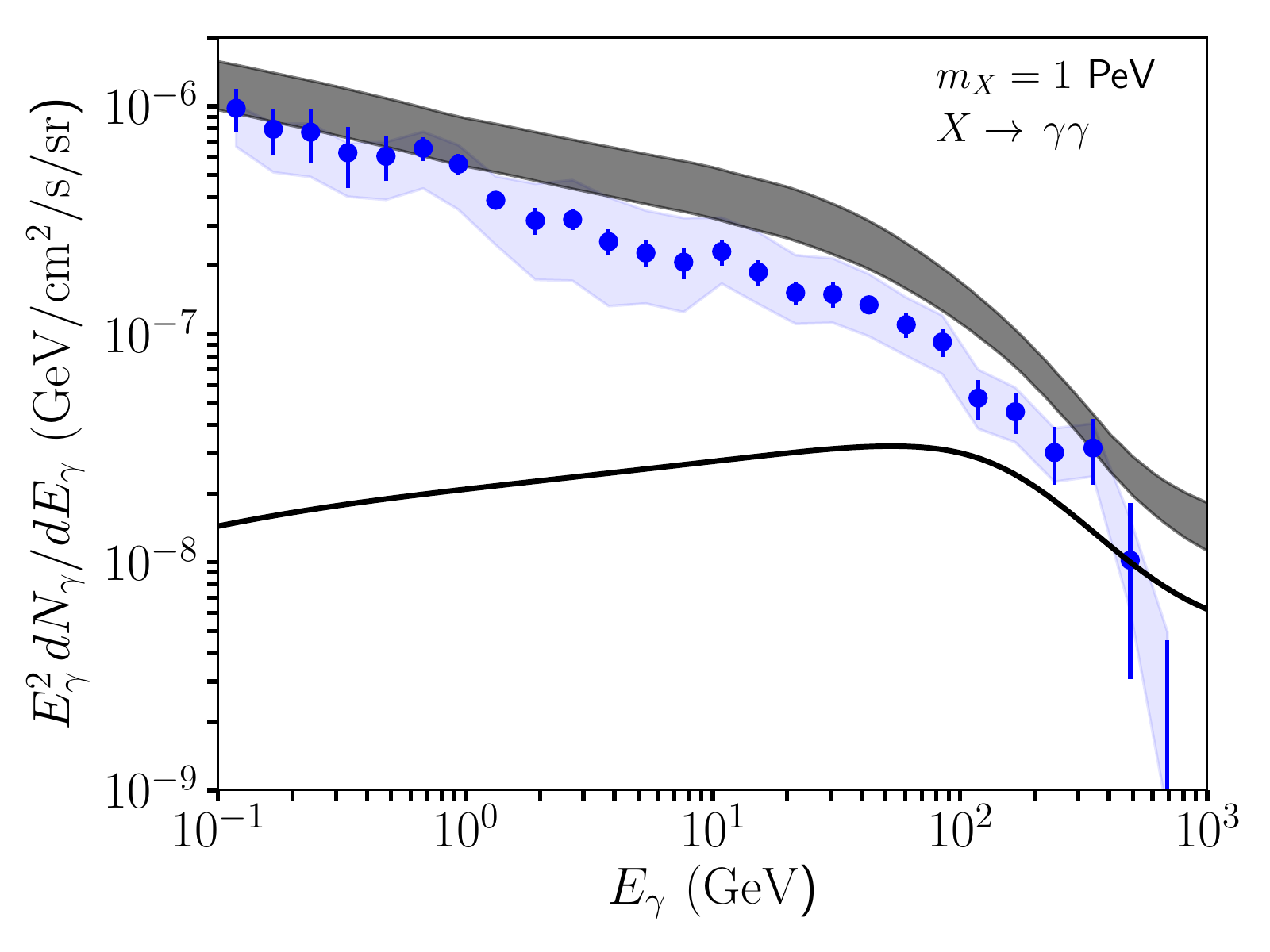} 
\includegraphics[scale=0.31]{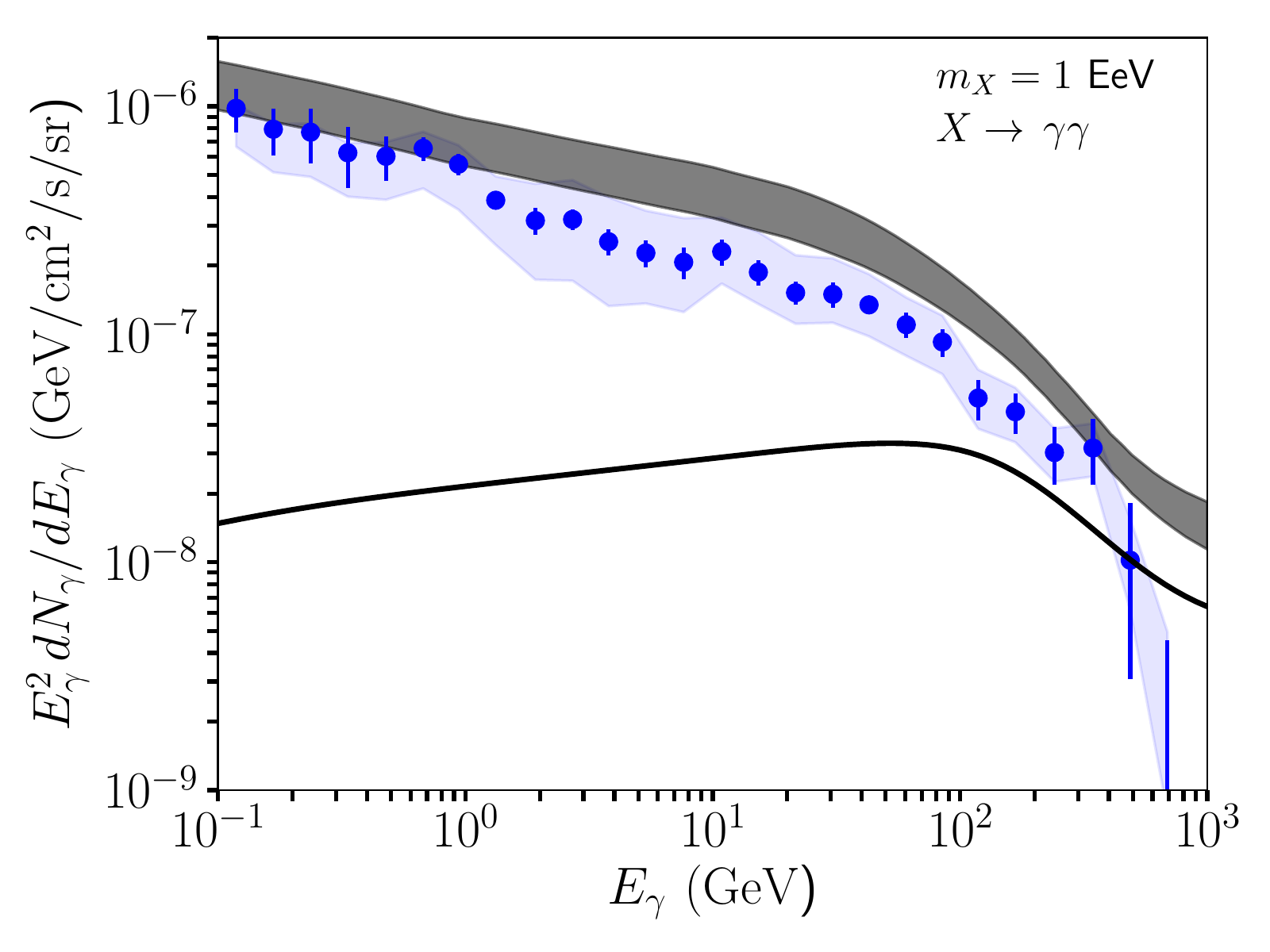} \\
\includegraphics[scale=0.31]{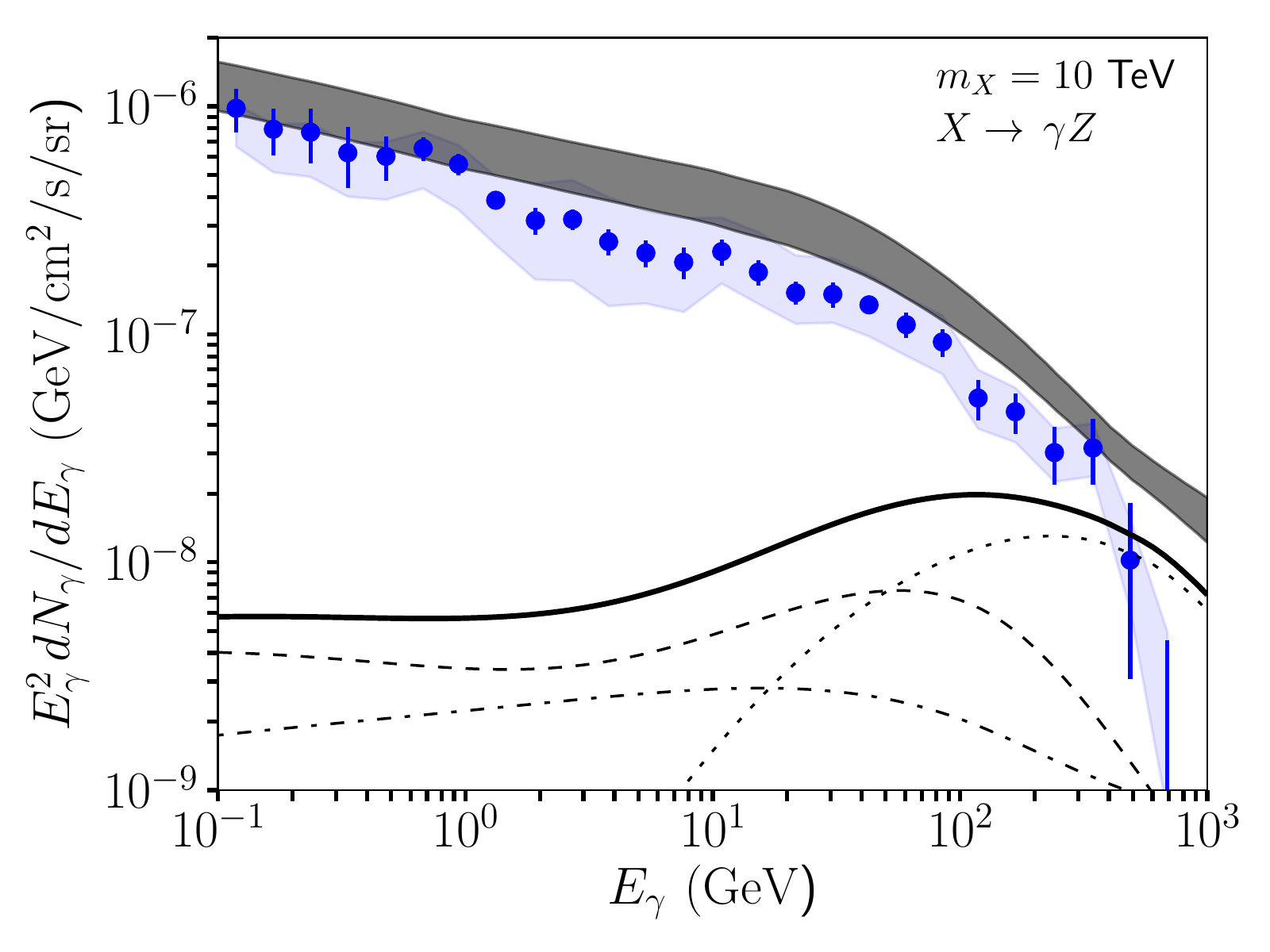} 
\includegraphics[scale=0.31]{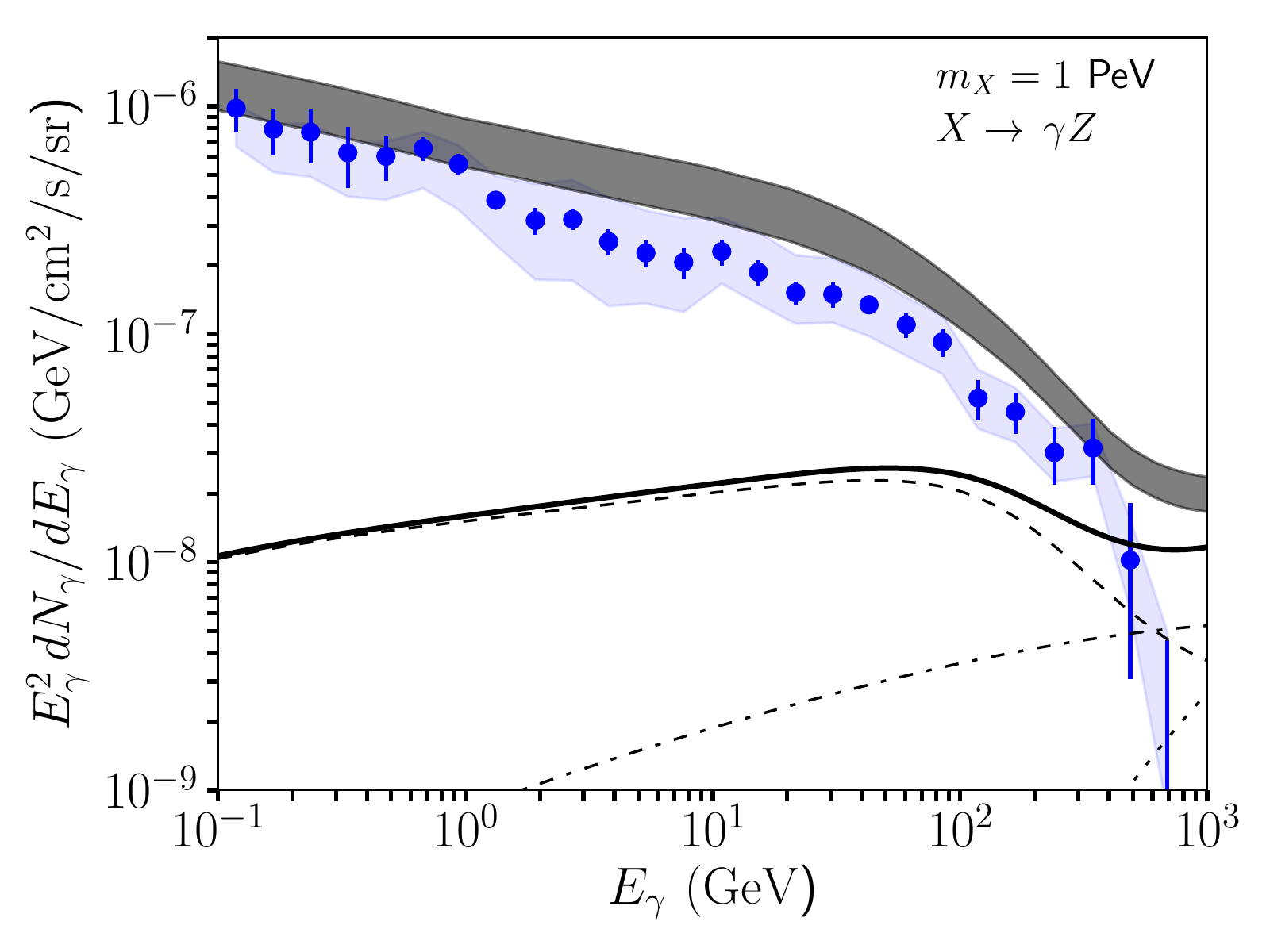} 
\includegraphics[scale=0.31]{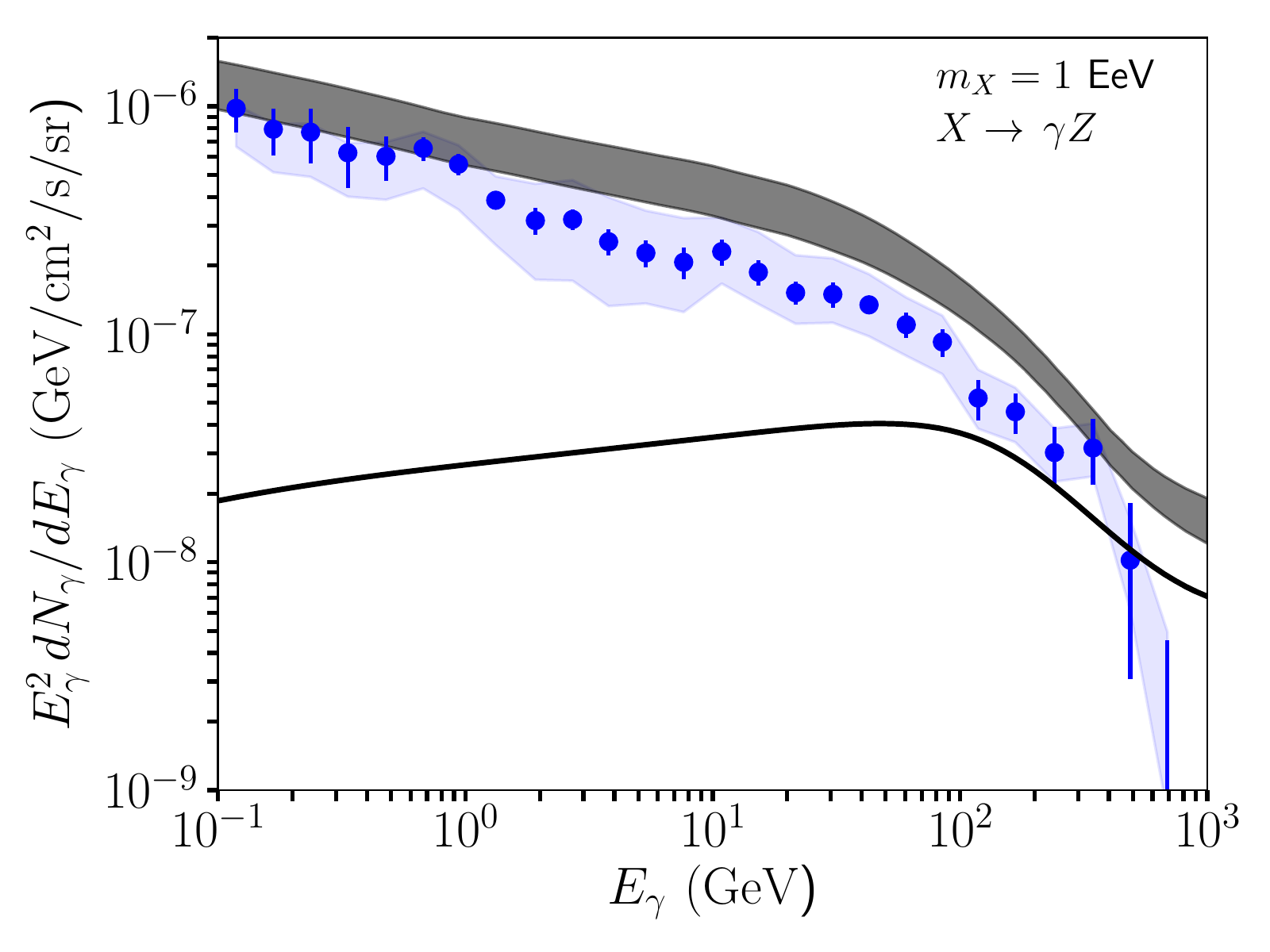} \\
\includegraphics[scale=0.31]{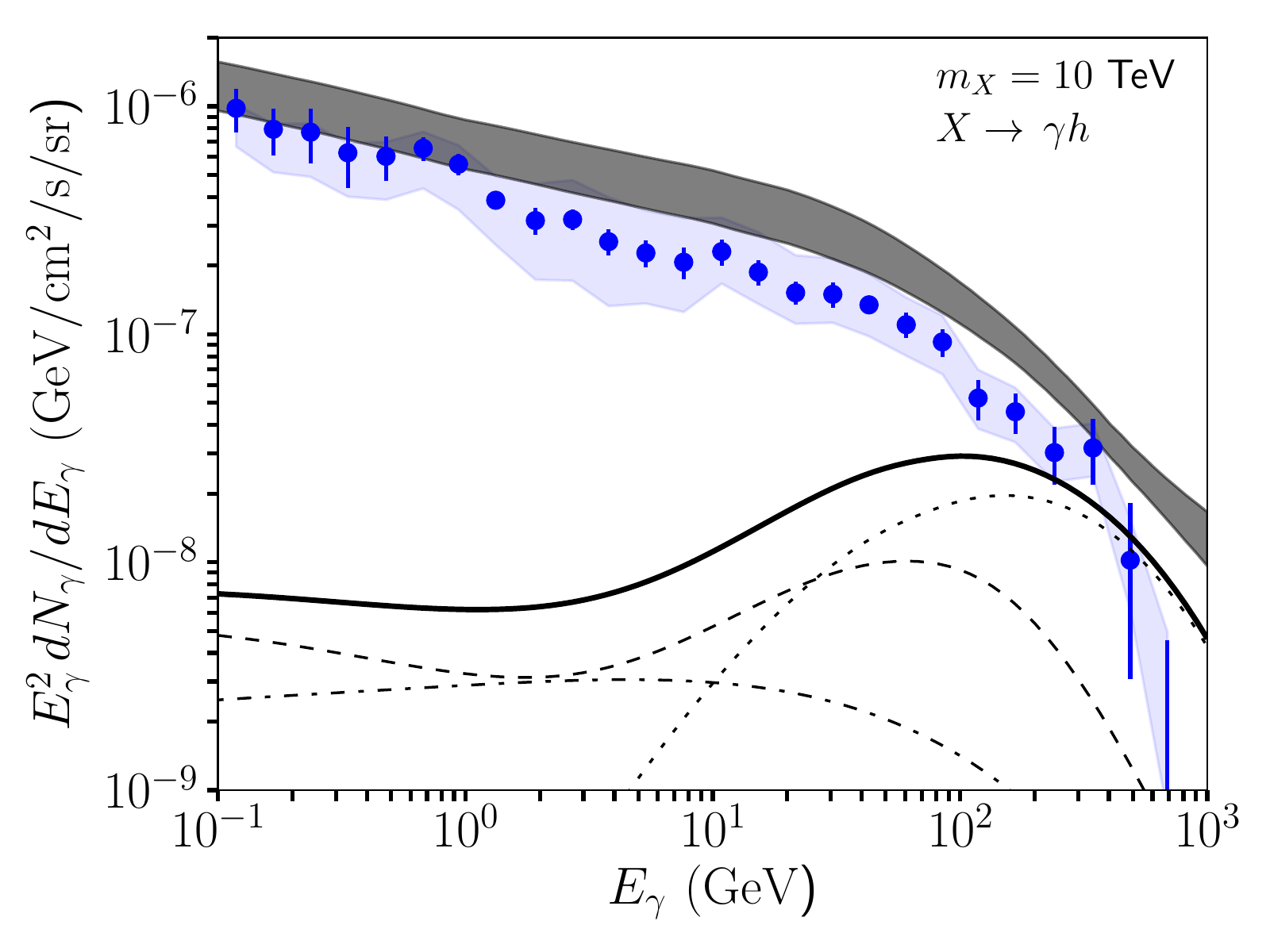} 
\includegraphics[scale=0.31]{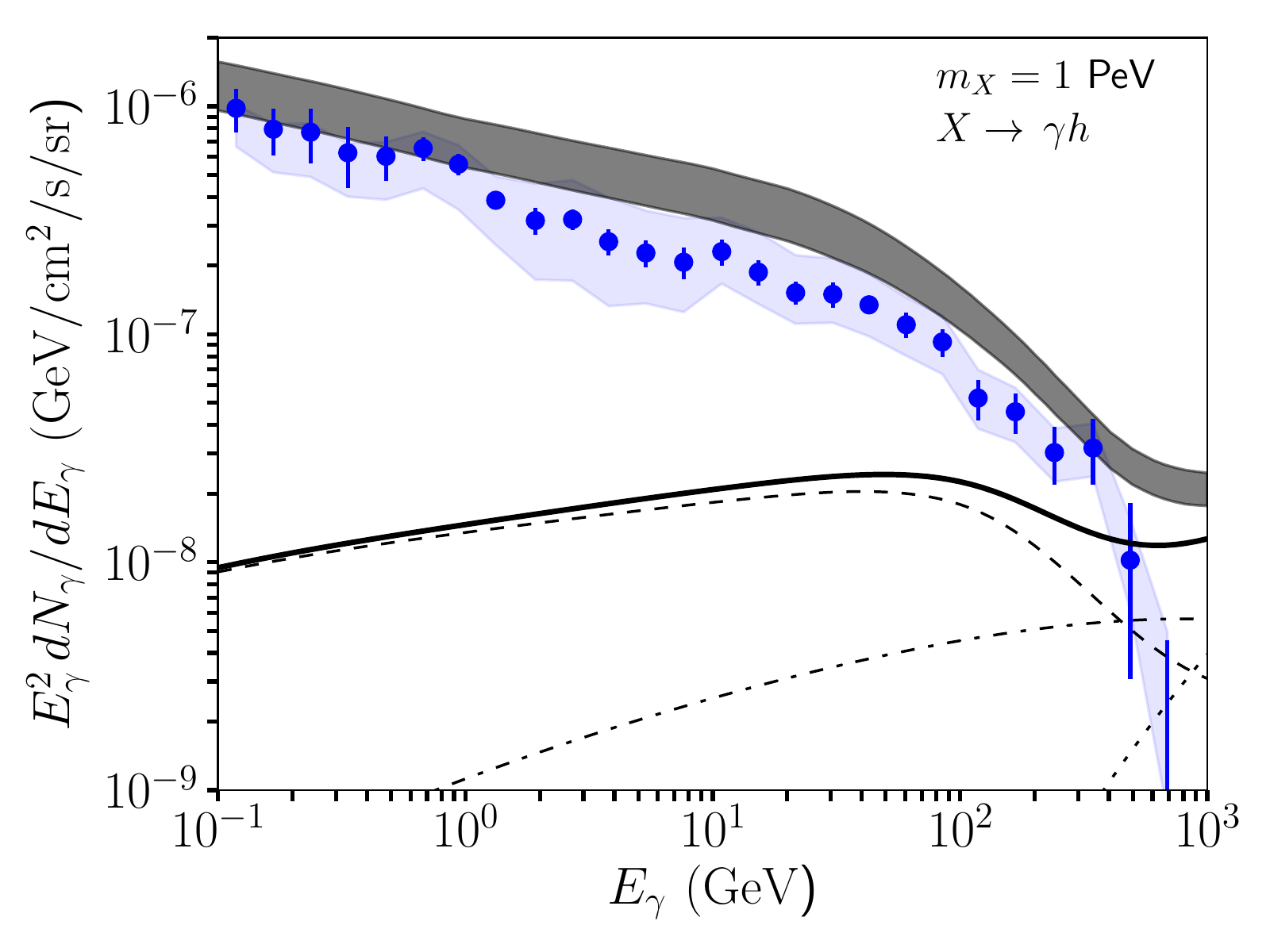} 
\includegraphics[scale=0.31]{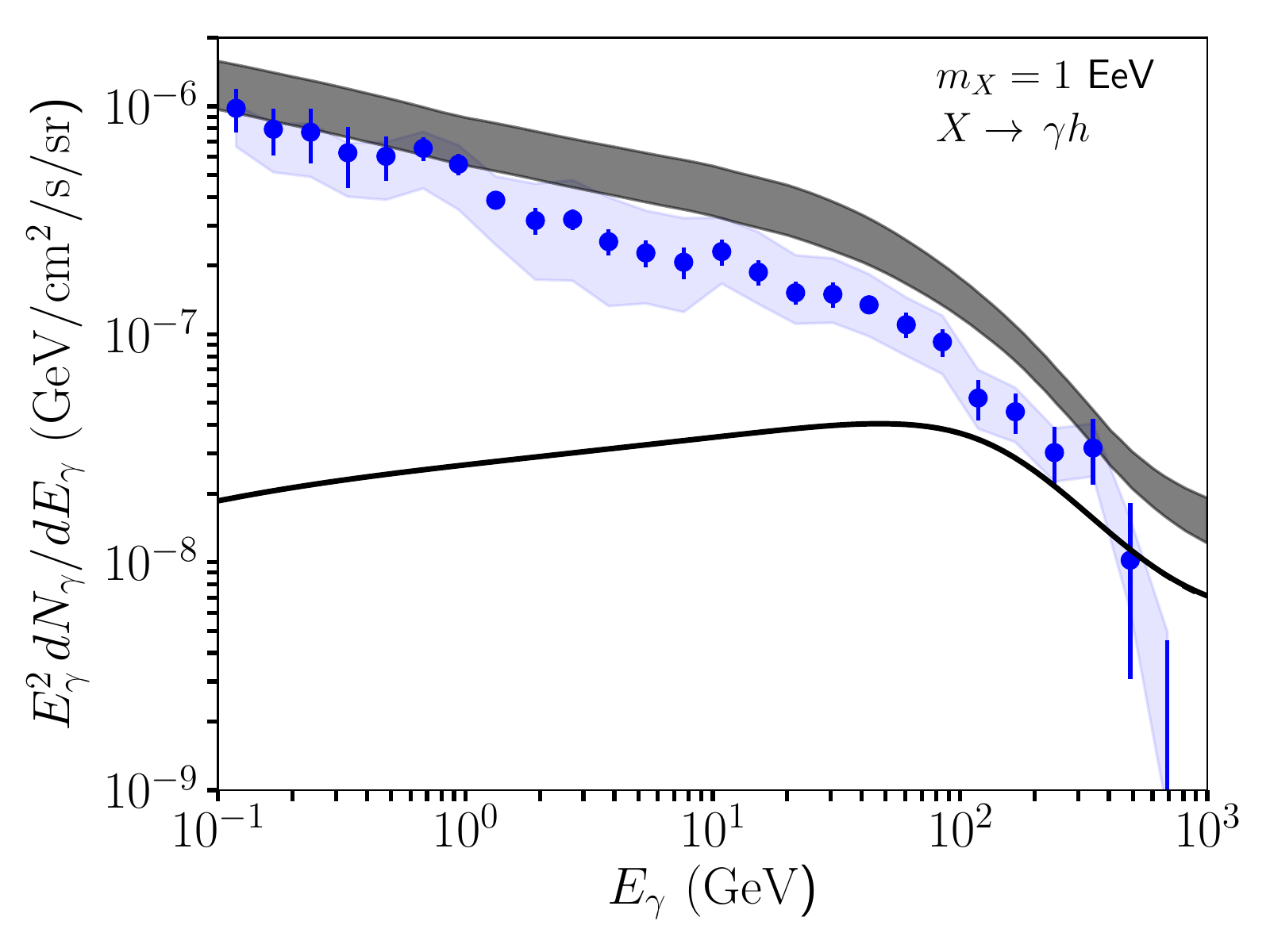} 
\caption{As in Fig.~\ref{bb}, but for the case of decays to $\gamma \gamma$, $\gamma Z$ or $\gamma h$.}
\label{GammaGamma}
\end{figure}

Following the approach described in the previous section, we have calculated the total gamma-ray spectrum from dark matter decay taking place in the Galaxy and throughout the universe, including the effects of pair production and inverse Compton scattering. In Figs.~\ref{bb}-\ref{GammaGamma}, we show these results for a variety of dark matter decay channels and masses. In each case, we plot the spectrum for a dark matter lifetime that corresponds to the 95\% confidence level upper limit (assuming uncorrelated errors; see in Sec.~\ref{igrbsec}). 

In Figs.~\ref{bb}-\ref{tt} we show the contribution to the gamma-ray spectrum from dark matter decaying to quark pairs. In the case of relatively light dark matter particles ($m_X \lsim 300$ GeV), the spectrum is dominated by prompt emission, while heavier dark matter particles generate large contributions through both Galactic inverse Compton scattering and cosmological cascades. For very high masses ($m_X\gsim 1$ PeV), the observed emission is almost entirely dominated by cosmological cascades. The spectrum that results from dark matter decays to light quarks peaks at slightly higher energies than in the case of heavy quarks, with spectra from decays to $s\bar{s}$ and $c\bar{c}$ (not shown) falling in between these extremes.

As expected, the gamma-ray spectrum is somewhat different in the case of dark matter particles that decay to charged leptons (see Figs.~\ref{tautau}-\ref{ee}). In the case of decays to $\tau^+ \tau^-$, the large branching fraction to final states containing energetic pions leads to a spectrum that peaks at rather high energies, often within an order of magnitude of the dark matter particle's mass. Decays to $\mu^+ \mu^-$ and $e^+ e^-$ final states produce a gamma-ray spectrum that is dominated by a combination of Inverse Compton emission from the Galactic Halo and cosmological cascades~\cite{Belikov:2009cx,Profumo:2009uf}. In Figs.~\ref{WWZZ} and~\ref{hhhZ}, we show the gamma-ray spectra that result from dark matter annihilating to bosonic final states, $W^+ W^-$, $ZZ$, $hh$ or $hZ$. These spectra generally resemble those predicted in the cases of decay to quark pairs.

Lastly, we consider in Fig.~\ref{GammaGamma} the case of dark matter that decays to a two-body final state that includes one or two photons, leading to the presence of a mono-energetic gamma-ray line. For dark matter particles with a mass below $\sim$1 TeV, the decays produce a line feature in an energy range that is directly measurable by Fermi. In this case, the analysis procedure followed here is far from optimal. In particular, such an analysis would ideally be conducted with energy bins that are comparable in size or smaller than the energy resolution of the instrument (see Ref.~\cite{Ackermann:2015lka}). With this in mind, we show results only for the case of heavy dark matter particles, which are constrained in our analysis through the emission produced through cosmological cascades. We note that our ability to constrain such lines is suppressed for masses below several tens of TeV due to the increasingly transparent nature of the universe to gamma rays at these energies.

\begin{figure}[h]
\includegraphics[scale=0.47]{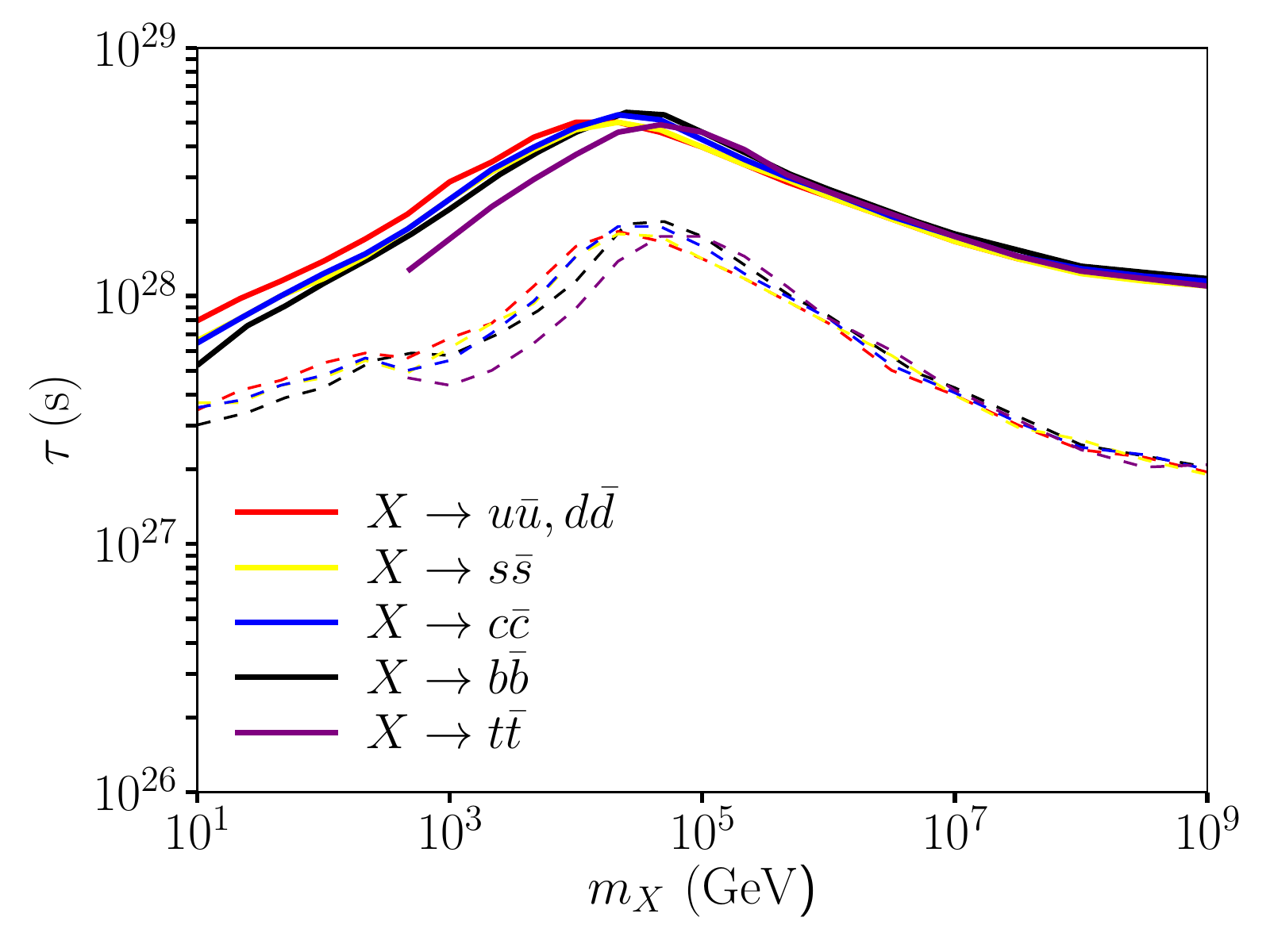} 
\includegraphics[scale=0.47]{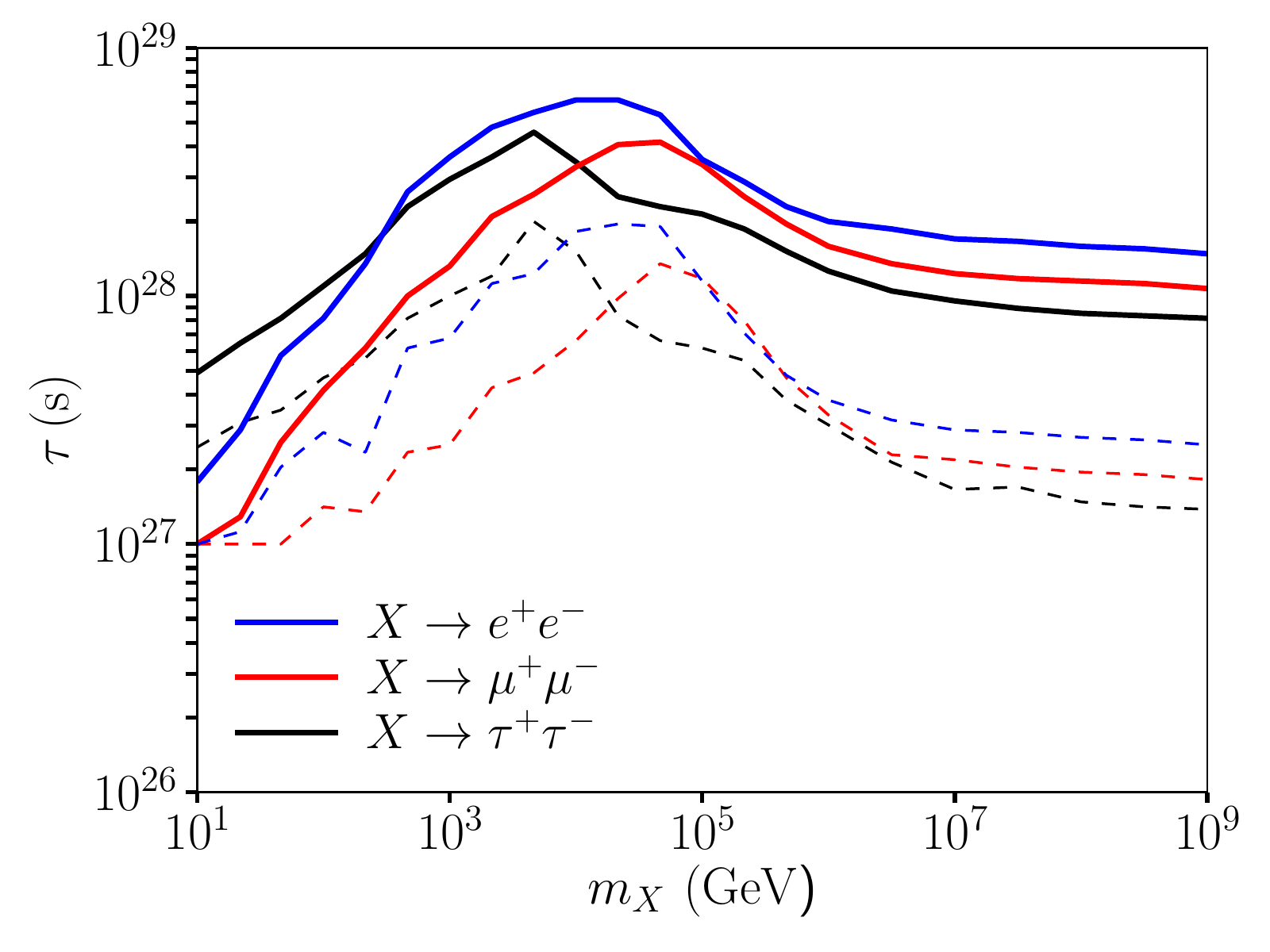} \\
\includegraphics[scale=0.47]{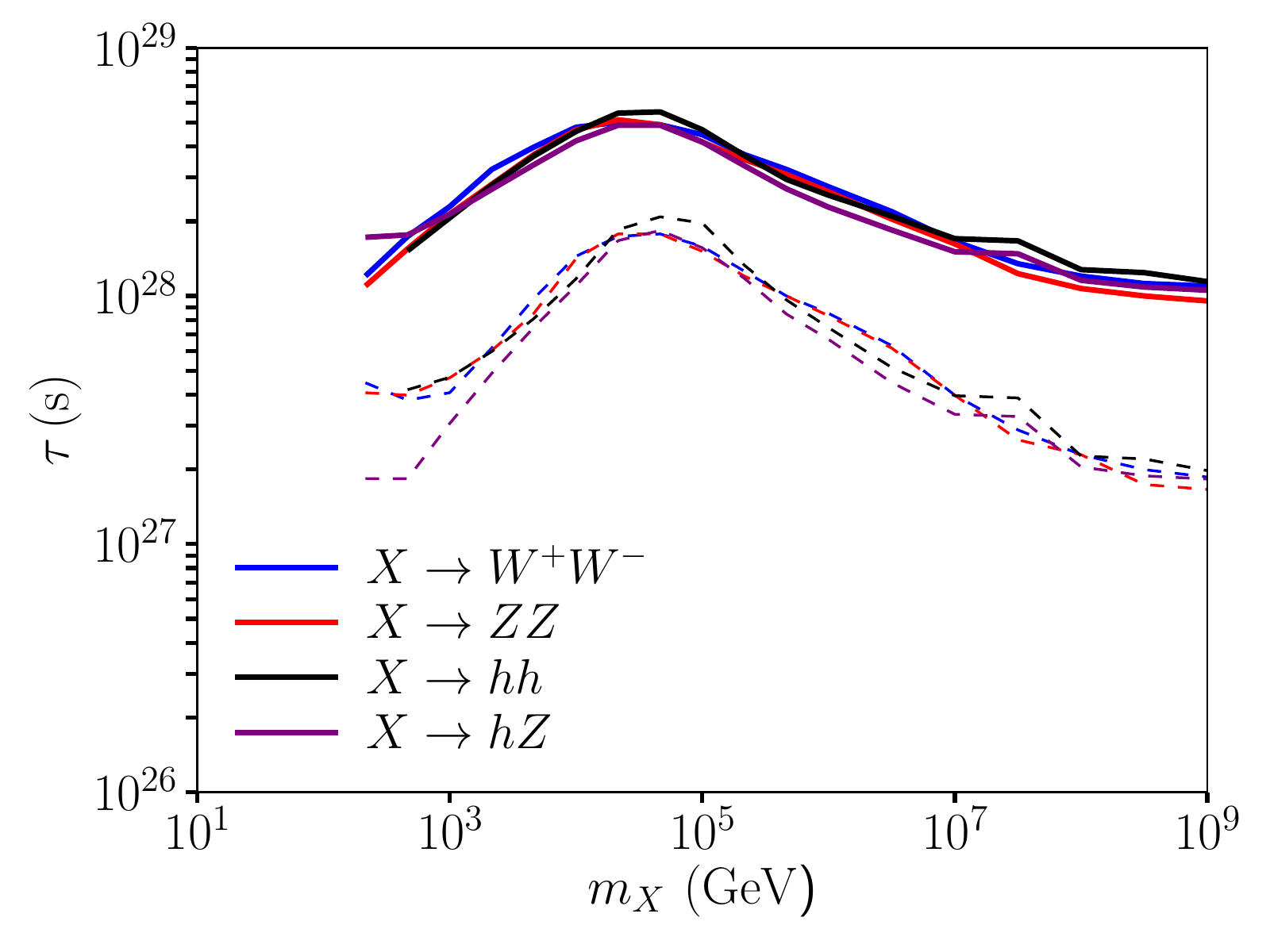}   
\includegraphics[scale=0.47]{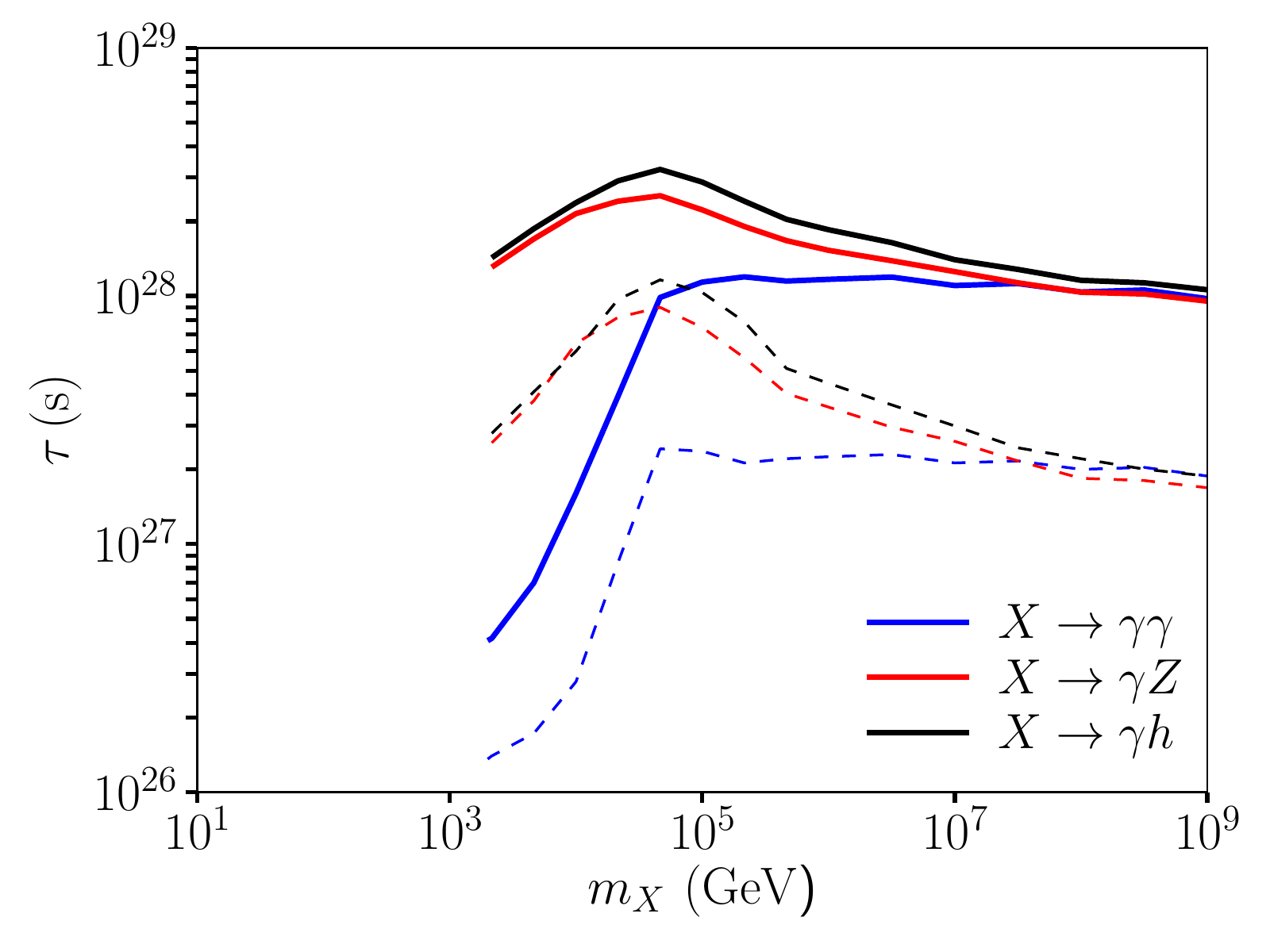}   
\caption{Lower limits on the dark matter lifetime (95\% confidence level), for a range of decay channels and masses. The solid curves treat the systematic errors (shown as a blue band around the error bars in the preceding figures) as entirely independent and uncorrelated. At the other extreme, the dashed curve has been derived assuming that the systematic errors are fully correlated, moving upward or downward together in unison. Given that the value of the total $\chi^2$ is unrealistically low in the later case ($\chi^2 = 9.4$ or less, over $26-1$ degrees-of-freedom), we consider the solid curves to represent our primary results.}
\label{limits}
\end{figure}

In Fig.~\ref{limits} we present our constraints on the dark matter's lifetime, which constitute the main results of this study. Across a wide range of masses and decay channels, we find that we can exclude lifetimes shorter than $\tau_X \sim (1-5)\times 10^{28}$ s. This range is reasonably well motivated within the context of decays that might be induced by physics at a high-scale. For example, a process that corresponds to a dimension-6 GUT scale ($M_{\rm GUT} \sim 2 \times 10^{16}$ GeV) or Planck scale ($M_{\rm Pl} \sim 2 \times 10^{18}$ GeV) operator would lead to a lifetime on the order of $\tau_X \sim M^4_{\rm GUT}/m^5_X \sim 10^{28} \, {\rm s} \times (400 \, {\rm GeV}/m_X)^5$ or $\tau_X \sim M^4_{\rm Pl}/m^5_X \sim 10^{28} \, {\rm s} \times (16 \, {\rm TeV}/m_X)^5$.

Traditionally, studies of dark matter have often focused on candidates with masses in the range of $\sim$MeV to $\sim$100 TeV. For the case of an annihilating thermal relic,  this range is particularly well motivated~\cite{Steigman:2012nb,Kolb:1990vq,Griest:1989wd} (see, however, Refs.~\cite{Gelmini:2006pq,Gelmini:2006pw,Berlin:2016gtr,Berlin:2016vnh,Davoudiasl:2015vba,Merle:2015oja,Merle:2013wta,Kane:2015jia}). In contrast, there is no clear mass range that is particularly favored by theoretical considerations in the case of decaying dark matter, so we instead chose to consider candidates with a wide range of masses, extending from 10 GeV (corresponding to roughly the minimum mass decaying particle detectable by Fermi) to 1 EeV. We also note that since electromagnetic cascades dominate the emission in the case of a very heavy decaying particle, the constraints on the dark matter's lifetime are approximately independent of mass at $m_X \gsim 0.1-1$ EeV. The limits presented here can thus be applied to decaying particles with masses well above 1 EeV.  Lastly, we point out that the results presented in this section can be easily rescaled for the case of multicomponent dark matter. If one considers an unstable particle species that makes up a fraction of the dark matter, $f_{\rm dm}$, the constraints presented in Fig.~\ref{limits} will be simply scaled downward by the inverse of this factor.



\section{Comparisons With Other Constraints}
\label{comparison}

A wide range of astrophysical observations have been used to place constraints on decaying dark matter. In this section, we compare the results of this study to the constraints derived by other groups, using various sets of data. We begin by comparing our results to other analyses of diffuse gamma-ray data. The most similar work in this respect is that of Cohen {\it et al.}~\cite{Cohen:2016uyg}, which presents an analysis of the diffuse gamma-ray emission observed by Fermi within $45^{\circ}$ of the Galactic Center, excluding the region within $20^{\circ}$ of the Galactic Plane. The constraints presented here are generally slightly stronger (within a factor of a few) than those of Cohen {\it et al.} (for example, compare our Fig.~\ref{bb} to Fig.~1 of Ref.~\cite{Cohen:2016uyg}), a fact that we attribute primarily to our use of multi-wavelength information to restrict the astrophysical contributions to the IGRB. We also note that by adopting the entire high-latitude sky as our region-of-interest, the results of our analysis are more robust to uncertainties associated with diffuse Galactic backgrounds, the Milky Way's dark matter halo profile, and features of Galactic cosmic-ray transport. Our constraints are also generally more stringent than those presented in Refs.~\cite{Ando:2015qda} and~\cite{Liu:2016ngs}.

If the dark matter decays to final states that produce a mono-energetic spectral feature ($\gamma \gamma$, $\gamma Z$, $\gamma h$), the analysis procedure employed here is far from optimal. In Ref.~\cite{Ackermann:2015lka} the Fermi Collaboration derived constraints on the dark matter lifetime to gamma-ray lines at a level of approximately $\tau_X \gsim 10^{29}$. Our analysis is only competitive for such final states if the spectral feature in question falls above the energy range measured by Fermi (see Fig.~\ref{GammaGamma} and the lower right frame of Fig.~\ref{limits}). We also note that for extremely heavy dark matter particles, $m_X\gsim 1$ EeV, searches for ultra high-energy photons using extensive air shower experiments can provide constraints that are even more stringent than those presented here~\cite{Kalashev:2016cre}.

\begin{figure}[t]
\includegraphics[scale=0.47]{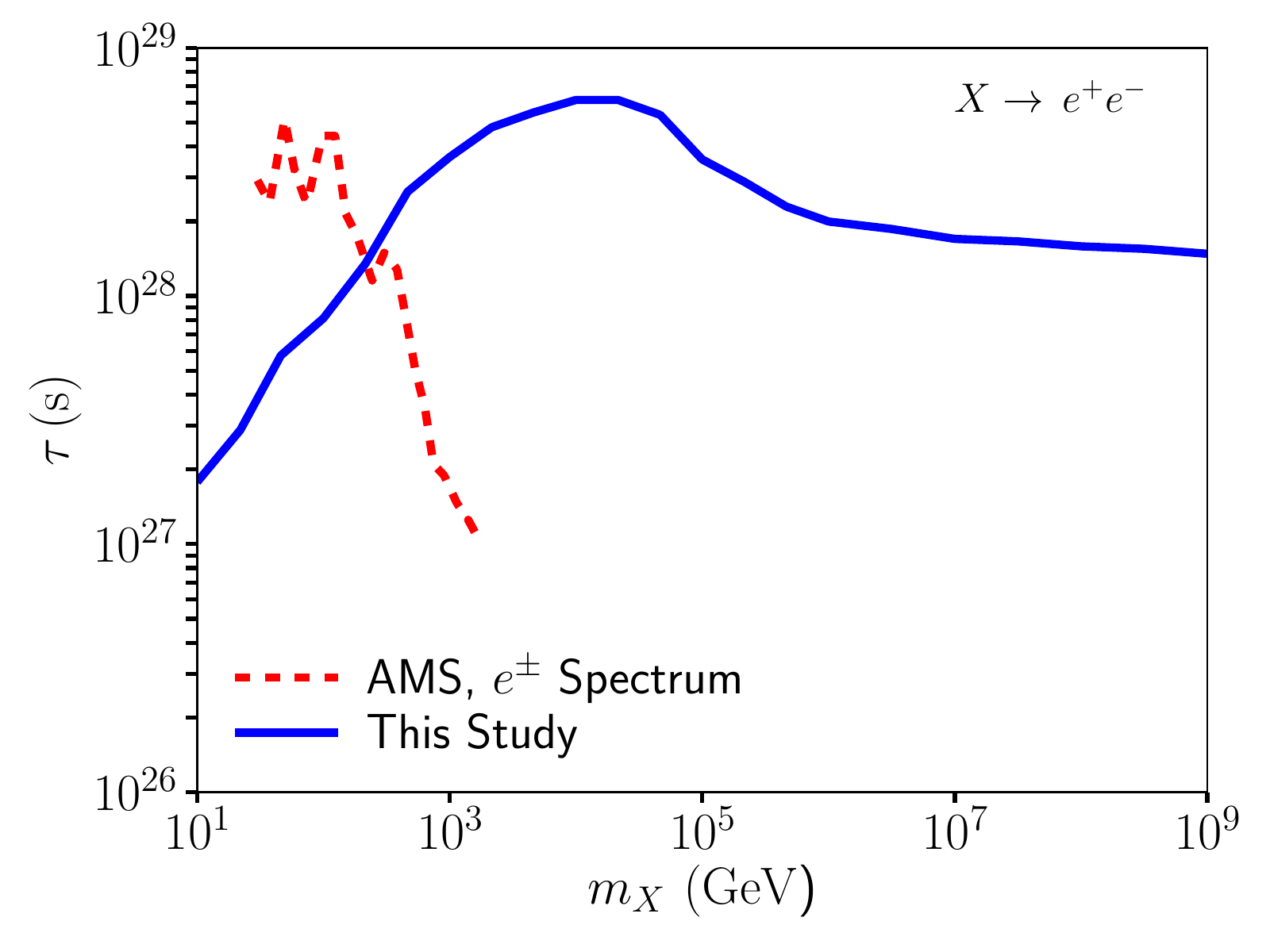} 
\includegraphics[scale=0.47]{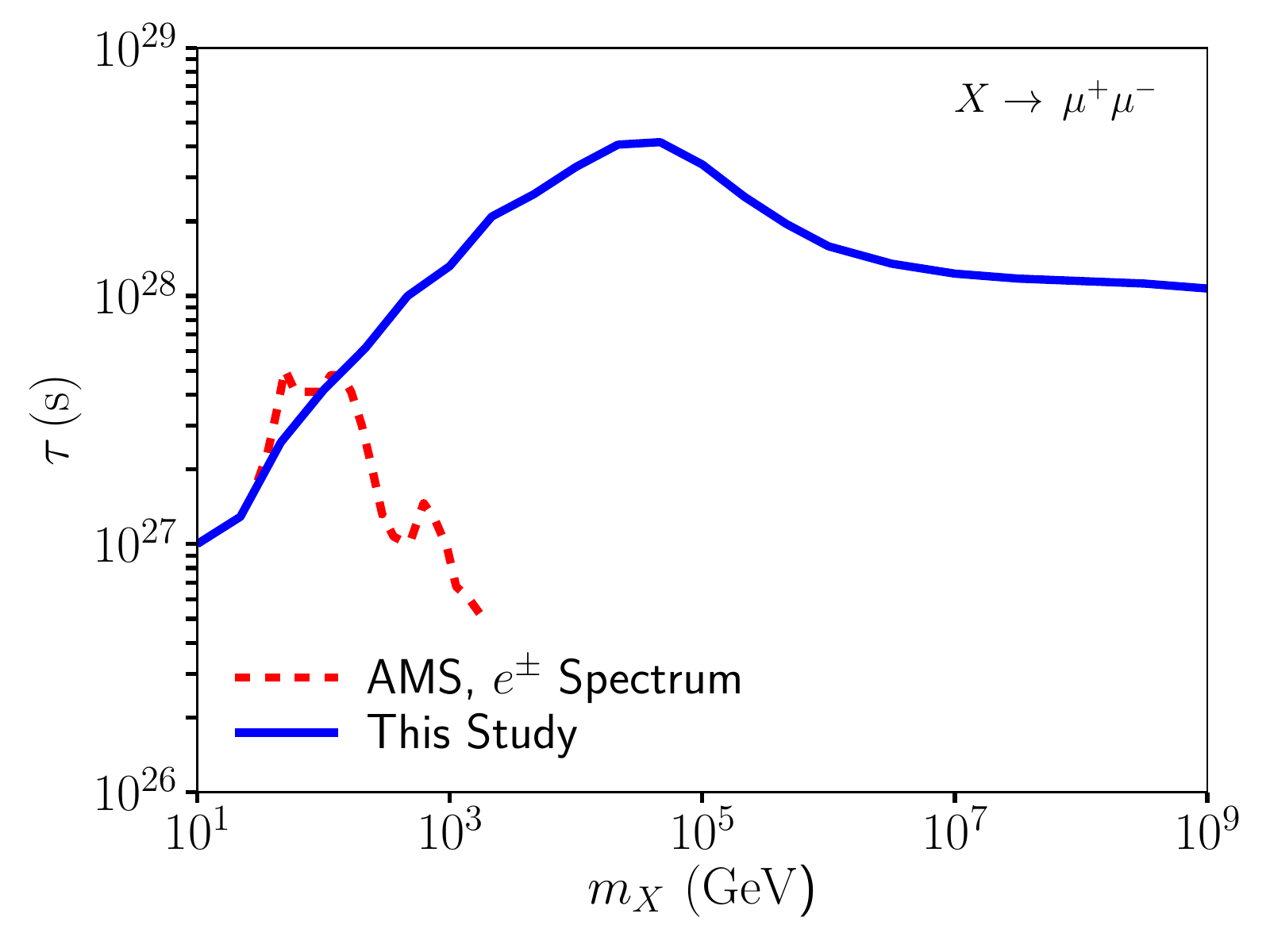} 
\caption{Constraints on the dark matter lifetime derived in this study for the case of decays to $e^+ e^-$ (left) or $ \mu^+ \mu^-$ (right) compared to those derived on an analysis of the cosmic-ray electron spectrum as measured by AMS~\cite{Ibarra:2013zia}.}
\label{limitAMS}
\end{figure}  

For dark matter particles which decay or annihilate to charged leptons, strong constraints can in some cases be derived from the cosmic-ray $e^{\pm}$ spectrum, as measured by the AMS experiment~\cite{Bergstrom:2013jra,Ibarra:2013zia}. In Fig.~\ref{limitAMS}, we compare the limits derived in this study to those based on the AMS $e^{\pm}$ spectrum~\cite{Ibarra:2013zia}. In the case of decays directly to $e^+ e^-$ or $\mu^+ \mu^-$, the cosmic-ray based constraint is competitive or more stringent for masses less than a few hundred GeV. For all other decay modes we have considered here, the constraints derived from the IGRB are more stringent than those based on the AMS $e^{\pm}$ spectrum.


\section{Summary and Conclusion}
\label{conclusion}

In this study, we have used the spectrum and intensity of the isotropic gamma-ray background (IGRB) as measured by the Fermi Collaboration~\cite{Ackermann:2014usa} to constrain any contribution from decaying dark matter particles, and to place limits on the dark matter's lifetime (see Fig.~\ref{limits}). We have included in our calculations contributions from both prompt photon production, and from the inverse Compton scattering of electrons and positions. We have also included the effects of pair production through scattering with cosmological radiation fields, and have calculated the full evolution of the electromagnetic cascades that result from such interactions. We also make use of recent multi-wavelength studies~\cite{hooper2016radio,Linden:2016fdd} to constrain the contributions to the IGRB from various astrophysical source classes, enabling us to more strongly restrict the presence of any contribution from decaying dark matter particles. 

The results described in this paper generally restrict the lifetime of the dark matter to be greater than $\tau_X \sim(1-5) \times 10^{28}$ s over a wide range of masses (10 GeV to 1 EeV and above) and decay channels. Over much of this parameter space, these limits are the most stringent presented to date and are quite robust to uncertainties associated with diffuse Galactic backgrounds, the Milky Way's dark matter halo profile, and cosmic-ray diffusion parameters.

\bigskip

\textbf{Acknowledgments.} We would like to thank Keith Bechtol, Markus Ackermann and Tim Cohen for helpful discussions. This manuscript has been authored by Fermi Research Alliance, LLC under Contract No. DE-AC02-07CH11359 with the U.S. Department of Energy, Office of Science, Office of High Energy Physics. The United States Government retains and the publisher, by accepting the article for publication, acknowledges that the United States Government retains a non-exclusive, paid-up, irrevocable, world-wide license to publish or reproduce the published form of this manuscript, or allow others to do so, for United States Government purposes.

\bibliography{decayigrb}
 \bibliographystyle{JHEP}

\end{document}